\documentclass[aps, prb, twocolumn, nofootinbib, superscriptaddress,10pt]{revtex4-2}
\usepackage{amsmath,amsthm,amsfonts,amssymb,amscd,bbold,bm}
\usepackage{graphicx}
\usepackage{dcolumn}
\usepackage{epstopdf}
\usepackage{graphicx}
\usepackage{lipsum}
\usepackage{pgf}
\usepackage{float}
\usepackage{hyperref}
\usepackage{tabularx}
\usepackage{subfigure}
\usepackage{physics}
\usepackage{float}

\begin{document}
%====================
%====================
\title{Dephasing of planar Ge hole spin qubits due to \textit{1/f} charge noise}

\author{Zhanning Wang}
\affiliation{School of Physics, The University of New South Wales, Sydney 2052, Australia}
\author{Sina Gholizadeh}
\affiliation{School of Physics, The University of New South Wales, Sydney 2052, Australia}
\affiliation{Australian Research Council Centre of Excellence in Low-Energy Electronics Technologies, UNSW Node, The University of New South Wales, Sydney 2052, Australia}
\author{Xuedong Hu}
\affiliation{Department of Physics, University at Buffalo, SUNY, Buffalo, NY 14260-1500}
\author{S.~Das Sarma}
\affiliation{Condensed Matter Theory Center and Joint Quantum Institute, Department of Physics, University of Maryland, College Park, Maryland 20742-4111 USA}
\author{Dimitrie Culcer}
\affiliation{School of Physics, The University of New South Wales, Sydney 2052, Australia}
\affiliation{Australian Research Council Centre of Excellence in Low-Energy Electronics Technologies, UNSW Node, The University of New South Wales, Sydney 2052, Australia}
\date{\today}
%====================
%====================

%====================
%====================
\begin{abstract}
Hole spin qubits in Ge, investigated for all-electrical spin manipulation because of their large spin-orbit coupling, are exposed to charge noise, leading to decoherence.
Here, we employ a model of $1/f$ noise due to individual fluctuators and determine the dephasing time $T_2^*$ as a function of qubit properties.
$T_2^*$ decreases with increasing magnetic field and is an order of magnitude longer for out-of-plane fields than for in-plane fields for the same Zeeman energy.
$T_2^*$ shows little variation as a function of the top gate field and is a complex function of the dot radius.
Our results should help experiments enhance coherence in hole qubit architectures.
\end{abstract}
\maketitle
%====================
%====================

%====================
%====================
\section{Introduction}
\label{S1-Introduction}
Hole spin qubits in Ge are excellent candidates for scalable, all-electrical quantum computing platforms because of their strong spin-orbit coupling and small effective mass \cite{Lada2018, Watzinger2018, Hendrickx2020N, Hendrickx2020NC, Hendrickx2021, Froning2021NN, Chatterjee2021, Jirovec2021NM, Scappucci2021, Ke2022, Yinan2023, Tidjani2023, Hartmann2023, Ivlev2024, Del2024, Strohbeen2024}.
Isotopic purification, which eliminates the nuclear hyperfine coupling \cite{Itoh1993, Kohei2003}, reduces an important source of decoherence, while the absence of piezoelectric interactions with phonons further improves coherence \cite{Yu2010}.
The strong hole spin-orbit interaction in Ge hole systems \cite{Rashba1988, Winkler2003, Yu2010, Winkler2008, Durnev2014, Marcellina2017}, and their anisotropic and tunable $g$-tensor \cite{Danneau2006, Ares2013PRL, Ares2013APL, Brauns2016, Watzinger2016, Voisin2016, Srinivasan2016, Miserev2017, Hung2017, Marcellina2018, Mizokuchi2018, Wei2020, Zhang2021, Liles2021, Qvist2022, Abadillo2023} make them ideal for electrical spin manipulation \cite{Bulaev2007, Kloeffel2011, Kloeffel2013, Chesi2014, Del2024}.
Experimental leaps in growth techniques leading to low disorder \cite{Dobbie2012}, and hence high mobility and low percolation density \cite{Sammak2019, Lodari2019, Lodari2021}, have enabled the fabrication of ultra-high quality quantum dots in Ge, while the low effective mass allows the formation of large dots for scaled-up technological applications \cite{Hendrickx2018, Lodari2019, Terrazos2021}.
In recent years Ge hole spin qubits have demonstrated electrical control of the underlying spin-orbit coupling \cite{Gao2020, Liu2022, Liu2023, Mena2023}, ultra-fast spin manipulation using the spin-orbit interaction \cite{Hendrickx2020N, Hendrickx2021, Froning2021PRR, Ke2022, Lawrie2023, Riggelen2024, Bosco2024}, long relaxation times \cite{Lawrie2020}, coherence sweet spots \cite{Hendrickx2024} (following a similar observation in Si \cite{Piot2022}), high-temperature operation \cite{Shimatani2020, Camenzind2021, Camenzind2022}, control and readout of multiple dots \cite{Brauns2016, Hendrickx2020N, Zhang2021, Hendrickx2021, Jirovec2021NM, Ke2022, Riggelen2022, Chien2023, Borsoi2024, Ungerer2023, Rooney2023, Sarkar2023, Xin2024, Wang2024, John2024}, as well as coupling to a superconducting resonator \cite{Li2018, Gang2020, Spethmann2024}.
This is a rapidly evolving subject with important implications for fundamental qubit research and quantum computing technologies.

Despite these advances, coherence remains a persistent problem for hole spin qubits, as their ability to couple to applied electric fields also exposes the qubits to electrical noise, such as phonons and background charge fluctuations \cite{Xuedong2006, Prechtel2016, Li2020, Bellentani2021, Froning2021, Shalak2023}.
For conduction electrons with spin-orbit coupling that is linear in electron momentum, it is known that electrical noise can only induce relaxation at the lowest order, thereby limiting its impact on qubit coherence \cite{Tahan2002, Itakura2003, Tahan2005, Tahan2014}.
However, for hole spins, symmetry dictates that a term of the form $\sigma_z(\bm{B} \cdot \bm{E})$ is permitted in a two-dimensional hole gas, where $\sigma_z$ is the out-of-plane spin component, $\bm{B}$ is the magnetic field, and $\bm{E}$ is the electric field \cite{Winkler2003}.
This indicates that in the presence of a fluctuating electric field $\bm{E}(t)$, pure dephasing is already present in a two-dimensional hole gas and could be further exacerbated in a quantum dot due to the large spin-orbit coupling, resulting in significant qubit decoherence.
In the meantime, while the Schrieffer-Wolff transformation to the qubit subspace may yield an effective spin Hamiltonian, the original multi-band Hamiltonian is dense and contains large off-diagonal matrix elements, raising questions about the convergence and validity of the lowest-order effective Hamiltonian \cite{Sarkar2023, Mena2023}.
Therefore, in this study, we aim to deepen our understanding of the coherence of a hole spin qubit in a planar Ge quantum dot by directly solving the multi-band Hamiltonian numerically with a relatively large basis for the hole.

Understanding coherence in hole spin qubits faces two main challenges.
The first is the lack of a comprehensive noise model for semiconductor spin qubits, where, to our knowledge, there are no studies of the effective magnetic noise spectrum on the spin degree of freedom that relate it to the properties of individual charge fluctuators.
Whereas the superconducting qubit literature contains abundant examples of such models \cite{Cywinski2008, Lutchyn2008,Bergli2009,Paladino2014} we wish to stress an important difference between semiconductor spin qubits and superconducting qubits.
To determine the dephasing properties of a semiconductor spin qubit, we require the change in the qubit Larmor frequency due to a charge defect potential, that is, an energy splitting affected by the spin-orbit interaction.
In order to obtain this, we need to diagonalize the quantum dot Hamiltonian together with the defect potential, then evaluate the difference between the Larmor frequencies with and without the defect.
The defect potential alters the confinement potential, which leads to a relative change in the two qubit levels via spin-orbit coupling.
This is an interesting contrast with superconducting qubits, where charge noise directly affects the gate voltage on a Cooper pair box or the voltage fluctuations on the smallest junction in the flux qubit, which is why capacitive shunting is so effective in screening out the charge noise induced decoherence in a transmon qubit \cite{Koch2007} or a noise-resistant flux qubit \cite{You2007}.
Here, on the other hand, charge noise affects the spin qubits indirectly via spin-orbit coupling.
Hence, whereas one can obtain significant insight into the effect of $1/f$ noise from the superconducting literature, including the correct procedure for constructing a model \cite{Paladino2014, Bergli2006}, its effect on semiconductor qubits must be considered from the ground up, in particular in view of the strong spin-orbit interaction characterizing hole qubits.

At present, electrical noise in a semiconductor quantum dot is typically measured through carrier energy level fluctuations, such as via transport measurements.
On the other hand, electrical noise causes spin qubit decoherence through fluctuating contributions to the confinement potential, resulting in variations in the qubit Larmor frequency: simply knowing the magnitude of the energy level fluctuations is insufficient for understanding the quantum coherence of the spin, and the required information cannot be obtained from the bare spectrum of $1/f$ voltage or current fluctuations \cite{Dutta1981, Shnirman2005, Paladino2014} -- one requires the functional form of the perturbing noise potential in real space.
In principle, fluctuations arising from random telegraph noise (RTN) due to an individual Coulomb potential can be described accurately once the confinement potential is specified -- allowing for complications inherent to modeling \cite{Shalak2023}.
On the other hand, no equivalent description exists for $1/f$ charge noise, which typically dominates in semiconductors.
This is because the traditional formulation of $1/f$ noise takes the strength of an individual fluctuator as a given parameter, then proceeds to average over an ensemble of fluctuators with a distribution of switching times.
The functional form of the perturbing noise potential in real space, relating the strength of fluctuators to their distance away from the qubit, has never been considered explicitly in the context of indirect spin magnetic noise arising from background charge fluctuations.
Considering such a functional form explicitly would enable one to determine the spin qubit energy level fluctuations induced by an ensemble of background charge fluctuators.
Hence, what is required in the field is a model describing fluctuations induced by $1/f$ noise in the qubit levels while accounting for the properties of individual fluctuators, so that coherence can be studied systematically as a function of qubit parameters.
The first steps in this direction were recently taken for Si electrons \cite{Focke2023, Shehata2023} and Ge holes \,\cite{Martinez2024}.

The second main challenge is the complexity of its spin-orbit interactions and the fact that interface roughness \cite{Martinez2022PRA}, alloy disorder \cite{JunWei2015}, inhomogeneous lattice strain \cite{Rooney2023}, and other atomistic-scale potential variations can alter the $g$-tensor of the hole spin \cite{Hendrickx2020N, Borsoi2024}, potentially making significant contributions to spin dynamics \cite{Martinez2022, Abadillo2023, Shalak2023, Martinez2024, Mauro2024}.
Decoherence mechanisms in hole systems are therefore considerably more intricate than those in spin-1/2 conduction electrons \cite{Bermeister2014}, as the spin-3/2 nature of the valence band introduces phenomena with no counterpart in conduction electron systems \cite{Culcer2006, Winkler2008, Liu2018, Abadillo2018, Cullen2021}.
On the other hand, this complexity also creates the possibility of sweet spots and sweet lines for qubit coherence in various systems \cite{Kloeffel2011, Salfi2016, Salfi2016PRL, Abadillo2018, Kloeffel2018, Marcellina2018, Terrazos2021, Wang2021, Bosco2021PRX, Bosco2022, Ke2022, Malkoc2022, Mauro2024}, adding to the intriguing prospects of hole spin qubits.

In light of these observations, the key questions concerning hole spin qubit coherence are as follows: (i) How does the dephasing time $T_2^*$ depend on qubit parameters in a realistic model of $1/f$ noise? (ii) Are the limitations to qubit coherence fundamental, or can they be mitigated through targeted engineering?
In this work, we address these questions by studying a realistic model of $1/f$ noise and determining its effect on Ge hole spin qubits.
We consider the quasi-static regime, where the noise fluctuates very slowly such that it can be treated as a constant over a single qubit operation or qubit state measurement \cite{Shnirman2002, Makhlin2004, Shnirman2005}.
We first diagonalize the full Hamiltonian for a Ge hole quantum dot affected by a single charge defect. Next, we build up the $1/f$ spectrum of qubit Zeeman splitting fluctuations based on a distribution of charge defects across the sample.
This approach enables us to directly relate the charge fluctuators that generate the $1/f$ noise to qubit and defect properties, transitioning from microscopic random telegraph noise sources from individual charge defects to the macroscopic $1/f$ power spectrum.
Specifically, we investigate the dephasing time $T_2^*$ as a function of the magnetic field magnitude $B$ and its orientation, quantum dot size, and top gate electric field.
We find that the matrix element responsible for dephasing is $\propto B$, so $T_2^*$ increases at lower magnetic fields for both in-plane and out-of-plane orientations.
However, for the same qubit Zeeman energy (in the absence of charge defects), $T_2^*$ is an order of magnitude longer in an out-of-plane magnetic field than in an in-plane magnetic field.
$T_2^*$ decreases slightly with increasing gate electric field regardless of the magnetic field orientation, although the variation as a function of gate field is very weak in the range considered ($0-2$\,MV/m), suggesting the bulk of the dephasing stems from the spin-orbit coupling inherent in the Luttinger Hamiltonian.
The dependence of $T_2^*$ on the dot radius is more complex due to the interplay between spin-orbit coupling, random disorder potential, and orbital magnetic field terms.

The outline of this paper is as follows. In Sec.\,\ref{S2-SS1-Qubit Hamiltonian} we introduce the Hamiltonian for a Ge hole quantum dot spin qubit starting from the Luttinger-Kohn Hamiltonian.
This is followed by details of the noise model, covering RTN due to single charge defects and the resulting $1/f$ noise from an ensemble of charge defects in subsection Sec.\,\ref{S2-SS2-Noise Model}.
The results, obtained by numerical diagonalizations of the full Ge hole spin qubit Hamiltonian affected by charge defects, are presented in Sec.\,\ref{S3-Results and discussions}.
Here, we study the dephasing rate $1/T_2^*$ in both out-of-plane and in-plane magnetic fields, as well as its dependence on the gate electric field and quantum dot geometries. The final section, Sec.\,\ref{S4-Conclusions and Outlooks}, summarizes the key findings of this work and provides an outlook on extending our methodologies to include interface roughness, inhomogeneous strains, and Si quantum dot hole spin qubits.
%====================
%====================

%====================
%====================
\section{Model and methodology}
\label{S2-Model and methodology}
In this section, we introduce the model Hamiltonian that defines Ge hole quantum dot spin qubits, as detailed in Sec.\,\ref{S2-SS1-Qubit Hamiltonian}, along with the numerical diagonalization method employed.
Following this, in Sec.\,\ref{S2-SS2-Noise Model}, we describe an in-plane Coulomb potential model for charge defects to explore the origins of $1/f$ noise caused by an ensemble of random telegraph noise sources and derive the dephasing rate for such an ensemble under varying electric and magnetic field conditions.

%====================
%====================
\subsection{Qubit Hamiltonian}
\label{S2-SS1-Qubit Hamiltonian}
The total Hamiltonian for a single Ge hole quantum dot qubit is given by $H = H_{\text{LK}} + H_{\text{BP}} + V + H_{\text{Z}}$, where $H_{\text{LK}}$ is the four-band Luttinger-Kohn Hamiltonian, $H_{\text{BP}}$ is the Bir-Pikus strain Hamiltonian, $V$ is the confinement potential, and $H_{\text{Z}}$ is the Zeeman Hamiltonian.
The two heavy-hole bands and two light-hole bands are denoted by $\left|\frac{3}{2}, \frac{3}{2}\right\rangle$, $\left|\frac{3}{2}, -\frac{3}{2}\right\rangle$ and $\left|\frac{3}{2}, \frac{1}{2}\right\rangle$, $\left|\frac{3}{2}, -\frac{1}{2}\right\rangle$, respectively.
The split-off band is neglected due to the large split-off gap in Ge (around $325$\,meV).
In this basis, the four-band Luttinger-Kohn Hamiltonian is given by:
\begin{equation}
H_{\text{LK}}=\mqty[
    P+Q & 0 & -L & M  \\
    0 & P+Q& M^\dag & -L^\dag \\
    L^\dag & M & P-Q & 0 \\
    M^\dag & -L & 0 & P-Q] \,,
\end{equation}
where $P=\gamma_1(p_x^2+p_y^2+p_z^2)/(2m_0)$, $Q=\gamma_2(p_x^2+p_y^2-2p_z^2)/(2m_0)$, $L=-\sqrt{3}\gamma_2\{p_-,p_z\}/m_0$, $M=-\sqrt{3}(\bar{\gamma}p_-^2-\delta p_+^2)/(2m_0)$.
The Luttinger-Kohn parameters $\gamma_1$, $\gamma_2$, and $\gamma_3$ are $13.18$, $4.24$, and $5.69$, respectively.
$m_0$ denotes the bare electron mass, $\bar{\gamma} = (\gamma_2 + \gamma_3)/2$, and $\delta = (\gamma_3 - \gamma_2)/2$.
The canonical momentum is $\bm{p} = -i\hbar\nabla + e\bm{A}$, where $\bm{A} = -\bm{r} \times \bm{B} / 2$ and $\bm{B}$ represents the magnetic field.
Additionally, we define $p_\pm = p_x \pm i p_y$, and use the anti-commutator $\{A, B\} = AB + BA$ to correctly incorporate the orbital magnetic field.
The confinement potential $V$ consists of a parabolic confinement in the $xy$-plane and an infinite well confinement along the z-axis, as well as a gate electric field along the $z$:
\begin{small}
\begin{equation}
V = \frac{1}{2}m_0(\omega_{0,x}^2 x^2 + \omega_{0,y}^2 y^2) + \begin{cases}
    eFz\,, & -L/2 < z < L/2\,, \\
    \infty\,, & \text{otherwise}
\end{cases} \,.
\end{equation} 
\end{small}
%==========
\begin{figure}[tbp]
\centering
\includegraphics[width=1\columnwidth]{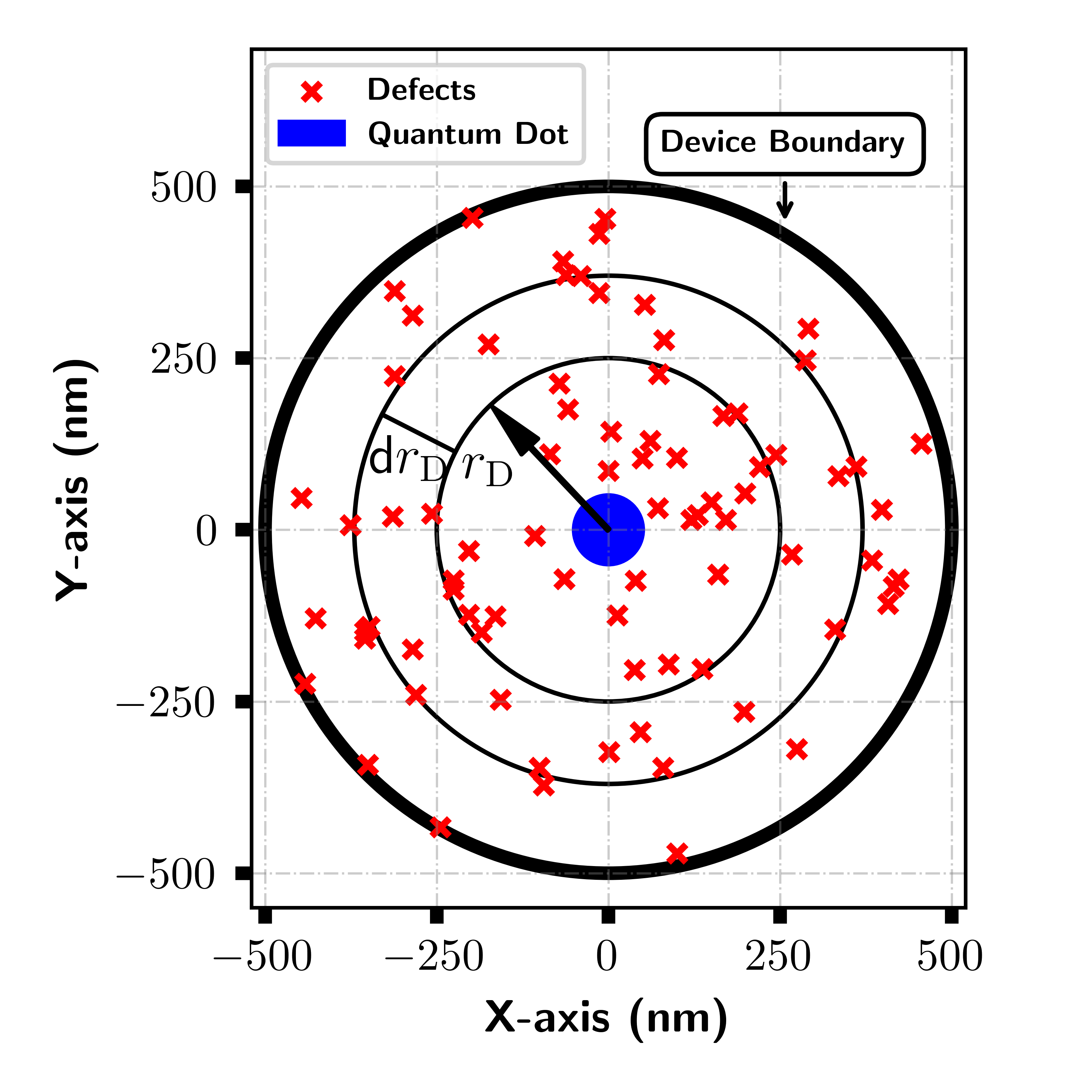}
\caption{A schematic planar view of the defect distribution across the device.
The size of the device in the xy-plane is $500$\,nm in radius, with its boundary indicated by the out-most circle.
The shaded circular at the center represents the quantum dot.
The crosses, denoting Coulomb-type charge defects, are randomly distributing over the sample.
The distance from a defect to the quantum dot center is indicated by $r_{\text{D}}$.
The differential distance between two concentric circles of defects is denoted as $\dd{r_{\text{D}}}$ (not to scale).}
\label{Fig: QuantumDots and Defects}
\end{figure}
%==========
The confinement frequencies along the $x$-axis and $y$-axis, which control the quantum dot ellipticity, are $\omega_{0,x}$ and $\omega_{0,y}$, respectively.
The applied gate electric field $F$ ranges from $0$ to $2$\,MV/m.
The Zeeman Hamiltonian is given by $H_Z = 2\mu_B(\kappa \bm{J} + q \bm{J}_3) \cdot \bm{B}$, where $\bm{J} = (J_x, J_y, J_z)$ and the anisotropic term $\bm{J}_3 = (J_x^3, J_y^3, J_z^3)$.
The constants $\kappa$ and $q$ are $3.14$ and $0.07$.
A uniaxial strain in the Ge heterostructure is modeled by the Bir-Pikus Hamiltonian, given by $H_{\text{BP}} = \operatorname{diag}[P_\epsilon + Q_\epsilon, \, P_\epsilon + Q_\epsilon, \, P_\epsilon - Q_\epsilon, \, P_\epsilon - Q_\epsilon]$, where $P_\epsilon = -a_v(\varepsilon_{xx} + \varepsilon_{yy} + \varepsilon_{zz})$ and $Q_\epsilon = -b_v(\varepsilon_{xx} + \varepsilon_{yy} - 2 \varepsilon_{zz})/2$.
The deformation potential constants are $a_v = 2$\,eV and $b_v = -2.3$\,eV for Ge.
The magnitude of uniaxial strain is estimated as $\epsilon_{xx} = \epsilon_{yy} = -0.006$, leading to $\epsilon_{zz} = 0.0042$, as reported in Ref.\,\cite{Terrazos2021}.
In this work, we do not consider shear strains, such as the $S_\epsilon$ and $L_\epsilon$ terms, which appear in the off-diagonal elements of the Bir-Pikus Hamiltonian, nor inhomogeneous strain profiles, which lead to strain gradient discussed in Refs.\,\cite{Abadillo2023, Mena2023, Mauro2024arxiv, Wang2024}.
These studies have shown that inhomogeneous strain fields can induce a linear Rashba spin-orbit coupling due to atomistic potentials, resulting in large $g$-factor modulations and enhancing electric dipole spin resonance (EDSR) Rabi oscillations by an order of magnitude.
We note that the methodologies and noise model we develop here remain qualitatively unchanged and thus can be applied under these more complex conditions.
Furthermore, since the device-dependent parameters may vary in wider ranges, further opportunities may arise to improve the coherence of a Ge planar qubit.

To obtain the energy spectrum of the quantum dot Hamiltonian $H$, we use a numerical approach introduced in Ref.\,\cite{Sarkar2023}, where the convergence and accuracy of the method are validated across different gauges and numbers of basis states.
The full Hamiltonian is projected onto the states $\Psi_{n_x, n_y, n_z}\chi$, where $\Psi_{n_x, n_y, n_z} = \phi_{n_x} \phi_{n_y} \phi_{n_z}$ consists of three wavefunctions along the $x$-, $y$-, and $z$-directions, respectively, and $\chi$ is the spinor for the $J = 3/2$ states.
The in-plane wavefunctions $\phi_{n_x}$ and $\phi_{n_y}$ are solutions to the harmonic oscillator with frequencies $\omega_{0,x}$ and $\omega_{0,y}$, while the out-of-plane wavefunction $\phi_{n_z}$ is a sinusoidal wave in an infinitely high square well symmetrically located between $-L/2$ and $L/2$.
Here, $n_x$, $n_y$, and $n_z$ label the energy levels in each spatial direction.
After numerically diagonalizing the full Hamiltonian $H$, the ground state of $H$ is denoted by $\ket{\mathbb{0}}$ with energy $E_{\mathbb{0}}$, and the first excited state is denoted by $\ket{\mathbb{1}}$ with energy $E_{\mathbb{1}}$.
These two levels define the Ge hole spin qubit.
The qubit Zeeman energy splitting is given by $\Delta_0 = E_{\mathbb{1}} - E_{\mathbb{0}}$, and the effective Hamiltonian is $H_{\text{Qubit}} = \Delta_0 \sigma_z / 2$, where $\sigma_z$ is the Pauli-Z matrix.
%====================
%====================

%====================
%====================
\subsection{Noise Model}
\label{S2-SS2-Noise Model}
The strong spin-orbit coupling, which enables ultra-fast electrical manipulation of the Ge hole spin qubit, also exposes it to electrical noise from charge defects, resulting in fluctuations in the qubit energy splitting.
Here we design a noise model based on an aggregate of random telegraphic noise sources in the form of single Coulomb type trapped charges.
Charge defects are commonly formed during the semiconductor quantum dot fabrication process, and there could be a wide variety of them.
Their spatial distribution is also an open question under active experimental explorations.
Here we focus on a planar charge defect distribution, as shown in Fig.\,\ref{Fig: QuantumDots and Defects}.
We believe such a simple two-dimensional model can catch the essence of charge fluctuations in semiconductor nanostructures: charge traps tend to form around the interface between an oxide layer under the conducting gates and the dielectric materials between the active quantum well and the oxides.
However, considering that growth-direction confinement is much stronger than in-plane direction for the planar quantum dots, in-plane electric field fluctuations should be the main source of noise affecting the qubit, and this effect is fully accounted for in our model.
The Coulomb potential of an individual defect takes the form:
\begin{equation}\label{Eq: Defect Coulomb Potential}
    U_{\text{D}}(\bm{r}_{\text{D}})= \frac{e^2}{4\pi\epsilon_0\epsilon_r} \frac{1}{\norm{\bm{r}-\bm{r}_D}} \,.
\end{equation}
Here, $\epsilon_0$ is the vacuum permittivity, and $\epsilon_r = 15.8$ is the relative permittivity in Ge.
We note that while alternative models for the potential due to a single charge defect are possible, such as a screened Thomas-Fermi potential \cite{Sarkar2023}, the choice of the unscreened two-dimensional Coulomb potential allows for the diagonalization of the total Hamiltonian with minimal computational cost, retaining all possible couplings between basis states $\Psi_{n_x, n_y, n_z}\chi$.  
This approach provides better accuracy than previous work on single RTN noise, such as Refs.\,\cite{Wang2021, Sarkar2023, Wang2024}, where only the difference between two diagonal matrix elements of the screened single charge defect potential is evaluated between $\ket{\mathbb{1}}$ and $\ket{\mathbb{0}}$.

For a single charge defect at $\bm{r}_{\text{D}}$, the fluctuation in the qubit Zeeman splitting is denoted by $\delta \varepsilon_Z(t, \bm{r}_{\text{D}})$, which can be expressed as $\delta \varepsilon_Z = \Delta_1(\bm{r}_{\text{D}}) - \Delta_0$.
Here, $\Delta_1(\bm{r}_{\text{D}})$ represents the energy gap between the two lowest levels of the qubit when affected by a single charge defect at $\bm{r}_{\text{D}}$.
To evaluate $\Delta_1(\bm{r}_{\text{D}})$, we project the Hamiltonian $\tilde{H} = H + U_{\text{D}}(\bm{r}_{\text{D}})$ onto the basis states $\ket{n_x, n_y, n_z, \chi}$, as described in Sec.\,\ref{S2-SS1-Qubit Hamiltonian}.
The energy difference between the ground state and the first excited state of $\tilde{H}$ can then be obtained by numerical diagonalization.
Repeating this procedure for an ensemble of charge defects allows us to determine the distribution of $\delta \varepsilon_Z(\bm{r}_{\text{D}})$.
An individual charge trap at $\bm{r}_{\text{D}}$ capturing and releasing a carrier generates RTN.
Such fluctuations can be modeled as a tunneling two-level system having the following form:
\begin{equation}
    U(t, \bm{r}_{\text{D}}) = (-1)^{N(t)} \times \norm{U(\bm{r}_{\text{D}})} \,.
\end{equation}
Here, $N(t)$ follows a Poisson distribution, taking values of 0 or 1. The potential switches between $\pm \norm{U(\bm{r}_{\text{D}})}$ with a switching time parameter $\tau$, which represents the average time interval for the activation of a single charge defect at $\bm{r}_{\text{D}}$.
The macroscopic behavior of this tunneling two-level system generates random telegraph noise characterized by the autocorrelation function $C(t) = \norm{U(\bm{r}_{\text{D}})}^2 e^{-t / \tau}$, resulting in a Lorentzian-type noise power spectral density \cite{Paladino2014}.
Consequently, the fluctuations in the qubit Zeeman splitting will exhibit a similar noise power spectral density due to a single random telegraph noise source, denoted as $S^{\text{RTN}}_\delta$:
\begin{equation}\label{Eq: S_RTN}
    S^{\text{RTN}}_\delta(\omega) = \frac{\delta\varepsilon_Z^2 \, \tau}{1 + \omega^2 \tau^2} \,.
\end{equation}
With our basic understanding of a source of RTN, we are ready to examine the aggregate effects of an ensemble of RTN sources from a planar charge trap distribution across the device.
To obtain the noise power spectral density for this ensemble of charge defects, we perform two averages of the single random telegraph noise power spectral density in Eq.\,\eqref{Eq: S_RTN}.
The first is a spatial average over the defect locations $\bm{r}_{\text{D}}$, and the second is an average over the switching times $\tau$ of each individual random telegraph noise source.
These two averages are assumed to be independent: the switching of the randomly distributed traps are unrelated to each other.
We begin by taking the spatial average over the defect locations $\bm{r}_{\text{D}}$ by integrating $\delta\varepsilon_Z^2(\bm{r}_{\text{D}})$ across the device, using a characteristic charge defect density $\rho(\bm{r}_{\text{D}})$ as a weighting factor:
\begin{equation}
    \expval{\delta\varepsilon_Z^2} = \int \delta\varepsilon_Z^2(\bm{r}_{\text{D}}) \rho(\bm{r}_{\text{D}}) \, \dd[2]{r_{\text{D}}} \,.
\end{equation}
%==========
\begin{figure}[tbp]
\centering
\includegraphics[width=1\columnwidth]{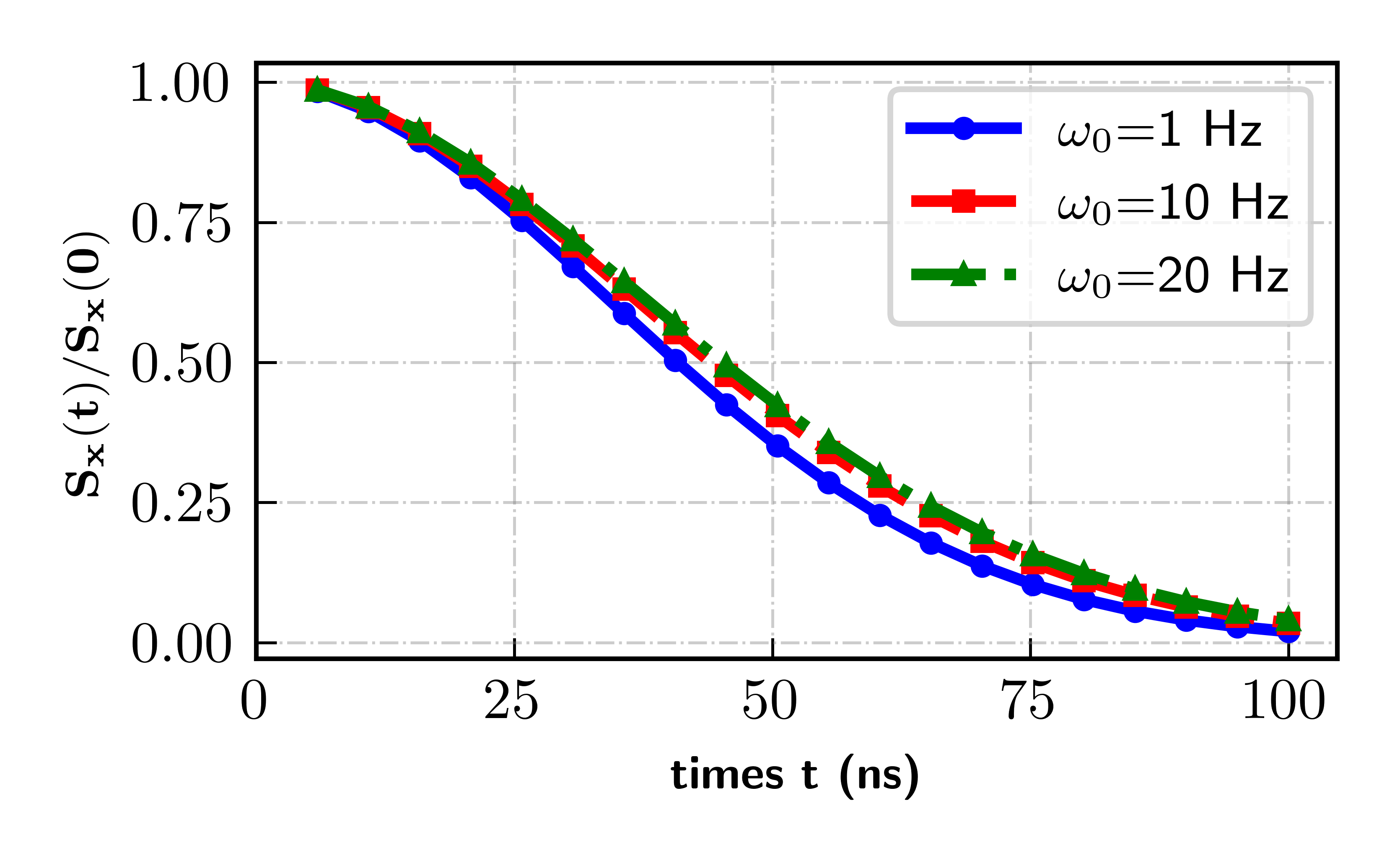}
\caption{
The free induction decay factor $S_x(t)/S_x(0)=e^{-\chi(t)}$ is shown as a function of time for a fixed magnetic field $B_x=0.3$\,T and gate electric field $F=2$\,MV/m, for different frequency cut-offs $\omega_0$.
The solid blue line with circular markers represents a cut-off frequency of 1\,Hz, the solid red line with square markers represents a cut-off frequency of 10\,Hz, and the dashed green line with triangular markers represents a cut-off frequency of 20\,Hz.
Within the range of cut-off frequencies considered, the shape of the free induction decay does not exhibit any significant difference in trend.
}
\label{Fig: Ratio vs t Bx03 omega}
\end{figure}
%==========
We note that this integral can be evaluated numerically by sampling the integrand over the device, as described in the previous subsection.
Here, $\delta\varepsilon_Z^2(\bm{r}_{\text{D}})$ is already known, and the charge defect density distribution $\rho(\bm{r}_{\text{D}})$ is device-specific. We consider the general case of a uniform distribution $n_D$ across the device such that $\rho(\bm{r}_{\text{D}}) \rightarrow n_D$.
Next, we need to take the average of the switching time $\tau$, which characterizes a single RTN source.
In semiconductor quantum dot systems, the position-dependent ensemble of individual RTN sources gives rise to the $1/f$ noise, which has also been explored in various solid-state platforms \cite{Afonin2002, Bergli2006, Cywinski2008, Lutchyn2008, Bergli2009, Pekola2015}.
A standard assumption in the literature, as discussed in Ref.\,\cite{VanDerZiel1950, Shnirman2005}, is that the switching time of a single RTN source is broadly distributed over $\tau_{\text{min}}$ to $\tau_{\text{max}}$ in a uniform logarithmic form.
$\tau_{\text{min}}$ represents the fastest switching time, limited by intrinsic scattering mechanisms \cite{Landauer1994,Pierre2018}, while $\tau_{\text{max}}$, the slowest switching time, is mostly determined by the experimental measurement timescale \cite{Machlup1954, Dutta1981}.
Additionally, these two switching times naturally set up a cut-off when performing the average over switching times (following a logarithmic distribution), which avoids divergences \cite{Paladino2014}.
In semiconductors, based on experimental observations e.g., Refs.\,\onlinecite{Petit2018, Connors2019, Rojas-Arias2023}, the range of switching times considered is typically $\tau_{\text{min}} \approx 1 \, \mu$s and $\tau_{\text{max}} \approx 1$ \,s.
Nevertheless, as outlined in Ref.\,\onlinecite{VanDerZiel1950}, the lower and upper cut-offs can in principle be extended to cover any desired range.
We can then evaluate the power spectral density by averaging over the ensemble of random telegraph noise sources:
\begin{equation}
    S_\delta(\omega) = \int_{\tau_{\text{min}}}^{\tau_{\text{max}}} p(\tau) \frac{\expval{\delta\varepsilon_Z^2} \tau}{1 + \omega^2 \tau^2} \, \dd{\tau} \,.
\end{equation}
Finally, for an ensemble of charge defects, the power spectral density takes on the $1/f$ form:
\begin{equation}\label{Eq: 1/f Spectral density}
    S_\delta(\omega) = \frac{\alpha \expval{\delta\varepsilon_Z^2}}{\omega} \,,
\end{equation}
where $\alpha$ is a dimensionless quantity which is sample dependent and can be calibrated using experimental values, as we discuss below.
Next, we focus on the case where $1/f$ noise-induced dephasing is of Gaussian type, in which cumulant terms higher than the second order vanish \cite{deSousa2009}.
%==========
\begin{figure}[tbp]
\centering
\includegraphics[width=1\columnwidth]{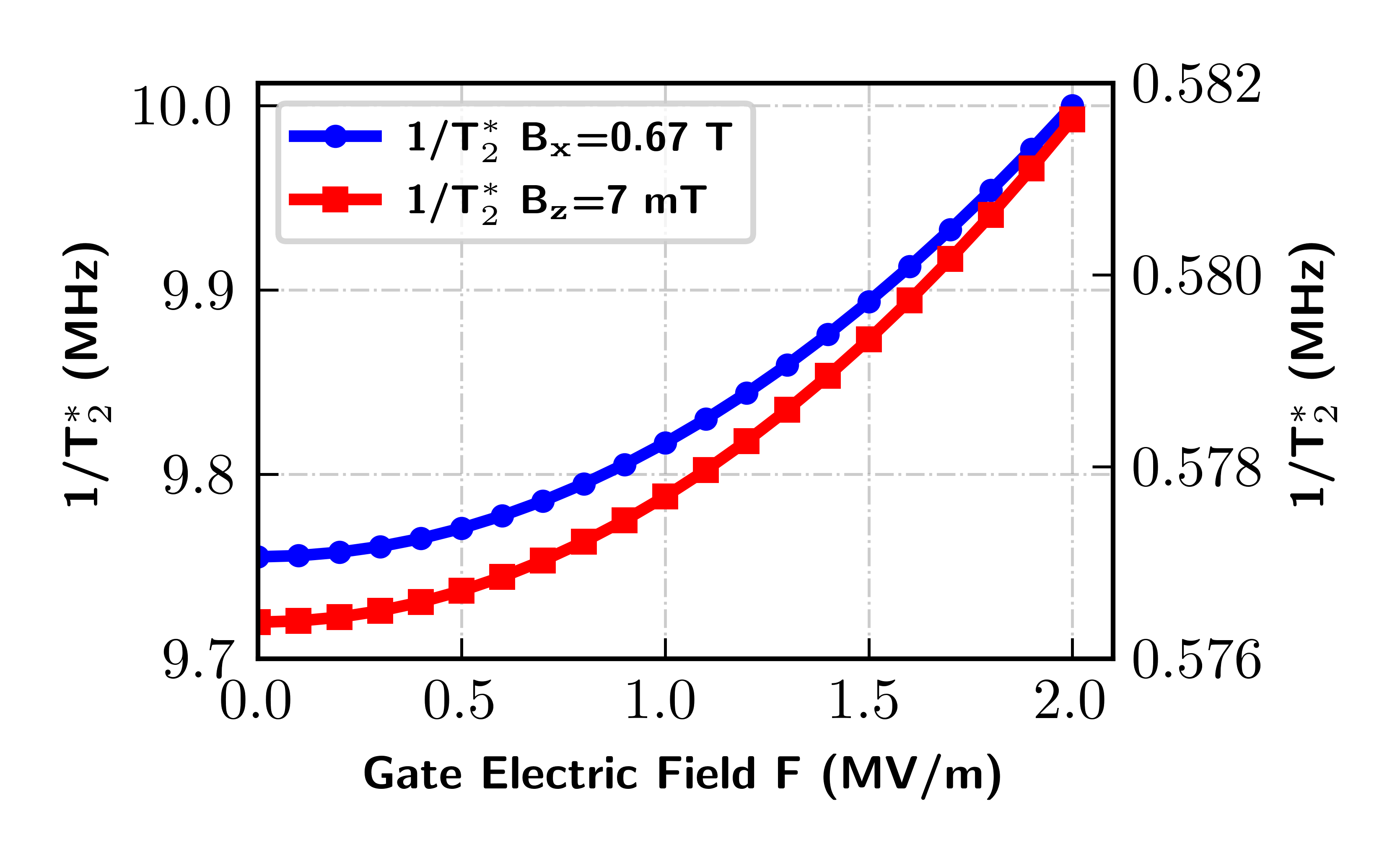}
\caption{
$1/T_{2}^*$ as a function of the gate electric field $F$ in an in-plane magnetic field (the blue curve with circular markers) and an out-of-plane magnetic field (the red curve with square markers).
The parameters used to generate this plot are calibrated by setting $1/T_{2}^*$=10\,MHz at $B_x$=0.67\,T, and $F$=2\,MV/m.
The selection of calibration parameters, the magnetic field magnitude, and the range of the gate electric field are detailed in the subsequent text.
}
\label{Fig: 50nm In-Plane Ensemble Dephasing Rate Total}
\end{figure}
%==========
Based on this property, the free induction decay factor can be expressed as $e^{-\chi(t)}$, as shown in Refs.\,\cite{deSousa2003, Culcer2009, Bermeister2014}, where
\begin{equation} \label{Eq: Decay Factor}
    \chi(t) = \frac{1}{2\hbar^2} \int_{\omega_0}^\infty S_\delta(\omega) \pqty{\frac{\sin (\omega t/2)}{\omega/2}}^2 \dd{\omega} \,.
\end{equation}
Here, $\omega_0$ is an experimentally determined parameter for the low-frequency cutoff.
Substituting Eq.\,\eqref{Eq: 1/f Spectral density} into Eq.\,\eqref{Eq: Decay Factor}, we obtain:
\begin{equation}
\begin{aligned}
    \chi(t) =& \frac{\alpha}{2 \hbar^2} \expval{\delta \varepsilon_Z^2} \times \\
    & \frac{1 - \cos (\omega_0 t) + \omega_0 t\left[\sin (\omega_0 t) - \omega_0 t \operatorname{Ci}(\omega_0 t)\right]}{\omega_0^2} \,,
\end{aligned}
\end{equation}
where $\operatorname{Ci}(\omega_0 t)$ is the Cosine integral function with argument $\omega_0 t$.
The low-frequency cut-off $\omega_0$ is estimated as the inverse of the measurement time, which typically corresponds to 1-10\,Hz. For $\omega_0 t \ll 1$, we have:
\begin{equation}
    \chi(t) = \pqty{\frac{t}{T_2^*}}^2 \ln(\frac{1}{\omega_0 t}) \, ,
\end{equation}
where
\begin{equation}\label{Eq: Dephasing Rate}
    \frac{1}{T_2^*} = \frac{1}{\hbar} \sqrt{\frac{\alpha}{2} \expval{\delta \varepsilon_Z^2}} \,.
\end{equation}
In Fig.\,\ref{Fig: Ratio vs t Bx03 omega}, we present the induction decay factor evaluated under different cut-off frequencies $\omega_0$.
We observe that for a fixed magnetic field and electric field, the induction decay factor does not change significantly with respect to the cut-off frequency, and the decay is primarily determined by $T_2^*$.
Moreover, the trend in the dephasing time as a function of external fields will not be affected by the cut-off frequency within the range of typical hole quantum dot devices measurements \cite{Petit2018, Connors2019, Rojas-Arias2023}, since this is assumed to be the same during each experimental run.
Since $\alpha$ is unknown we calibrate our results using an experimentally relevant estimate: we consider $T_2^* = 100$\,ns at $B_{\parallel} = 0.67$\,T for a circular dot with a radius of 50\,nm under a gate electric field of $F = 2$\,MV/m, which approximates the experimental observations in Ref.\,\cite{Hendrickx2020NC}.
%==========
\begin{figure}[tbp]
\centering
\includegraphics[width=1\columnwidth]{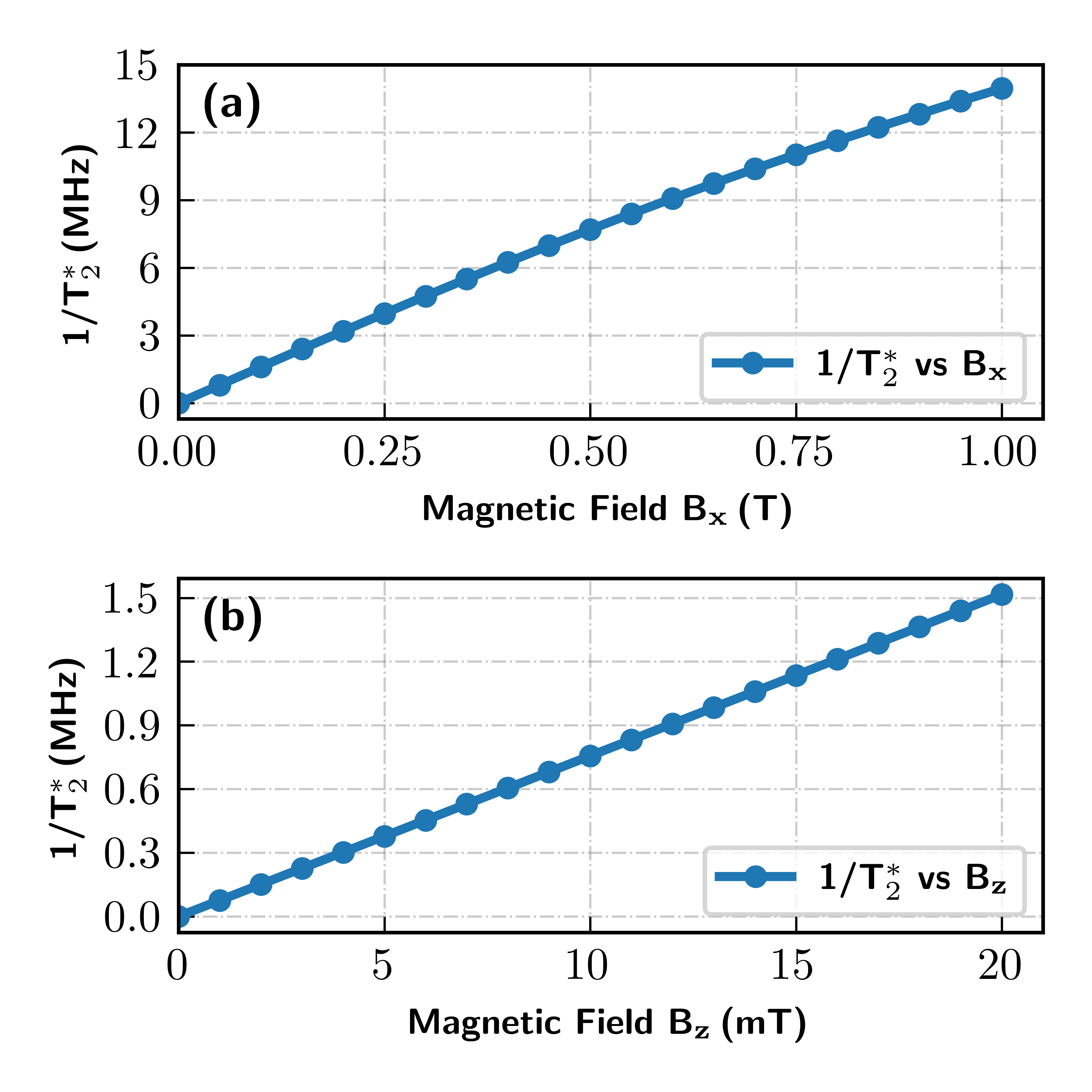}
\caption{
$1/T_{2}^*$ as a function of the magnetic field magnitude for different magnetic field orientations.
(a) $1/T_{2}^*$ as a function of $B_x$ at a fixed gate electric field $F$=2\,MV/m.
(b) $1/T_{2}^*$ as a function of $B_z$ at a fixed gate electric field $F$=2\,MV/m.
}
\label{Fig: 50nm OP-Plane Ensemble Dephasing Rate Total}
\end{figure}
%==========
%====================
%====================

%====================
%====================
\section{Results and discussion}
\label{S3-Results and discussions}
The dephasing rate $1/T_2^*$ is calculated for in-plane magnetic field $B_x$ and out-of-plane magnetic field $B_z$ separately as a function of the gate electric field $F$, magnitude of the magnetic field $B_x$ and $B_z$, and the quantum dot radius $a_x$ and $a_y$.
These results are obtained by numerically diagonalizing $H$ and $\tilde{H}$ to get $\delta\varepsilon_Z(\bm{r}_{\text{D}})$, the energy fluctuations due to defects.
This is then averaged over the sample following the methodology described above.

Fig.\,\ref{Fig: 50nm In-Plane Ensemble Dephasing Rate Total} presents the dependence of $1/T_2^*$ on the top gate electric field for a circular dot, for both in-plane and out-of-plane magnetic fields.
Given the vast difference in magnitude between the in-plane and out-of-plane $g$-factors, the range of the magnetic field is different for $B_x$ and $B_z$.
To compare the parameter-dependence of the dephasing rate $1/T_2^*$ for different magnetic field orientations, the magnitude of $B$ is calculated by equating the qubit Zeeman splitting $\Delta_0$ for the two cases, ${\bm B} \parallel \hat{\bm x}$ and ${\bm B} \parallel \hat{\bm z}$.
%==========
\begin{figure}[tbp]
\centering
\includegraphics[width=1\columnwidth]{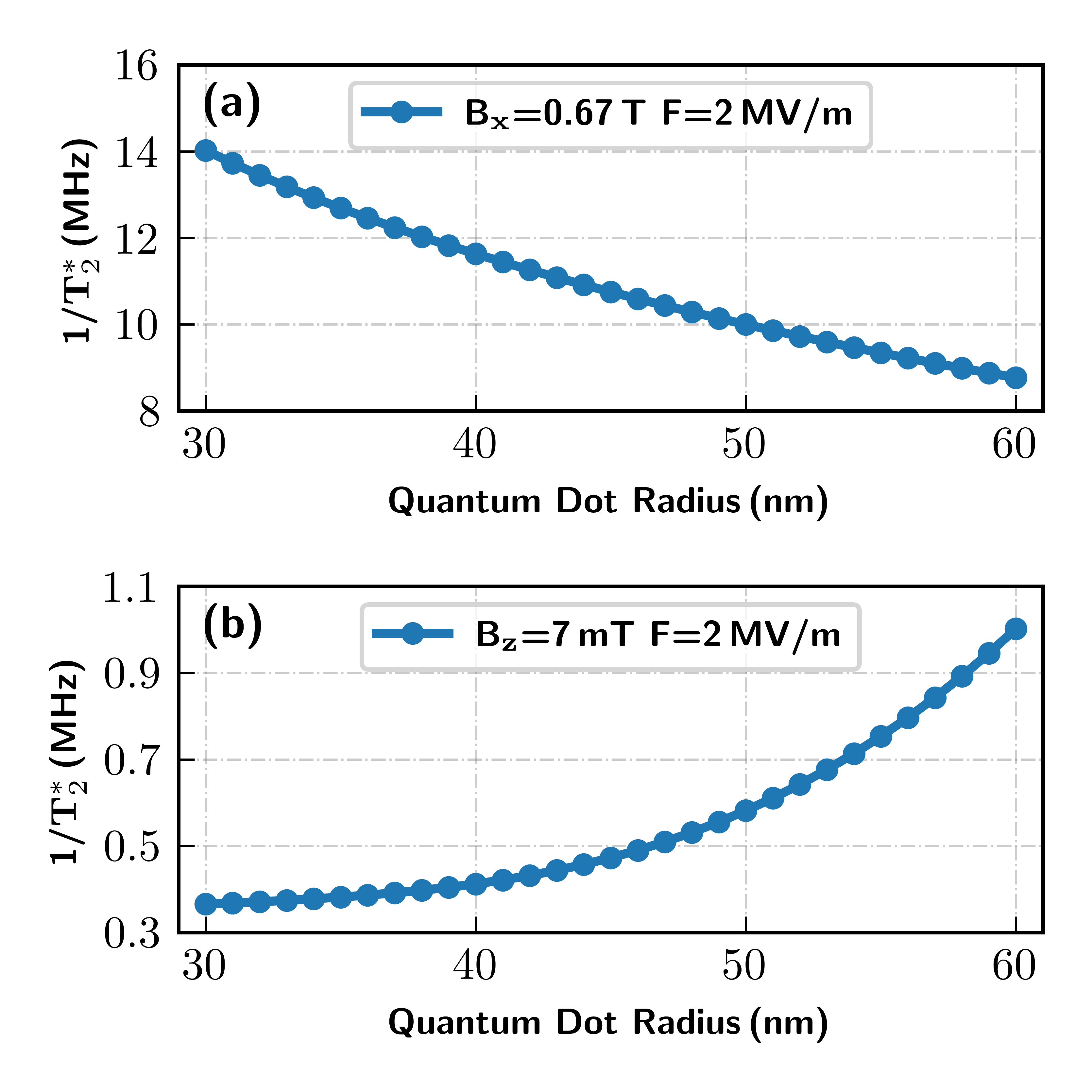}
\caption{
$1/T_{2}^*$ as a function of the quantum dot radius for different magnetic field orientations.
The form of the distribution of the Coulomb-type charge defects remains the same.
As the quantum dot size increases, the allowed range of defects changes to avoid any defects in the dot.
(a) $1/T_2^*$ as a function of quantum dot radius at $F$=2\,MV/m, $B_x$=0.67\,T.
(b) $1/T_2^*$ as a function of the quantum dot radius at $F$=2\,MV/m, $B_z$=7\,mT.
}
\label{Fig: Total Ensemble T2-rate vs Dotsize}
\end{figure}
%==========
For $B_x$, we choose the experimental value of 0.67\,T \cite{Hendrickx2020NC}.
Due to the large out-of-plane Land{\'e} $g$-factor of Ge, the value of $B_z$ that corresponds to the same Zeeman energy is 7\,mT.
In both cases, the Zeeman energy is considerably smaller than the orbital splitting.
We note that the dephasing rate increases monotonically as a function of $F$ for a fixed magnetic field in both cases, as the gate electric field $F$ enhances the Rashba spin-orbit coupling.
However, in the range of gate electric fields considered, the dephasing rate only exhibits a small variation as a function of the top gate field.
This implies that most of the dephasing is due to the interplay between the charge defect potentials and the spin-orbit terms present in the Luttinger-Kohn Hamiltonian.

The dephasing rate increases with magnetic field as seen in Fig.\,\ref{Fig: 50nm OP-Plane Ensemble Dephasing Rate Total}.
This trend is expected: the dephasing in the effective qubit Hamiltonian separates spin-up and spin-down states and must break time reversal, hence, the dephasing rate increases with increasing magnetic field and vanishes at $B$=0\,T reflecting the recovery of Kramers degeneracy.
At the same time, the magnitude of the dephasing time in an in-plane magnetic field is an order of magnitude larger than in the out-of-plane magnetic field.
Moreover, we have checked that, when the magnetic field is in the plane, most of the dephasing rate is due to the orbital magnetic field terms coupling the in-plane and out-of-plane dynamics.
%==========
\begin{figure}[tbp]
\centering
\includegraphics[width=1\columnwidth]{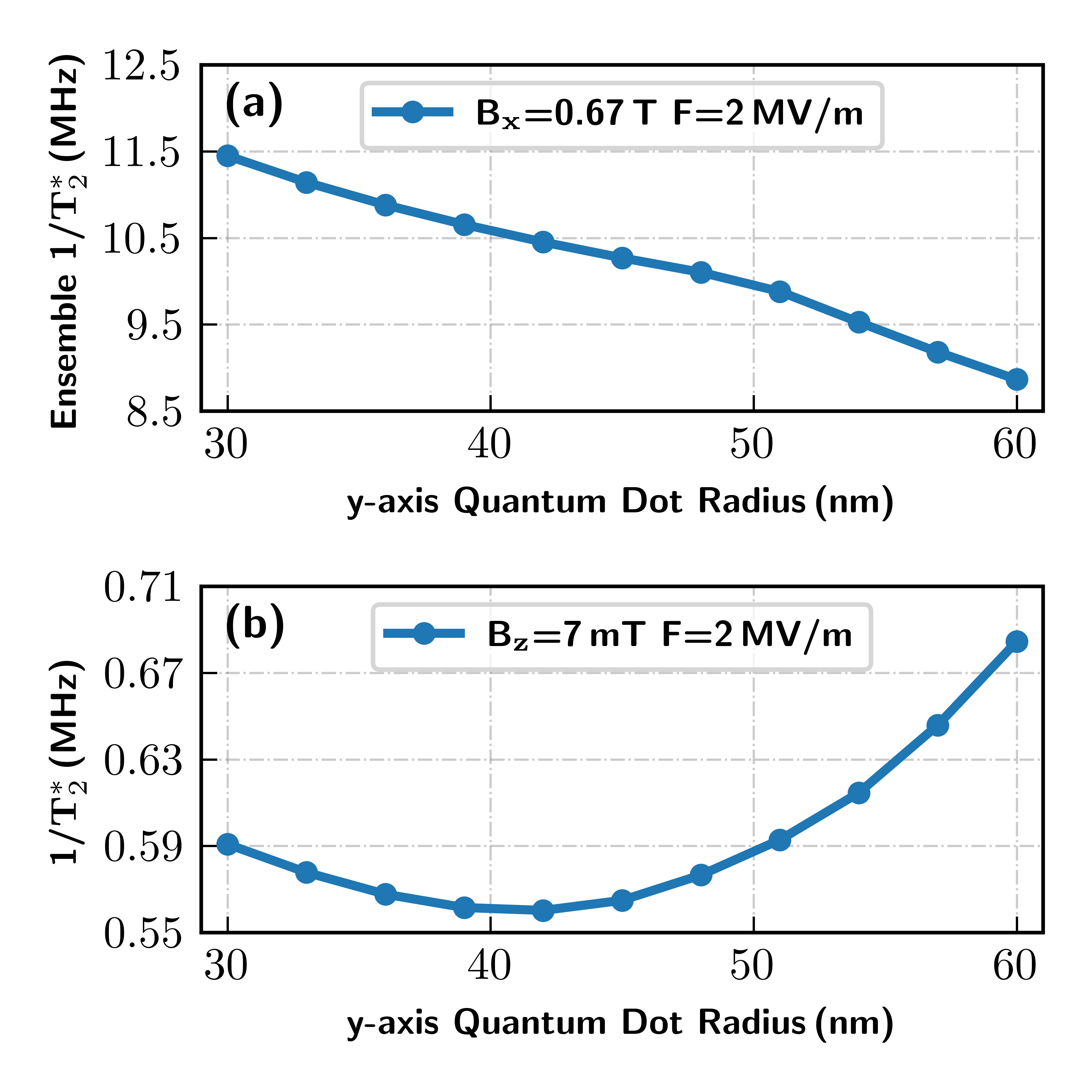}
\caption{
$1/T_2^*$ as a function of the dot anisotropy for a fixed gate electric field $F$=2\,MV/m in different magnetic field orientations.
The quantum dot radius along the x-axis is fixed to be $a_x$=50\,nm, the quantum dot radius along the y-axis $a_y$ varies from 30\,nm to 60\,nm.
(a) $1/T_2^*$ as a function of $a_y$ at $F$=2\,MV/m, $B_x$=0.67\,T.
(b) $1/T_2^*$ as a function of $a_y$ at $F$=2\,MV/m, $B_z$=7\,mT.
}
\label{Fig: Total Ensemble T2-rate vs Anisotropy}
\end{figure}
%==========
Fig.\,\ref{Fig: Total Ensemble T2-rate vs Dotsize} presents $1/T_2^*$ for a circular quantum dot as a function of the dot radius in different magnetic field orientations for a fixed gate electric field $F$=2\,MV/m.
The trend differs markedly depending on the magnetic field orientation.
As the dot radius increases, $1/T_2^*$ decreases for a fixed in-plane magnetic field $B_x$=0.67\,T, while it decreases then increases for a perpendicular field of $B_z=$7\,mT.
This difference reveals distinct mechanisms by which the magnetic field enhances the dephasing time as a function of the quantum dot size.
For ${\bm B} \parallel \hat{\bm x}$, both the orbital terms (captured by the Peierls substitution $-i\hbar\nabla + e\bm{A}$) and the Zeeman contributions (captured by $H_{\text{Z}}$) lead to a decreasing $1/T_2^*$ with $a_x$, resulting in the monotonically decreasing trend shown in Fig.\,\ref{Fig: Total Ensemble T2-rate vs Dotsize}-(a).
In the $B_z$ case, the Zeeman terms cause $1/T_2^*$ to decrease with increasing $a_x$, but the dominant orbital terms show an increasing trend, leading to the monotonically increasing trend in Fig.\,\ref{Fig: Total Ensemble T2-rate vs Dotsize}-(b).
These results reflect the complex interplay of the hole spin-orbit interaction, the random defect potential, and the orbital magnetic field contributions, highlighting the different role played by orbital terms and Zeeman terms in different magnetic field orientations.
Whereas the confinement energy scales as $1/a^2$, the effective Rashba spin-orbit coupling for holes has terms linear and cubic in the wave vector, which scale as $1/a$ and $1/a^3$, respectively.
The effective defect potential decreases with increasing $a$ but cannot be expressed in closed form, while the 2D Rashba model has a very limited range of applicability.
Hence the observed trends as a function of radius cannot be encapsulated in a simple expression.
Similar trends are observed with respect to the dot anisotropy, where $a_x$ is kept constant and $a_y$ is varied.
Fig.\,\ref{Fig: Total Ensemble T2-rate vs Anisotropy} considers the anisotropy of the quantum dot.

In this context, we note that the strongest change in $T_2^*$ occurs in response to altering the magnetic field.
One surmises that, aside from pulse sequences, the most effective way to enhance the coherence time is to work at the smallest possible magnetic field.
Interestingly, for the experimentally achievable gate electric field range, the dephasing time does not change significantly.
Finally, whereas the dependence of $1/T_2^*$ on dot geometry reflects the strong spin-orbit coupling in spin-3/2 systems, the variation with dot size is relatively small and one expects similar dephasing times across qubits of different sizes.
Essential materials-dependent questions remain open concerning the robustness of coherence properties against roughness, their variation with roughness parameters such as the correlation length and the spatial amplitude, their variation across samples, and whether they can be maintained even in the presence of multiple qubits.
Important factors affecting the dephasing time are inhomogeneous strain, spatially non-uniform shear strain \cite{Abadillo2023, Corley2023, Mauro2024arxiv}, and interface inversion asymmetry \cite{Durnev2014}.
Modeling of realistic hole dots must also consider a substantial defect distribution in the vicinity of the gates above the qubit.
This work has considered the ideal case of a flat interface, deferring rough interfaces and alternative defect distributions to future studies when detailed sample-dependent materials information becomes available.

The results in Figs.\,\ref{Fig: Total Ensemble T2-rate vs Dotsize} and \ref{Fig: Total Ensemble T2-rate vs Anisotropy} show that the dependence of $T_2^*$ on the dot radius is quite complex.
Figure \ref{Fig: Total Ensemble T2-rate vs Dotsize} shows that when the applied magnetic field is in-plane, $T_2^*$ increases with dot radius over the range studied ($30-60$\,nm).
In contrast, when the magnetic field is out of plane, $T_2^*$ decreases with increasing dot radius.
The effects of quantum dot ellipticity, which plays a key role in enhancing Rashba spin-orbit couplings, are shown in Fig.\,\ref{Fig: Total Ensemble T2-rate vs Anisotropy}, where we fix the dot size along the $x$-axis ($a_x$) and vary the size along the $y$-axis ($a_y$).
Now, with an in-plane magnetic field, $T_2^*$ increases monotonically with $a_y$, while with an out-of-plane magnetic field, $T_2^*$ exhibits a minimum as a function of $a_y$, though the variations remain small.
These contrasting behaviors are clear evidence of the complex interplay between spin-orbit coupling, random disorder potential, and orbital magnetic field terms.

We note that, whereas sweet spots as well as hot spots have been predicted as a function of gate electric field in the single charge defect case in Refs.\,\cite{Wang2021, Bosco2021PRB, Malkoc2022, Sarkar2023}, these do not appear in our study of $1/f$ noise. Previous studies, whose approach to noise was limited in scope and preliminary \cite{Wang2021, Sarkar2023}, showed that sweet spots or hot spots appear by first diagonalizing the Hamiltonian for the unperturbed qubit, then considering the \textit{diagonal} matrix elements of the defect potential between the qubit ground and excited states.
This approach, when applied to random telegraph noise, results in clear minima (sweet spots) or maxima (hot spots) in the dephasing rate as a function of the gate electric field, with sweet spots appearing in a perpendicular magnetic field and hot spots in an in-plane magnetic field. These extrema were seen at large electric fields, of the order of tens of MV/m, which are an order of magnitude larger than the fields accessed experimentally at present. Importantly, our work goes beyond existing studies of noise effects on hole spin qubits by fully diagonalizing the qubit Hamiltonian in the presence of a defect, therefore accounting for \textit{all matrix elements} of the defect potential. When the qubit Hamiltonian is diagonalized together with the defect potential, sweet spots or hot spots are no longer clearly seen as a function of the top gate field, either for a single defect or for an ensemble of defects, demonstrating the importance of the off-diagonal matrix elements of the defect potential.
Nevertheless, we note that, whereas the present approach is complete with respect to the defect potential, it is still focused on a flat interface.
In view of the fact that several experiments have in fact reported sweet spots as in Refs.\,\cite{Froning2021NN, Piot2022, Hendrickx2024, Geyer2024}, accurate modeling of experimental work requires the inclusion of interface roughness, which will be considered in future work.
%====================
%====================

%====================
%====================
\section{Conclusion and Outlook}
\label{S4-Conclusions and Outlooks}
We have studied the coherence properties of Ge hole quantum dots exposed to $1/f$ noise focusing on trends in the dephasing time $T_2^\ast$. We have derived a relationship between the properties of the $1/f$ noise power spectral and $T_2^\ast$, showing that the dephasing rate (i) decreases with decreasing magnetic field and vanishes as the field approaches zero, (ii) increases as a function of the top gate field, and (iii) shows opposing trends as a function of dot radius and anisotropy for different magnetic field orientations.
The key message, however, is that its dependence on system parameters is relatively weak.

The method can be generalized to include other sources of $1/f$ noise such as non-planar defect distributions and interacting two-level systems \cite{Mickelsen2023}, as well as the effect of $1/f$ noise on entanglement strategies \cite{Mutter2021}. Moreover, at the moment the only way to study the effect of different defect distribution is to run this code for a multitude of distributions. Future work can explore strategies for reversing the process and using noise as a spectroscopic tool.

We have focused on charges in a 2D plane since these constitute the prevailing understanding of the layout in a semiconductor heterostructure. In a forthcoming publication we will consider defects located above the dot, which are equally important, yet involve integrations over additional degrees of freedom. We note also parallel developments in Si holes where many of our considerations are likely to be applicable \cite{Zwanenburg2013, Wei2020, Shimatani2020, Piot2022, Yinan2023}.
%====================
%====================
\section{Acknowledgments}
\label{S5-Acknowledgments}
This work is supported by ARC Centre of Excellence in Future Low-Energy Electronics Technologies CE170100039 (SG and DC) and the Laboratory for Physical Sciences (SDS).
%====================
%====================
\bibliography{References_Ge}

\begin{thebibliography}{135}%
\makeatletter
\providecommand \@ifxundefined [1]{%
 \@ifx{#1\undefined}
}%
\providecommand \@ifnum [1]{%
 \ifnum #1\expandafter \@firstoftwo
 \else \expandafter \@secondoftwo
 \fi
}%
\providecommand \@ifx [1]{%
 \ifx #1\expandafter \@firstoftwo
 \else \expandafter \@secondoftwo
 \fi
}%
\providecommand \natexlab [1]{#1}%
\providecommand \enquote  [1]{``#1''}%
\providecommand \bibnamefont  [1]{#1}%
\providecommand \bibfnamefont [1]{#1}%
\providecommand \citenamefont [1]{#1}%
\providecommand \href@noop [0]{\@secondoftwo}%
\providecommand \href [0]{\begingroup \@sanitize@url \@href}%
\providecommand \@href[1]{\@@startlink{#1}\@@href}%
\providecommand \@@href[1]{\endgroup#1\@@endlink}%
\providecommand \@sanitize@url [0]{\catcode `\\12\catcode `\$12\catcode `\&12\catcode `\#12\catcode `\^12\catcode `\_12\catcode `\%12\relax}%
\providecommand \@@startlink[1]{}%
\providecommand \@@endlink[0]{}%
\providecommand \url  [0]{\begingroup\@sanitize@url \@url }%
\providecommand \@url [1]{\endgroup\@href {#1}{\urlprefix }}%
\providecommand \urlprefix  [0]{URL }%
\providecommand \Eprint [0]{\href }%
\providecommand \doibase [0]{https://doi.org/}%
\providecommand \selectlanguage [0]{\@gobble}%
\providecommand \bibinfo  [0]{\@secondoftwo}%
\providecommand \bibfield  [0]{\@secondoftwo}%
\providecommand \translation [1]{[#1]}%
\providecommand \BibitemOpen [0]{}%
\providecommand \bibitemStop [0]{}%
\providecommand \bibitemNoStop [0]{.\EOS\space}%
\providecommand \EOS [0]{\spacefactor3000\relax}%
\providecommand \BibitemShut  [1]{\csname bibitem#1\endcsname}%
\let\auto@bib@innerbib\@empty
%</preamble>
\bibitem [{\citenamefont {Vuku{\v s}i{\'c}}\ \emph {et~al.}(2018)\citenamefont {Vuku{\v s}i{\'c}}, \citenamefont {Kuku{\v c}ka}, \citenamefont {Watzinger}, \citenamefont {Milem}, \citenamefont {Sch{\"a}ffler},\ and\ \citenamefont {Katsaros}}]{Lada2018}%
  \BibitemOpen
  \bibfield  {author} {\bibinfo {author} {\bibfnamefont {L.}~\bibnamefont {Vuku{\v s}i{\'c}}}, \bibinfo {author} {\bibfnamefont {J.}~\bibnamefont {Kuku{\v c}ka}}, \bibinfo {author} {\bibfnamefont {H.}~\bibnamefont {Watzinger}}, \bibinfo {author} {\bibfnamefont {J.~M.}\ \bibnamefont {Milem}}, \bibinfo {author} {\bibfnamefont {F.}~\bibnamefont {Sch{\"a}ffler}},\ and\ \bibinfo {author} {\bibfnamefont {G.}~\bibnamefont {Katsaros}},\ }\bibfield  {title} {\bibinfo {title} {Single-shot readout of hole spins in ge},\ }\href {https://doi.org/10.1021/acs.nanolett.8b03217} {\bibfield  {journal} {\bibinfo  {journal} {Nano Letters}\ }\textbf {\bibinfo {volume} {18}},\ \bibinfo {pages} {7141} (\bibinfo {year} {2018})}\BibitemShut {NoStop}%
\bibitem [{\citenamefont {Watzinger}\ \emph {et~al.}(2018)\citenamefont {Watzinger}, \citenamefont {Kuku{\v c}ka}, \citenamefont {Vuku{\v s}i{\'c}}, \citenamefont {Gao}, \citenamefont {Wang}, \citenamefont {Sch{\"a}ffler}, \citenamefont {Zhang},\ and\ \citenamefont {Katsaros}}]{Watzinger2018}%
  \BibitemOpen
  \bibfield  {author} {\bibinfo {author} {\bibfnamefont {H.}~\bibnamefont {Watzinger}}, \bibinfo {author} {\bibfnamefont {J.}~\bibnamefont {Kuku{\v c}ka}}, \bibinfo {author} {\bibfnamefont {L.}~\bibnamefont {Vuku{\v s}i{\'c}}}, \bibinfo {author} {\bibfnamefont {F.}~\bibnamefont {Gao}}, \bibinfo {author} {\bibfnamefont {T.}~\bibnamefont {Wang}}, \bibinfo {author} {\bibfnamefont {F.}~\bibnamefont {Sch{\"a}ffler}}, \bibinfo {author} {\bibfnamefont {J.-J.}\ \bibnamefont {Zhang}},\ and\ \bibinfo {author} {\bibfnamefont {G.}~\bibnamefont {Katsaros}},\ }\bibfield  {title} {\bibinfo {title} {A germanium hole spin qubit},\ }\href {https://doi.org/10.1038/s41467-018-06418-4} {\bibfield  {journal} {\bibinfo  {journal} {Nature Communications}\ }\textbf {\bibinfo {volume} {9}},\ \bibinfo {pages} {3902} (\bibinfo {year} {2018})}\BibitemShut {NoStop}%
\bibitem [{\citenamefont {Hendrickx}\ \emph {et~al.}(2020{\natexlab{a}})\citenamefont {Hendrickx}, \citenamefont {Franke}, \citenamefont {Sammak}, \citenamefont {Scappucci},\ and\ \citenamefont {Veldhorst}}]{Hendrickx2020N}%
  \BibitemOpen
  \bibfield  {author} {\bibinfo {author} {\bibfnamefont {N.~W.}\ \bibnamefont {Hendrickx}}, \bibinfo {author} {\bibfnamefont {D.~P.}\ \bibnamefont {Franke}}, \bibinfo {author} {\bibfnamefont {A.}~\bibnamefont {Sammak}}, \bibinfo {author} {\bibfnamefont {G.}~\bibnamefont {Scappucci}},\ and\ \bibinfo {author} {\bibfnamefont {M.}~\bibnamefont {Veldhorst}},\ }\bibfield  {title} {\bibinfo {title} {Fast two-qubit logic with holes in germanium},\ }\href {https://doi.org/10.1038/s41586-019-1919-3} {\bibfield  {journal} {\bibinfo  {journal} {Nature}\ }\textbf {\bibinfo {volume} {577}},\ \bibinfo {pages} {487} (\bibinfo {year} {2020}{\natexlab{a}})}\BibitemShut {NoStop}%
\bibitem [{\citenamefont {Hendrickx}\ \emph {et~al.}(2020{\natexlab{b}})\citenamefont {Hendrickx}, \citenamefont {Lawrie}, \citenamefont {Petit}, \citenamefont {Sammak}, \citenamefont {Scappucci},\ and\ \citenamefont {Veldhorst}}]{Hendrickx2020NC}%
  \BibitemOpen
  \bibfield  {author} {\bibinfo {author} {\bibfnamefont {N.~W.}\ \bibnamefont {Hendrickx}}, \bibinfo {author} {\bibfnamefont {W.~I.~L.}\ \bibnamefont {Lawrie}}, \bibinfo {author} {\bibfnamefont {L.}~\bibnamefont {Petit}}, \bibinfo {author} {\bibfnamefont {A.}~\bibnamefont {Sammak}}, \bibinfo {author} {\bibfnamefont {G.}~\bibnamefont {Scappucci}},\ and\ \bibinfo {author} {\bibfnamefont {M.}~\bibnamefont {Veldhorst}},\ }\bibfield  {title} {\bibinfo {title} {A single-hole spin qubit},\ }\href {https://doi.org/10.1038/s41467-020-17211-7} {\bibfield  {journal} {\bibinfo  {journal} {Nature Communications}\ }\textbf {\bibinfo {volume} {11}},\ \bibinfo {pages} {3478} (\bibinfo {year} {2020}{\natexlab{b}})}\BibitemShut {NoStop}%
\bibitem [{\citenamefont {Hendrickx}\ \emph {et~al.}(2021)\citenamefont {Hendrickx}, \citenamefont {Lawrie}, \citenamefont {Russ}, \citenamefont {van Riggelen}, \citenamefont {de~Snoo}, \citenamefont {Schouten}, \citenamefont {Sammak}, \citenamefont {Scappucci},\ and\ \citenamefont {Veldhorst}}]{Hendrickx2021}%
  \BibitemOpen
  \bibfield  {author} {\bibinfo {author} {\bibfnamefont {N.~W.}\ \bibnamefont {Hendrickx}}, \bibinfo {author} {\bibfnamefont {W.~I.~L.}\ \bibnamefont {Lawrie}}, \bibinfo {author} {\bibfnamefont {M.}~\bibnamefont {Russ}}, \bibinfo {author} {\bibfnamefont {F.}~\bibnamefont {van Riggelen}}, \bibinfo {author} {\bibfnamefont {S.~L.}\ \bibnamefont {de~Snoo}}, \bibinfo {author} {\bibfnamefont {R.~N.}\ \bibnamefont {Schouten}}, \bibinfo {author} {\bibfnamefont {A.}~\bibnamefont {Sammak}}, \bibinfo {author} {\bibfnamefont {G.}~\bibnamefont {Scappucci}},\ and\ \bibinfo {author} {\bibfnamefont {M.}~\bibnamefont {Veldhorst}},\ }\bibfield  {title} {\bibinfo {title} {A four-qubit germanium quantum processor},\ }\href {https://doi.org/10.1038/s41586-021-03332-6} {\bibfield  {journal} {\bibinfo  {journal} {Nature}\ }\textbf {\bibinfo {volume} {591}},\ \bibinfo {pages} {580} (\bibinfo {year} {2021})}\BibitemShut {NoStop}%
\bibitem [{\citenamefont {Froning}\ \emph {et~al.}(2021{\natexlab{a}})\citenamefont {Froning}, \citenamefont {Camenzind}, \citenamefont {van~der Molen}, \citenamefont {Li}, \citenamefont {Bakkers}, \citenamefont {Zumb{\"u}hl},\ and\ \citenamefont {Braakman}}]{Froning2021NN}%
  \BibitemOpen
  \bibfield  {author} {\bibinfo {author} {\bibfnamefont {F.~N.~M.}\ \bibnamefont {Froning}}, \bibinfo {author} {\bibfnamefont {L.~C.}\ \bibnamefont {Camenzind}}, \bibinfo {author} {\bibfnamefont {O.~A.~H.}\ \bibnamefont {van~der Molen}}, \bibinfo {author} {\bibfnamefont {A.}~\bibnamefont {Li}}, \bibinfo {author} {\bibfnamefont {E.~P. A.~M.}\ \bibnamefont {Bakkers}}, \bibinfo {author} {\bibfnamefont {D.~M.}\ \bibnamefont {Zumb{\"u}hl}},\ and\ \bibinfo {author} {\bibfnamefont {F.~R.}\ \bibnamefont {Braakman}},\ }\bibfield  {title} {\bibinfo {title} {Ultrafast hole spin qubit with gate-tunable spin--orbit switch functionality},\ }\href {https://doi.org/10.1038/s41565-020-00828-6} {\bibfield  {journal} {\bibinfo  {journal} {Nature Nanotechnology}\ }\textbf {\bibinfo {volume} {16}},\ \bibinfo {pages} {308} (\bibinfo {year} {2021}{\natexlab{a}})}\BibitemShut {NoStop}%
\bibitem [{\citenamefont {Chatterjee}\ \emph {et~al.}(2021)\citenamefont {Chatterjee}, \citenamefont {Stevenson}, \citenamefont {De~Franceschi}, \citenamefont {Morello}, \citenamefont {de~Leon},\ and\ \citenamefont {Kuemmeth}}]{Chatterjee2021}%
  \BibitemOpen
  \bibfield  {author} {\bibinfo {author} {\bibfnamefont {A.}~\bibnamefont {Chatterjee}}, \bibinfo {author} {\bibfnamefont {P.}~\bibnamefont {Stevenson}}, \bibinfo {author} {\bibfnamefont {S.}~\bibnamefont {De~Franceschi}}, \bibinfo {author} {\bibfnamefont {A.}~\bibnamefont {Morello}}, \bibinfo {author} {\bibfnamefont {N.~P.}\ \bibnamefont {de~Leon}},\ and\ \bibinfo {author} {\bibfnamefont {F.}~\bibnamefont {Kuemmeth}},\ }\bibfield  {title} {\bibinfo {title} {Semiconductor qubits in practice},\ }\href {https://doi.org/10.1038/s42254-021-00283-9} {\bibfield  {journal} {\bibinfo  {journal} {Nature Reviews Physics}\ }\textbf {\bibinfo {volume} {3}},\ \bibinfo {pages} {157} (\bibinfo {year} {2021})}\BibitemShut {NoStop}%
\bibitem [{\citenamefont {Jirovec}\ \emph {et~al.}(2021)\citenamefont {Jirovec}, \citenamefont {Hofmann}, \citenamefont {Ballabio}, \citenamefont {Mutter}, \citenamefont {Tavani}, \citenamefont {Botifoll}, \citenamefont {Crippa}, \citenamefont {Kukucka}, \citenamefont {Sagi}, \citenamefont {Martins}, \citenamefont {Saez-Mollejo}, \citenamefont {Prieto}, \citenamefont {Borovkov}, \citenamefont {Arbiol}, \citenamefont {Chrastina}, \citenamefont {Isella},\ and\ \citenamefont {Katsaros}}]{Jirovec2021NM}%
  \BibitemOpen
  \bibfield  {author} {\bibinfo {author} {\bibfnamefont {D.}~\bibnamefont {Jirovec}}, \bibinfo {author} {\bibfnamefont {A.}~\bibnamefont {Hofmann}}, \bibinfo {author} {\bibfnamefont {A.}~\bibnamefont {Ballabio}}, \bibinfo {author} {\bibfnamefont {P.~M.}\ \bibnamefont {Mutter}}, \bibinfo {author} {\bibfnamefont {G.}~\bibnamefont {Tavani}}, \bibinfo {author} {\bibfnamefont {M.}~\bibnamefont {Botifoll}}, \bibinfo {author} {\bibfnamefont {A.}~\bibnamefont {Crippa}}, \bibinfo {author} {\bibfnamefont {J.}~\bibnamefont {Kukucka}}, \bibinfo {author} {\bibfnamefont {O.}~\bibnamefont {Sagi}}, \bibinfo {author} {\bibfnamefont {F.}~\bibnamefont {Martins}}, \bibinfo {author} {\bibfnamefont {J.}~\bibnamefont {Saez-Mollejo}}, \bibinfo {author} {\bibfnamefont {I.}~\bibnamefont {Prieto}}, \bibinfo {author} {\bibfnamefont {M.}~\bibnamefont {Borovkov}}, \bibinfo {author} {\bibfnamefont {J.}~\bibnamefont {Arbiol}}, \bibinfo {author} {\bibfnamefont {D.}~\bibnamefont {Chrastina}}, \bibinfo {author} {\bibfnamefont {G.}~\bibnamefont
  {Isella}},\ and\ \bibinfo {author} {\bibfnamefont {G.}~\bibnamefont {Katsaros}},\ }\bibfield  {title} {\bibinfo {title} {A singlet-triplet hole spin qubit in planar ge},\ }\href {https://doi.org/10.1038/s41563-021-01022-2} {\bibfield  {journal} {\bibinfo  {journal} {Nature Materials}\ }\textbf {\bibinfo {volume} {20}},\ \bibinfo {pages} {1106} (\bibinfo {year} {2021})}\BibitemShut {NoStop}%
\bibitem [{\citenamefont {Scappucci}\ \emph {et~al.}(2021)\citenamefont {Scappucci}, \citenamefont {Kloeffel}, \citenamefont {Zwanenburg}, \citenamefont {Loss}, \citenamefont {Myronov}, \citenamefont {Zhang}, \citenamefont {De~Franceschi}, \citenamefont {Katsaros},\ and\ \citenamefont {Veldhorst}}]{Scappucci2021}%
  \BibitemOpen
  \bibfield  {author} {\bibinfo {author} {\bibfnamefont {G.}~\bibnamefont {Scappucci}}, \bibinfo {author} {\bibfnamefont {C.}~\bibnamefont {Kloeffel}}, \bibinfo {author} {\bibfnamefont {F.~A.}\ \bibnamefont {Zwanenburg}}, \bibinfo {author} {\bibfnamefont {D.}~\bibnamefont {Loss}}, \bibinfo {author} {\bibfnamefont {M.}~\bibnamefont {Myronov}}, \bibinfo {author} {\bibfnamefont {J.-J.}\ \bibnamefont {Zhang}}, \bibinfo {author} {\bibfnamefont {S.}~\bibnamefont {De~Franceschi}}, \bibinfo {author} {\bibfnamefont {G.}~\bibnamefont {Katsaros}},\ and\ \bibinfo {author} {\bibfnamefont {M.}~\bibnamefont {Veldhorst}},\ }\bibfield  {title} {\bibinfo {title} {The germanium quantum information route},\ }\href {https://doi.org/10.1038/s41578-020-00262-z} {\bibfield  {journal} {\bibinfo  {journal} {Nature Reviews Materials}\ }\textbf {\bibinfo {volume} {6}},\ \bibinfo {pages} {926} (\bibinfo {year} {2021})}\BibitemShut {NoStop}%
\bibitem [{\citenamefont {Wang}\ \emph {et~al.}(2022)\citenamefont {Wang}, \citenamefont {Xu}, \citenamefont {Gao}, \citenamefont {Liu}, \citenamefont {Ma}, \citenamefont {Zhang}, \citenamefont {Wang}, \citenamefont {Cao}, \citenamefont {Wang}, \citenamefont {Zhang}, \citenamefont {Culcer}, \citenamefont {Hu}, \citenamefont {Jiang}, \citenamefont {Li}, \citenamefont {Guo},\ and\ \citenamefont {Guo}}]{Ke2022}%
  \BibitemOpen
  \bibfield  {author} {\bibinfo {author} {\bibfnamefont {K.}~\bibnamefont {Wang}}, \bibinfo {author} {\bibfnamefont {G.}~\bibnamefont {Xu}}, \bibinfo {author} {\bibfnamefont {F.}~\bibnamefont {Gao}}, \bibinfo {author} {\bibfnamefont {H.}~\bibnamefont {Liu}}, \bibinfo {author} {\bibfnamefont {R.-L.}\ \bibnamefont {Ma}}, \bibinfo {author} {\bibfnamefont {X.}~\bibnamefont {Zhang}}, \bibinfo {author} {\bibfnamefont {Z.}~\bibnamefont {Wang}}, \bibinfo {author} {\bibfnamefont {G.}~\bibnamefont {Cao}}, \bibinfo {author} {\bibfnamefont {T.}~\bibnamefont {Wang}}, \bibinfo {author} {\bibfnamefont {J.-J.}\ \bibnamefont {Zhang}}, \bibinfo {author} {\bibfnamefont {D.}~\bibnamefont {Culcer}}, \bibinfo {author} {\bibfnamefont {X.}~\bibnamefont {Hu}}, \bibinfo {author} {\bibfnamefont {H.-W.}\ \bibnamefont {Jiang}}, \bibinfo {author} {\bibfnamefont {H.-O.}\ \bibnamefont {Li}}, \bibinfo {author} {\bibfnamefont {G.-C.}\ \bibnamefont {Guo}},\ and\ \bibinfo {author} {\bibfnamefont {G.-P.}\ \bibnamefont {Guo}},\ }\bibfield
  {title} {\bibinfo {title} {Ultrafast coherent control of a hole spin qubit in a germanium quantum dot},\ }\href {https://doi.org/10.1038/s41467-021-27880-7} {\bibfield  {journal} {\bibinfo  {journal} {Nature Communications}\ }\textbf {\bibinfo {volume} {13}},\ \bibinfo {pages} {206} (\bibinfo {year} {2022})}\BibitemShut {NoStop}%
\bibitem [{\citenamefont {Fang}\ \emph {et~al.}(2023)\citenamefont {Fang}, \citenamefont {Philippopoulos}, \citenamefont {Culcer}, \citenamefont {Coish},\ and\ \citenamefont {Chesi}}]{Yinan2023}%
  \BibitemOpen
  \bibfield  {author} {\bibinfo {author} {\bibfnamefont {Y.}~\bibnamefont {Fang}}, \bibinfo {author} {\bibfnamefont {P.}~\bibnamefont {Philippopoulos}}, \bibinfo {author} {\bibfnamefont {D.}~\bibnamefont {Culcer}}, \bibinfo {author} {\bibfnamefont {W.~A.}\ \bibnamefont {Coish}},\ and\ \bibinfo {author} {\bibfnamefont {S.}~\bibnamefont {Chesi}},\ }\bibfield  {title} {\bibinfo {title} {Recent advances in hole-spin qubits},\ }\href {https://doi.org/10.1088/2633-4356/acb87e} {\bibfield  {journal} {\bibinfo  {journal} {Materials for Quantum Technology}\ }\textbf {\bibinfo {volume} {3}},\ \bibinfo {pages} {012003} (\bibinfo {year} {2023})}\BibitemShut {NoStop}%
\bibitem [{\citenamefont {Tidjani}\ \emph {et~al.}(2023)\citenamefont {Tidjani}, \citenamefont {Tosato}, \citenamefont {Ivlev}, \citenamefont {D\'eprez}, \citenamefont {Oosterhout}, \citenamefont {Stehouwer}, \citenamefont {Sammak}, \citenamefont {Scappucci},\ and\ \citenamefont {Veldhorst}}]{Tidjani2023}%
  \BibitemOpen
  \bibfield  {author} {\bibinfo {author} {\bibfnamefont {H.}~\bibnamefont {Tidjani}}, \bibinfo {author} {\bibfnamefont {A.}~\bibnamefont {Tosato}}, \bibinfo {author} {\bibfnamefont {A.}~\bibnamefont {Ivlev}}, \bibinfo {author} {\bibfnamefont {C.}~\bibnamefont {D\'eprez}}, \bibinfo {author} {\bibfnamefont {S.}~\bibnamefont {Oosterhout}}, \bibinfo {author} {\bibfnamefont {L.}~\bibnamefont {Stehouwer}}, \bibinfo {author} {\bibfnamefont {A.}~\bibnamefont {Sammak}}, \bibinfo {author} {\bibfnamefont {G.}~\bibnamefont {Scappucci}},\ and\ \bibinfo {author} {\bibfnamefont {M.}~\bibnamefont {Veldhorst}},\ }\bibfield  {title} {\bibinfo {title} {Vertical gate-defined double quantum dot in a strained germanium double quantum well},\ }\href {https://doi.org/10.1103/PhysRevApplied.20.054035} {\bibfield  {journal} {\bibinfo  {journal} {Phys. Rev. Appl.}\ }\textbf {\bibinfo {volume} {20}},\ \bibinfo {pages} {054035} (\bibinfo {year} {2023})}\BibitemShut {NoStop}%
\bibitem [{\citenamefont {Hartmann}\ \emph {et~al.}(2023)\citenamefont {Hartmann}, \citenamefont {Bernier}, \citenamefont {Pierre}, \citenamefont {Barnes}, \citenamefont {Mazzocchi}, \citenamefont {Krawczyk}, \citenamefont {Lima}, \citenamefont {Kiyooka},\ and\ \citenamefont {Franceschi}}]{Hartmann2023}%
  \BibitemOpen
  \bibfield  {author} {\bibinfo {author} {\bibfnamefont {J.-M.}\ \bibnamefont {Hartmann}}, \bibinfo {author} {\bibfnamefont {N.}~\bibnamefont {Bernier}}, \bibinfo {author} {\bibfnamefont {F.}~\bibnamefont {Pierre}}, \bibinfo {author} {\bibfnamefont {J.-P.}\ \bibnamefont {Barnes}}, \bibinfo {author} {\bibfnamefont {V.}~\bibnamefont {Mazzocchi}}, \bibinfo {author} {\bibfnamefont {J.}~\bibnamefont {Krawczyk}}, \bibinfo {author} {\bibfnamefont {G.}~\bibnamefont {Lima}}, \bibinfo {author} {\bibfnamefont {E.}~\bibnamefont {Kiyooka}},\ and\ \bibinfo {author} {\bibfnamefont {S.~D.}\ \bibnamefont {Franceschi}},\ }\bibfield  {title} {\bibinfo {title} {Epitaxy of group-iv semiconductors for quantum electronics},\ }\href {https://doi.org/10.1149/11101.0053ecst} {\bibfield  {journal} {\bibinfo  {journal} {ECS Transactions}\ }\textbf {\bibinfo {volume} {111}},\ \bibinfo {pages} {53} (\bibinfo {year} {2023})}\BibitemShut {NoStop}%
\bibitem [{\citenamefont {Ivlev}\ \emph {et~al.}(2024)\citenamefont {Ivlev}, \citenamefont {Tidjani}, \citenamefont {Oosterhout}, \citenamefont {Sammak}, \citenamefont {Scappucci},\ and\ \citenamefont {Veldhorst}}]{Ivlev2024}%
  \BibitemOpen
  \bibfield  {author} {\bibinfo {author} {\bibfnamefont {A.~S.}\ \bibnamefont {Ivlev}}, \bibinfo {author} {\bibfnamefont {H.}~\bibnamefont {Tidjani}}, \bibinfo {author} {\bibfnamefont {S.~D.}\ \bibnamefont {Oosterhout}}, \bibinfo {author} {\bibfnamefont {A.}~\bibnamefont {Sammak}}, \bibinfo {author} {\bibfnamefont {G.}~\bibnamefont {Scappucci}},\ and\ \bibinfo {author} {\bibfnamefont {M.}~\bibnamefont {Veldhorst}},\ }\bibfield  {title} {\bibinfo {title} {{Coupled vertical double quantum dots at single-hole occupancy}},\ }\href {https://doi.org/10.1063/5.0198274} {\bibfield  {journal} {\bibinfo  {journal} {Applied Physics Letters}\ }\textbf {\bibinfo {volume} {125}},\ \bibinfo {pages} {023501} (\bibinfo {year} {2024})}\BibitemShut {NoStop}%
\bibitem [{\citenamefont {Del~Vecchio}\ and\ \citenamefont {Moutanabbir}(2024)}]{Del2024}%
  \BibitemOpen
  \bibfield  {author} {\bibinfo {author} {\bibfnamefont {P.}~\bibnamefont {Del~Vecchio}}\ and\ \bibinfo {author} {\bibfnamefont {O.}~\bibnamefont {Moutanabbir}},\ }\bibfield  {title} {\bibinfo {title} {Light-hole spin confined in germanium},\ }\href {https://doi.org/10.1103/PhysRevB.110.045409} {\bibfield  {journal} {\bibinfo  {journal} {Phys. Rev. B}\ }\textbf {\bibinfo {volume} {110}},\ \bibinfo {pages} {045409} (\bibinfo {year} {2024})}\BibitemShut {NoStop}%
\bibitem [{\citenamefont {Strohbeen}\ \emph {et~al.}(2024)\citenamefont {Strohbeen}, \citenamefont {Banerjee}, \citenamefont {Brook}, \citenamefont {Levy}, \citenamefont {Sarney}, \citenamefont {van Dijk}, \citenamefont {Orth}, \citenamefont {Mikalsen}, \citenamefont {Fatemi},\ and\ \citenamefont {Shabani}}]{Strohbeen2024}%
  \BibitemOpen
  \bibfield  {author} {\bibinfo {author} {\bibfnamefont {P.~J.}\ \bibnamefont {Strohbeen}}, \bibinfo {author} {\bibfnamefont {T.}~\bibnamefont {Banerjee}}, \bibinfo {author} {\bibfnamefont {A.~M.}\ \bibnamefont {Brook}}, \bibinfo {author} {\bibfnamefont {I.}~\bibnamefont {Levy}}, \bibinfo {author} {\bibfnamefont {W.~L.}\ \bibnamefont {Sarney}}, \bibinfo {author} {\bibfnamefont {J.}~\bibnamefont {van Dijk}}, \bibinfo {author} {\bibfnamefont {H.}~\bibnamefont {Orth}}, \bibinfo {author} {\bibfnamefont {M.}~\bibnamefont {Mikalsen}}, \bibinfo {author} {\bibfnamefont {V.}~\bibnamefont {Fatemi}},\ and\ \bibinfo {author} {\bibfnamefont {J.}~\bibnamefont {Shabani}},\ }\bibfield  {title} {\bibinfo {title} {{Molecular beam epitaxy growth of superconducting tantalum germanide}},\ }\href {https://doi.org/10.1063/5.0189597} {\bibfield  {journal} {\bibinfo  {journal} {Applied Physics Letters}\ }\textbf {\bibinfo {volume} {124}},\ \bibinfo {pages} {092102} (\bibinfo {year} {2024})}\BibitemShut {NoStop}%
\bibitem [{\citenamefont {Itoh}\ \emph {et~al.}(1993)\citenamefont {Itoh}, \citenamefont {Hansen}, \citenamefont {Haller}, \citenamefont {Farmer}, \citenamefont {Ozhogin}, \citenamefont {Rudnev},\ and\ \citenamefont {Tikhomirov}}]{Itoh1993}%
  \BibitemOpen
  \bibfield  {author} {\bibinfo {author} {\bibfnamefont {K.}~\bibnamefont {Itoh}}, \bibinfo {author} {\bibfnamefont {W.~L.}\ \bibnamefont {Hansen}}, \bibinfo {author} {\bibfnamefont {E.~E.}\ \bibnamefont {Haller}}, \bibinfo {author} {\bibfnamefont {J.~W.}\ \bibnamefont {Farmer}}, \bibinfo {author} {\bibfnamefont {V.~I.}\ \bibnamefont {Ozhogin}}, \bibinfo {author} {\bibfnamefont {A.}~\bibnamefont {Rudnev}},\ and\ \bibinfo {author} {\bibfnamefont {A.}~\bibnamefont {Tikhomirov}},\ }\bibfield  {title} {\bibinfo {title} {High purity isotopically enriched 70ge and 74ge single crystals: Isotope separation, growth, and properties},\ }\href {https://doi.org/10.1557/JMR.1993.1341} {\bibfield  {journal} {\bibinfo  {journal} {Journal of Materials Research}\ }\textbf {\bibinfo {volume} {8}},\ \bibinfo {pages} {1341} (\bibinfo {year} {1993})}\BibitemShut {NoStop}%
\bibitem [{\citenamefont {Itoh}\ \emph {et~al.}(2003)\citenamefont {Itoh}, \citenamefont {Kato}, \citenamefont {Uemura}, \citenamefont {Kaliteevskii}, \citenamefont {Godisov}, \citenamefont {Devyatych}, \citenamefont {Bulanov}, \citenamefont {Gusev}, \citenamefont {Kovalev}, \citenamefont {Sennikov}, \citenamefont {Pohl}, \citenamefont {Abrosimov},\ and\ \citenamefont {Riemann}}]{Kohei2003}%
  \BibitemOpen
  \bibfield  {author} {\bibinfo {author} {\bibfnamefont {K.~M.}\ \bibnamefont {Itoh}}, \bibinfo {author} {\bibfnamefont {J.}~\bibnamefont {Kato}}, \bibinfo {author} {\bibfnamefont {M.}~\bibnamefont {Uemura}}, \bibinfo {author} {\bibfnamefont {A.~K.}\ \bibnamefont {Kaliteevskii}}, \bibinfo {author} {\bibfnamefont {O.~N.}\ \bibnamefont {Godisov}}, \bibinfo {author} {\bibfnamefont {G.~G.}\ \bibnamefont {Devyatych}}, \bibinfo {author} {\bibfnamefont {A.~D.}\ \bibnamefont {Bulanov}}, \bibinfo {author} {\bibfnamefont {A.~V.}\ \bibnamefont {Gusev}}, \bibinfo {author} {\bibfnamefont {I.~D.}\ \bibnamefont {Kovalev}}, \bibinfo {author} {\bibfnamefont {P.~G.}\ \bibnamefont {Sennikov}}, \bibinfo {author} {\bibfnamefont {H.-J.}\ \bibnamefont {Pohl}}, \bibinfo {author} {\bibfnamefont {N.~V.}\ \bibnamefont {Abrosimov}},\ and\ \bibinfo {author} {\bibfnamefont {H.}~\bibnamefont {Riemann}},\ }\bibfield  {title} {\bibinfo {title} {High purity isotopically enriched 29si and 30si single crystals: Isotope separation, purification,
  and growth},\ }\href {https://doi.org/10.1143/JJAP.42.6248} {\bibfield  {journal} {\bibinfo  {journal} {Japanese Journal of Applied Physics}\ }\textbf {\bibinfo {volume} {42}},\ \bibinfo {pages} {6248} (\bibinfo {year} {2003})}\BibitemShut {NoStop}%
\bibitem [{\citenamefont {Yu}\ and\ \citenamefont {Cardona}(2010)}]{Yu2010}%
  \BibitemOpen
  \bibfield  {author} {\bibinfo {author} {\bibfnamefont {P.~Y.}\ \bibnamefont {Yu}}\ and\ \bibinfo {author} {\bibfnamefont {M.}~\bibnamefont {Cardona}},\ }\href {https://doi.org/10.1007/978-3-642-00710-1} {\emph {\bibinfo {title} {Fundamentals of Semiconductors: Physics and Materials Properties}}},\ \bibinfo {edition} {4th}\ ed.,\ Graduate Texts in Physics\ (\bibinfo  {publisher} {Springer Berlin, Heidelberg},\ \bibinfo {year} {2010})\ pp.\ \bibinfo {pages} {XXII, 778}\BibitemShut {NoStop}%
\bibitem [{\citenamefont {Rashba}\ and\ \citenamefont {Sherman}(1988)}]{Rashba1988}%
  \BibitemOpen
  \bibfield  {author} {\bibinfo {author} {\bibfnamefont {E.}~\bibnamefont {Rashba}}\ and\ \bibinfo {author} {\bibfnamefont {E.}~\bibnamefont {Sherman}},\ }\bibfield  {title} {\bibinfo {title} {Spin-orbital band splitting in symmetric quantum wells},\ }\href {https://doi.org/https://doi.org/10.1016/0375-9601(88)90140-5} {\bibfield  {journal} {\bibinfo  {journal} {Physics Letters A}\ }\textbf {\bibinfo {volume} {129}},\ \bibinfo {pages} {175} (\bibinfo {year} {1988})}\BibitemShut {NoStop}%
\bibitem [{\citenamefont {Winkler}(2003)}]{Winkler2003}%
  \BibitemOpen
  \bibfield  {author} {\bibinfo {author} {\bibfnamefont {R.}~\bibnamefont {Winkler}},\ }\href {https://doi.org/10.1007/b13586} {\emph {\bibinfo {title} {Spin-orbit Coupling Effects in Two-Dimensional Electron and Hole Systems}}},\ \bibinfo {edition} {1st}\ ed.,\ Springer Tracts in Modern Physics\ (\bibinfo  {publisher} {Springer Berlin, Heidelberg},\ \bibinfo {year} {2003})\ pp.\ \bibinfo {pages} {XII, 228}\BibitemShut {NoStop}%
\bibitem [{\citenamefont {Winkler}\ \emph {et~al.}(2008)\citenamefont {Winkler}, \citenamefont {Culcer}, \citenamefont {Papadakis}, \citenamefont {Habib},\ and\ \citenamefont {Shayegan}}]{Winkler2008}%
  \BibitemOpen
  \bibfield  {author} {\bibinfo {author} {\bibfnamefont {R.}~\bibnamefont {Winkler}}, \bibinfo {author} {\bibfnamefont {D.}~\bibnamefont {Culcer}}, \bibinfo {author} {\bibfnamefont {S.~J.}\ \bibnamefont {Papadakis}}, \bibinfo {author} {\bibfnamefont {B.}~\bibnamefont {Habib}},\ and\ \bibinfo {author} {\bibfnamefont {M.}~\bibnamefont {Shayegan}},\ }\bibfield  {title} {\bibinfo {title} {Spin orientation of holes in quantum wells},\ }\href {https://doi.org/10.1088/0268-1242/23/11/114017} {\bibfield  {journal} {\bibinfo  {journal} {Semiconductor Science and Technology}\ }\textbf {\bibinfo {volume} {23}},\ \bibinfo {pages} {114017} (\bibinfo {year} {2008})}\BibitemShut {NoStop}%
\bibitem [{\citenamefont {Durnev}\ \emph {et~al.}(2014)\citenamefont {Durnev}, \citenamefont {Glazov},\ and\ \citenamefont {Ivchenko}}]{Durnev2014}%
  \BibitemOpen
  \bibfield  {author} {\bibinfo {author} {\bibfnamefont {M.~V.}\ \bibnamefont {Durnev}}, \bibinfo {author} {\bibfnamefont {M.~M.}\ \bibnamefont {Glazov}},\ and\ \bibinfo {author} {\bibfnamefont {E.~L.}\ \bibnamefont {Ivchenko}},\ }\bibfield  {title} {\bibinfo {title} {Spin-orbit splitting of valence subbands in semiconductor nanostructures},\ }\href {https://doi.org/10.1103/PhysRevB.89.075430} {\bibfield  {journal} {\bibinfo  {journal} {Phys. Rev. B}\ }\textbf {\bibinfo {volume} {89}},\ \bibinfo {pages} {075430} (\bibinfo {year} {2014})}\BibitemShut {NoStop}%
\bibitem [{\citenamefont {Marcellina}\ \emph {et~al.}(2017)\citenamefont {Marcellina}, \citenamefont {Hamilton}, \citenamefont {Winkler},\ and\ \citenamefont {Culcer}}]{Marcellina2017}%
  \BibitemOpen
  \bibfield  {author} {\bibinfo {author} {\bibfnamefont {E.}~\bibnamefont {Marcellina}}, \bibinfo {author} {\bibfnamefont {A.~R.}\ \bibnamefont {Hamilton}}, \bibinfo {author} {\bibfnamefont {R.}~\bibnamefont {Winkler}},\ and\ \bibinfo {author} {\bibfnamefont {D.}~\bibnamefont {Culcer}},\ }\bibfield  {title} {\bibinfo {title} {Spin-orbit interactions in inversion-asymmetric two-dimensional hole systems: A variational analysis},\ }\href {https://doi.org/10.1103/PhysRevB.95.075305} {\bibfield  {journal} {\bibinfo  {journal} {Phys. Rev. B}\ }\textbf {\bibinfo {volume} {95}},\ \bibinfo {pages} {075305} (\bibinfo {year} {2017})}\BibitemShut {NoStop}%
\bibitem [{\citenamefont {Danneau}\ \emph {et~al.}(2006)\citenamefont {Danneau}, \citenamefont {Klochan}, \citenamefont {Clarke}, \citenamefont {Ho}, \citenamefont {Micolich}, \citenamefont {Simmons}, \citenamefont {Hamilton}, \citenamefont {Pepper}, \citenamefont {Ritchie},\ and\ \citenamefont {Z\"ulicke}}]{Danneau2006}%
  \BibitemOpen
  \bibfield  {author} {\bibinfo {author} {\bibfnamefont {R.}~\bibnamefont {Danneau}}, \bibinfo {author} {\bibfnamefont {O.}~\bibnamefont {Klochan}}, \bibinfo {author} {\bibfnamefont {W.~R.}\ \bibnamefont {Clarke}}, \bibinfo {author} {\bibfnamefont {L.~H.}\ \bibnamefont {Ho}}, \bibinfo {author} {\bibfnamefont {A.~P.}\ \bibnamefont {Micolich}}, \bibinfo {author} {\bibfnamefont {M.~Y.}\ \bibnamefont {Simmons}}, \bibinfo {author} {\bibfnamefont {A.~R.}\ \bibnamefont {Hamilton}}, \bibinfo {author} {\bibfnamefont {M.}~\bibnamefont {Pepper}}, \bibinfo {author} {\bibfnamefont {D.~A.}\ \bibnamefont {Ritchie}},\ and\ \bibinfo {author} {\bibfnamefont {U.}~\bibnamefont {Z\"ulicke}},\ }\bibfield  {title} {\bibinfo {title} {Zeeman splitting in ballistic hole quantum wires},\ }\href {https://doi.org/10.1103/PhysRevLett.97.026403} {\bibfield  {journal} {\bibinfo  {journal} {Phys. Rev. Lett.}\ }\textbf {\bibinfo {volume} {97}},\ \bibinfo {pages} {026403} (\bibinfo {year} {2006})}\BibitemShut {NoStop}%
\bibitem [{\citenamefont {Ares}\ \emph {et~al.}(2013{\natexlab{a}})\citenamefont {Ares}, \citenamefont {Golovach}, \citenamefont {Katsaros}, \citenamefont {Stoffel}, \citenamefont {Fournel}, \citenamefont {Glazman}, \citenamefont {Schmidt},\ and\ \citenamefont {De~Franceschi}}]{Ares2013PRL}%
  \BibitemOpen
  \bibfield  {author} {\bibinfo {author} {\bibfnamefont {N.}~\bibnamefont {Ares}}, \bibinfo {author} {\bibfnamefont {V.~N.}\ \bibnamefont {Golovach}}, \bibinfo {author} {\bibfnamefont {G.}~\bibnamefont {Katsaros}}, \bibinfo {author} {\bibfnamefont {M.}~\bibnamefont {Stoffel}}, \bibinfo {author} {\bibfnamefont {F.}~\bibnamefont {Fournel}}, \bibinfo {author} {\bibfnamefont {L.~I.}\ \bibnamefont {Glazman}}, \bibinfo {author} {\bibfnamefont {O.~G.}\ \bibnamefont {Schmidt}},\ and\ \bibinfo {author} {\bibfnamefont {S.}~\bibnamefont {De~Franceschi}},\ }\bibfield  {title} {\bibinfo {title} {Nature of tunable hole $g$ factors in quantum dots},\ }\href {https://doi.org/10.1103/PhysRevLett.110.046602} {\bibfield  {journal} {\bibinfo  {journal} {Phys. Rev. Lett.}\ }\textbf {\bibinfo {volume} {110}},\ \bibinfo {pages} {046602} (\bibinfo {year} {2013}{\natexlab{a}})}\BibitemShut {NoStop}%
\bibitem [{\citenamefont {Ares}\ \emph {et~al.}(2013{\natexlab{b}})\citenamefont {Ares}, \citenamefont {Katsaros}, \citenamefont {Golovach}, \citenamefont {Zhang}, \citenamefont {Prager}, \citenamefont {Glazman}, \citenamefont {Schmidt},\ and\ \citenamefont {De~Franceschi}}]{Ares2013APL}%
  \BibitemOpen
  \bibfield  {author} {\bibinfo {author} {\bibfnamefont {N.}~\bibnamefont {Ares}}, \bibinfo {author} {\bibfnamefont {G.}~\bibnamefont {Katsaros}}, \bibinfo {author} {\bibfnamefont {V.~N.}\ \bibnamefont {Golovach}}, \bibinfo {author} {\bibfnamefont {J.~J.}\ \bibnamefont {Zhang}}, \bibinfo {author} {\bibfnamefont {A.}~\bibnamefont {Prager}}, \bibinfo {author} {\bibfnamefont {L.~I.}\ \bibnamefont {Glazman}}, \bibinfo {author} {\bibfnamefont {O.~G.}\ \bibnamefont {Schmidt}},\ and\ \bibinfo {author} {\bibfnamefont {S.}~\bibnamefont {De~Franceschi}},\ }\bibfield  {title} {\bibinfo {title} {{SiGe quantum dots for fast hole spin Rabi oscillations}},\ }\href {https://doi.org/10.1063/1.4858959} {\bibfield  {journal} {\bibinfo  {journal} {Applied Physics Letters}\ }\textbf {\bibinfo {volume} {103}},\ \bibinfo {pages} {263113} (\bibinfo {year} {2013}{\natexlab{b}})}\BibitemShut {NoStop}%
\bibitem [{\citenamefont {Brauns}\ \emph {et~al.}(2016)\citenamefont {Brauns}, \citenamefont {Ridderbos}, \citenamefont {Li}, \citenamefont {Bakkers}, \citenamefont {van~der Wiel},\ and\ \citenamefont {Zwanenburg}}]{Brauns2016}%
  \BibitemOpen
  \bibfield  {author} {\bibinfo {author} {\bibfnamefont {M.}~\bibnamefont {Brauns}}, \bibinfo {author} {\bibfnamefont {J.}~\bibnamefont {Ridderbos}}, \bibinfo {author} {\bibfnamefont {A.}~\bibnamefont {Li}}, \bibinfo {author} {\bibfnamefont {E.~P. A.~M.}\ \bibnamefont {Bakkers}}, \bibinfo {author} {\bibfnamefont {W.~G.}\ \bibnamefont {van~der Wiel}},\ and\ \bibinfo {author} {\bibfnamefont {F.~A.}\ \bibnamefont {Zwanenburg}},\ }\bibfield  {title} {\bibinfo {title} {Anisotropic pauli spin blockade in hole quantum dots},\ }\href {https://doi.org/10.1103/PhysRevB.94.041411} {\bibfield  {journal} {\bibinfo  {journal} {Phys. Rev. B}\ }\textbf {\bibinfo {volume} {94}},\ \bibinfo {pages} {041411} (\bibinfo {year} {2016})}\BibitemShut {NoStop}%
\bibitem [{\citenamefont {Watzinger}\ \emph {et~al.}(2016)\citenamefont {Watzinger}, \citenamefont {Kloeffel}, \citenamefont {Vuku{\v s}i{\'c}}, \citenamefont {Rossell}, \citenamefont {Sessi}, \citenamefont {Kuku{\v c}ka}, \citenamefont {Kirchschlager}, \citenamefont {Lausecker}, \citenamefont {Truhlar}, \citenamefont {Glaser}, \citenamefont {Rastelli}, \citenamefont {Fuhrer}, \citenamefont {Loss},\ and\ \citenamefont {Katsaros}}]{Watzinger2016}%
  \BibitemOpen
  \bibfield  {author} {\bibinfo {author} {\bibfnamefont {H.}~\bibnamefont {Watzinger}}, \bibinfo {author} {\bibfnamefont {C.}~\bibnamefont {Kloeffel}}, \bibinfo {author} {\bibfnamefont {L.}~\bibnamefont {Vuku{\v s}i{\'c}}}, \bibinfo {author} {\bibfnamefont {M.~D.}\ \bibnamefont {Rossell}}, \bibinfo {author} {\bibfnamefont {V.}~\bibnamefont {Sessi}}, \bibinfo {author} {\bibfnamefont {J.}~\bibnamefont {Kuku{\v c}ka}}, \bibinfo {author} {\bibfnamefont {R.}~\bibnamefont {Kirchschlager}}, \bibinfo {author} {\bibfnamefont {E.}~\bibnamefont {Lausecker}}, \bibinfo {author} {\bibfnamefont {A.}~\bibnamefont {Truhlar}}, \bibinfo {author} {\bibfnamefont {M.}~\bibnamefont {Glaser}}, \bibinfo {author} {\bibfnamefont {A.}~\bibnamefont {Rastelli}}, \bibinfo {author} {\bibfnamefont {A.}~\bibnamefont {Fuhrer}}, \bibinfo {author} {\bibfnamefont {D.}~\bibnamefont {Loss}},\ and\ \bibinfo {author} {\bibfnamefont {G.}~\bibnamefont {Katsaros}},\ }\bibfield  {title} {\bibinfo {title} {Heavy-hole states in germanium hut wires},\
  }\href {https://doi.org/10.1021/acs.nanolett.6b02715} {\bibfield  {journal} {\bibinfo  {journal} {Nano Letters}\ }\textbf {\bibinfo {volume} {16}},\ \bibinfo {pages} {6879} (\bibinfo {year} {2016})}\BibitemShut {NoStop}%
\bibitem [{\citenamefont {Voisin}\ \emph {et~al.}(2016)\citenamefont {Voisin}, \citenamefont {Maurand}, \citenamefont {Barraud}, \citenamefont {Vinet}, \citenamefont {Jehl}, \citenamefont {Sanquer}, \citenamefont {Renard},\ and\ \citenamefont {De~Franceschi}}]{Voisin2016}%
  \BibitemOpen
  \bibfield  {author} {\bibinfo {author} {\bibfnamefont {B.}~\bibnamefont {Voisin}}, \bibinfo {author} {\bibfnamefont {R.}~\bibnamefont {Maurand}}, \bibinfo {author} {\bibfnamefont {S.}~\bibnamefont {Barraud}}, \bibinfo {author} {\bibfnamefont {M.}~\bibnamefont {Vinet}}, \bibinfo {author} {\bibfnamefont {X.}~\bibnamefont {Jehl}}, \bibinfo {author} {\bibfnamefont {M.}~\bibnamefont {Sanquer}}, \bibinfo {author} {\bibfnamefont {J.}~\bibnamefont {Renard}},\ and\ \bibinfo {author} {\bibfnamefont {S.}~\bibnamefont {De~Franceschi}},\ }\bibfield  {title} {\bibinfo {title} {Electrical control of g-factor in a few-hole silicon nanowire mosfet},\ }\href {https://doi.org/10.1021/acs.nanolett.5b02920} {\bibfield  {journal} {\bibinfo  {journal} {Nano Letters}\ }\textbf {\bibinfo {volume} {16}},\ \bibinfo {pages} {88} (\bibinfo {year} {2016})}\BibitemShut {NoStop}%
\bibitem [{\citenamefont {Srinivasan}\ \emph {et~al.}(2016)\citenamefont {Srinivasan}, \citenamefont {Hudson}, \citenamefont {Miserev}, \citenamefont {Yeoh}, \citenamefont {Klochan}, \citenamefont {Muraki}, \citenamefont {Hirayama}, \citenamefont {Sushkov},\ and\ \citenamefont {Hamilton}}]{Srinivasan2016}%
  \BibitemOpen
  \bibfield  {author} {\bibinfo {author} {\bibfnamefont {A.}~\bibnamefont {Srinivasan}}, \bibinfo {author} {\bibfnamefont {K.~L.}\ \bibnamefont {Hudson}}, \bibinfo {author} {\bibfnamefont {D.}~\bibnamefont {Miserev}}, \bibinfo {author} {\bibfnamefont {L.~A.}\ \bibnamefont {Yeoh}}, \bibinfo {author} {\bibfnamefont {O.}~\bibnamefont {Klochan}}, \bibinfo {author} {\bibfnamefont {K.}~\bibnamefont {Muraki}}, \bibinfo {author} {\bibfnamefont {Y.}~\bibnamefont {Hirayama}}, \bibinfo {author} {\bibfnamefont {O.~P.}\ \bibnamefont {Sushkov}},\ and\ \bibinfo {author} {\bibfnamefont {A.~R.}\ \bibnamefont {Hamilton}},\ }\bibfield  {title} {\bibinfo {title} {Electrical control of the sign of the $g$ factor in a gaas hole quantum point contact},\ }\href {https://doi.org/10.1103/PhysRevB.94.041406} {\bibfield  {journal} {\bibinfo  {journal} {Phys. Rev. B}\ }\textbf {\bibinfo {volume} {94}},\ \bibinfo {pages} {041406} (\bibinfo {year} {2016})}\BibitemShut {NoStop}%
\bibitem [{\citenamefont {Miserev}\ \emph {et~al.}(2017)\citenamefont {Miserev}, \citenamefont {Srinivasan}, \citenamefont {Tkachenko}, \citenamefont {Tkachenko}, \citenamefont {Farrer}, \citenamefont {Ritchie}, \citenamefont {Hamilton},\ and\ \citenamefont {Sushkov}}]{Miserev2017}%
  \BibitemOpen
  \bibfield  {author} {\bibinfo {author} {\bibfnamefont {D.~S.}\ \bibnamefont {Miserev}}, \bibinfo {author} {\bibfnamefont {A.}~\bibnamefont {Srinivasan}}, \bibinfo {author} {\bibfnamefont {O.~A.}\ \bibnamefont {Tkachenko}}, \bibinfo {author} {\bibfnamefont {V.~A.}\ \bibnamefont {Tkachenko}}, \bibinfo {author} {\bibfnamefont {I.}~\bibnamefont {Farrer}}, \bibinfo {author} {\bibfnamefont {D.~A.}\ \bibnamefont {Ritchie}}, \bibinfo {author} {\bibfnamefont {A.~R.}\ \bibnamefont {Hamilton}},\ and\ \bibinfo {author} {\bibfnamefont {O.~P.}\ \bibnamefont {Sushkov}},\ }\bibfield  {title} {\bibinfo {title} {Mechanisms for strong anisotropy of in-plane $g$-factors in hole based quantum point contacts},\ }\href {https://doi.org/10.1103/PhysRevLett.119.116803} {\bibfield  {journal} {\bibinfo  {journal} {Phys. Rev. Lett.}\ }\textbf {\bibinfo {volume} {119}},\ \bibinfo {pages} {116803} (\bibinfo {year} {2017})}\BibitemShut {NoStop}%
\bibitem [{\citenamefont {Hung}\ \emph {et~al.}(2017)\citenamefont {Hung}, \citenamefont {Marcellina}, \citenamefont {Wang}, \citenamefont {Hamilton},\ and\ \citenamefont {Culcer}}]{Hung2017}%
  \BibitemOpen
  \bibfield  {author} {\bibinfo {author} {\bibfnamefont {J.-T.}\ \bibnamefont {Hung}}, \bibinfo {author} {\bibfnamefont {E.}~\bibnamefont {Marcellina}}, \bibinfo {author} {\bibfnamefont {B.}~\bibnamefont {Wang}}, \bibinfo {author} {\bibfnamefont {A.~R.}\ \bibnamefont {Hamilton}},\ and\ \bibinfo {author} {\bibfnamefont {D.}~\bibnamefont {Culcer}},\ }\bibfield  {title} {\bibinfo {title} {Spin blockade in hole quantum dots: Tuning exchange electrically and probing zeeman interactions},\ }\href {https://doi.org/10.1103/PhysRevB.95.195316} {\bibfield  {journal} {\bibinfo  {journal} {Phys. Rev. B}\ }\textbf {\bibinfo {volume} {95}},\ \bibinfo {pages} {195316} (\bibinfo {year} {2017})}\BibitemShut {NoStop}%
\bibitem [{\citenamefont {Marcellina}\ \emph {et~al.}(2018)\citenamefont {Marcellina}, \citenamefont {Srinivasan}, \citenamefont {Miserev}, \citenamefont {Croxall}, \citenamefont {Ritchie}, \citenamefont {Farrer}, \citenamefont {Sushkov}, \citenamefont {Culcer},\ and\ \citenamefont {Hamilton}}]{Marcellina2018}%
  \BibitemOpen
  \bibfield  {author} {\bibinfo {author} {\bibfnamefont {E.}~\bibnamefont {Marcellina}}, \bibinfo {author} {\bibfnamefont {A.}~\bibnamefont {Srinivasan}}, \bibinfo {author} {\bibfnamefont {D.~S.}\ \bibnamefont {Miserev}}, \bibinfo {author} {\bibfnamefont {A.~F.}\ \bibnamefont {Croxall}}, \bibinfo {author} {\bibfnamefont {D.~A.}\ \bibnamefont {Ritchie}}, \bibinfo {author} {\bibfnamefont {I.}~\bibnamefont {Farrer}}, \bibinfo {author} {\bibfnamefont {O.~P.}\ \bibnamefont {Sushkov}}, \bibinfo {author} {\bibfnamefont {D.}~\bibnamefont {Culcer}},\ and\ \bibinfo {author} {\bibfnamefont {A.~R.}\ \bibnamefont {Hamilton}},\ }\bibfield  {title} {\bibinfo {title} {Electrical control of the zeeman spin splitting in two-dimensional hole systems},\ }\href {https://doi.org/10.1103/PhysRevLett.121.077701} {\bibfield  {journal} {\bibinfo  {journal} {Phys. Rev. Lett.}\ }\textbf {\bibinfo {volume} {121}},\ \bibinfo {pages} {077701} (\bibinfo {year} {2018})}\BibitemShut {NoStop}%
\bibitem [{\citenamefont {Mizokuchi}\ \emph {et~al.}(2018)\citenamefont {Mizokuchi}, \citenamefont {Maurand}, \citenamefont {Vigneau}, \citenamefont {Myronov},\ and\ \citenamefont {De~Franceschi}}]{Mizokuchi2018}%
  \BibitemOpen
  \bibfield  {author} {\bibinfo {author} {\bibfnamefont {R.}~\bibnamefont {Mizokuchi}}, \bibinfo {author} {\bibfnamefont {R.}~\bibnamefont {Maurand}}, \bibinfo {author} {\bibfnamefont {F.}~\bibnamefont {Vigneau}}, \bibinfo {author} {\bibfnamefont {M.}~\bibnamefont {Myronov}},\ and\ \bibinfo {author} {\bibfnamefont {S.}~\bibnamefont {De~Franceschi}},\ }\bibfield  {title} {\bibinfo {title} {Ballistic one-dimensional holes with strong g-factor anisotropy in germanium},\ }\href {https://doi.org/10.1021/acs.nanolett.8b01457} {\bibfield  {journal} {\bibinfo  {journal} {Nano Letters}\ }\textbf {\bibinfo {volume} {18}},\ \bibinfo {pages} {4861} (\bibinfo {year} {2018})}\BibitemShut {NoStop}%
\bibitem [{\citenamefont {Wei}\ \emph {et~al.}(2020)\citenamefont {Wei}, \citenamefont {Mizoguchi}, \citenamefont {Mizokuchi},\ and\ \citenamefont {Kodera}}]{Wei2020}%
  \BibitemOpen
  \bibfield  {author} {\bibinfo {author} {\bibfnamefont {H.}~\bibnamefont {Wei}}, \bibinfo {author} {\bibfnamefont {S.}~\bibnamefont {Mizoguchi}}, \bibinfo {author} {\bibfnamefont {R.}~\bibnamefont {Mizokuchi}},\ and\ \bibinfo {author} {\bibfnamefont {T.}~\bibnamefont {Kodera}},\ }\bibfield  {title} {\bibinfo {title} {Estimation of hole spin g-factors in p-channel silicon single and double quantum dots towards spin manipulation},\ }\href {https://doi.org/10.35848/1347-4065/ab6b7e} {\bibfield  {journal} {\bibinfo  {journal} {Japanese Journal of Applied Physics}\ }\textbf {\bibinfo {volume} {59}},\ \bibinfo {pages} {SGGI10} (\bibinfo {year} {2020})}\BibitemShut {NoStop}%
\bibitem [{\citenamefont {Zhang}\ \emph {et~al.}(2021)\citenamefont {Zhang}, \citenamefont {Liu}, \citenamefont {Gao}, \citenamefont {Xu}, \citenamefont {Wang}, \citenamefont {Zhang}, \citenamefont {Cao}, \citenamefont {Wang}, \citenamefont {Zhang}, \citenamefont {Hu}, \citenamefont {Li},\ and\ \citenamefont {Guo}}]{Zhang2021}%
  \BibitemOpen
  \bibfield  {author} {\bibinfo {author} {\bibfnamefont {T.}~\bibnamefont {Zhang}}, \bibinfo {author} {\bibfnamefont {H.}~\bibnamefont {Liu}}, \bibinfo {author} {\bibfnamefont {F.}~\bibnamefont {Gao}}, \bibinfo {author} {\bibfnamefont {G.}~\bibnamefont {Xu}}, \bibinfo {author} {\bibfnamefont {K.}~\bibnamefont {Wang}}, \bibinfo {author} {\bibfnamefont {X.}~\bibnamefont {Zhang}}, \bibinfo {author} {\bibfnamefont {G.}~\bibnamefont {Cao}}, \bibinfo {author} {\bibfnamefont {T.}~\bibnamefont {Wang}}, \bibinfo {author} {\bibfnamefont {J.}~\bibnamefont {Zhang}}, \bibinfo {author} {\bibfnamefont {X.}~\bibnamefont {Hu}}, \bibinfo {author} {\bibfnamefont {H.-O.}\ \bibnamefont {Li}},\ and\ \bibinfo {author} {\bibfnamefont {G.-P.}\ \bibnamefont {Guo}},\ }\bibfield  {title} {\bibinfo {title} {Anisotropic g-factor and spin--orbit field in a germanium hut wire double quantum dot},\ }\href {https://doi.org/10.1021/acs.nanolett.1c00263} {\bibfield  {journal} {\bibinfo  {journal} {Nano Letters}\ }\textbf {\bibinfo {volume}
  {21}},\ \bibinfo {pages} {3835} (\bibinfo {year} {2021})}\BibitemShut {NoStop}%
\bibitem [{\citenamefont {Liles}\ \emph {et~al.}(2021)\citenamefont {Liles}, \citenamefont {Martins}, \citenamefont {Miserev}, \citenamefont {Kiselev}, \citenamefont {Thorvaldson}, \citenamefont {Rendell}, \citenamefont {Jin}, \citenamefont {Hudson}, \citenamefont {Veldhorst}, \citenamefont {Itoh}, \citenamefont {Sushkov}, \citenamefont {Ladd}, \citenamefont {Dzurak},\ and\ \citenamefont {Hamilton}}]{Liles2021}%
  \BibitemOpen
  \bibfield  {author} {\bibinfo {author} {\bibfnamefont {S.~D.}\ \bibnamefont {Liles}}, \bibinfo {author} {\bibfnamefont {F.}~\bibnamefont {Martins}}, \bibinfo {author} {\bibfnamefont {D.~S.}\ \bibnamefont {Miserev}}, \bibinfo {author} {\bibfnamefont {A.~A.}\ \bibnamefont {Kiselev}}, \bibinfo {author} {\bibfnamefont {I.~D.}\ \bibnamefont {Thorvaldson}}, \bibinfo {author} {\bibfnamefont {M.~J.}\ \bibnamefont {Rendell}}, \bibinfo {author} {\bibfnamefont {I.~K.}\ \bibnamefont {Jin}}, \bibinfo {author} {\bibfnamefont {F.~E.}\ \bibnamefont {Hudson}}, \bibinfo {author} {\bibfnamefont {M.}~\bibnamefont {Veldhorst}}, \bibinfo {author} {\bibfnamefont {K.~M.}\ \bibnamefont {Itoh}}, \bibinfo {author} {\bibfnamefont {O.~P.}\ \bibnamefont {Sushkov}}, \bibinfo {author} {\bibfnamefont {T.~D.}\ \bibnamefont {Ladd}}, \bibinfo {author} {\bibfnamefont {A.~S.}\ \bibnamefont {Dzurak}},\ and\ \bibinfo {author} {\bibfnamefont {A.~R.}\ \bibnamefont {Hamilton}},\ }\bibfield  {title} {\bibinfo {title} {Electrical control of the $g$
  tensor of the first hole in a silicon mos quantum dot},\ }\href {https://doi.org/10.1103/PhysRevB.104.235303} {\bibfield  {journal} {\bibinfo  {journal} {Phys. Rev. B}\ }\textbf {\bibinfo {volume} {104}},\ \bibinfo {pages} {235303} (\bibinfo {year} {2021})}\BibitemShut {NoStop}%
\bibitem [{\citenamefont {Qvist}\ and\ \citenamefont {Danon}(2022)}]{Qvist2022}%
  \BibitemOpen
  \bibfield  {author} {\bibinfo {author} {\bibfnamefont {J.~H.}\ \bibnamefont {Qvist}}\ and\ \bibinfo {author} {\bibfnamefont {J.}~\bibnamefont {Danon}},\ }\bibfield  {title} {\bibinfo {title} {Anisotropic $g$-tensors in hole quantum dots: Role of transverse confinement direction},\ }\href {https://doi.org/10.1103/PhysRevB.105.075303} {\bibfield  {journal} {\bibinfo  {journal} {Phys. Rev. B}\ }\textbf {\bibinfo {volume} {105}},\ \bibinfo {pages} {075303} (\bibinfo {year} {2022})}\BibitemShut {NoStop}%
\bibitem [{\citenamefont {Abadillo-Uriel}\ \emph {et~al.}(2023)\citenamefont {Abadillo-Uriel}, \citenamefont {Rodr\'{\i}guez-Mena}, \citenamefont {Martinez},\ and\ \citenamefont {Niquet}}]{Abadillo2023}%
  \BibitemOpen
  \bibfield  {author} {\bibinfo {author} {\bibfnamefont {J.~C.}\ \bibnamefont {Abadillo-Uriel}}, \bibinfo {author} {\bibfnamefont {E.~A.}\ \bibnamefont {Rodr\'{\i}guez-Mena}}, \bibinfo {author} {\bibfnamefont {B.}~\bibnamefont {Martinez}},\ and\ \bibinfo {author} {\bibfnamefont {Y.-M.}\ \bibnamefont {Niquet}},\ }\bibfield  {title} {\bibinfo {title} {Hole-spin driving by strain-induced spin-orbit interactions},\ }\href {https://doi.org/10.1103/PhysRevLett.131.097002} {\bibfield  {journal} {\bibinfo  {journal} {Phys. Rev. Lett.}\ }\textbf {\bibinfo {volume} {131}},\ \bibinfo {pages} {097002} (\bibinfo {year} {2023})}\BibitemShut {NoStop}%
\bibitem [{\citenamefont {Bulaev}\ and\ \citenamefont {Loss}(2007)}]{Bulaev2007}%
  \BibitemOpen
  \bibfield  {author} {\bibinfo {author} {\bibfnamefont {D.~V.}\ \bibnamefont {Bulaev}}\ and\ \bibinfo {author} {\bibfnamefont {D.}~\bibnamefont {Loss}},\ }\bibfield  {title} {\bibinfo {title} {Electric dipole spin resonance for heavy holes in quantum dots},\ }\href {https://doi.org/10.1103/PhysRevLett.98.097202} {\bibfield  {journal} {\bibinfo  {journal} {Phys. Rev. Lett.}\ }\textbf {\bibinfo {volume} {98}},\ \bibinfo {pages} {097202} (\bibinfo {year} {2007})}\BibitemShut {NoStop}%
\bibitem [{\citenamefont {Kloeffel}\ \emph {et~al.}(2011)\citenamefont {Kloeffel}, \citenamefont {Trif},\ and\ \citenamefont {Loss}}]{Kloeffel2011}%
  \BibitemOpen
  \bibfield  {author} {\bibinfo {author} {\bibfnamefont {C.}~\bibnamefont {Kloeffel}}, \bibinfo {author} {\bibfnamefont {M.}~\bibnamefont {Trif}},\ and\ \bibinfo {author} {\bibfnamefont {D.}~\bibnamefont {Loss}},\ }\bibfield  {title} {\bibinfo {title} {Strong spin-orbit interaction and helical hole states in ge/si nanowires},\ }\href {https://doi.org/10.1103/PhysRevB.84.195314} {\bibfield  {journal} {\bibinfo  {journal} {Phys. Rev. B}\ }\textbf {\bibinfo {volume} {84}},\ \bibinfo {pages} {195314} (\bibinfo {year} {2011})}\BibitemShut {NoStop}%
\bibitem [{\citenamefont {Kloeffel}\ \emph {et~al.}(2013)\citenamefont {Kloeffel}, \citenamefont {Trif}, \citenamefont {Stano},\ and\ \citenamefont {Loss}}]{Kloeffel2013}%
  \BibitemOpen
  \bibfield  {author} {\bibinfo {author} {\bibfnamefont {C.}~\bibnamefont {Kloeffel}}, \bibinfo {author} {\bibfnamefont {M.}~\bibnamefont {Trif}}, \bibinfo {author} {\bibfnamefont {P.}~\bibnamefont {Stano}},\ and\ \bibinfo {author} {\bibfnamefont {D.}~\bibnamefont {Loss}},\ }\bibfield  {title} {\bibinfo {title} {Circuit qed with hole-spin qubits in ge/si nanowire quantum dots},\ }\href {https://doi.org/10.1103/PhysRevB.88.241405} {\bibfield  {journal} {\bibinfo  {journal} {Phys. Rev. B}\ }\textbf {\bibinfo {volume} {88}},\ \bibinfo {pages} {241405} (\bibinfo {year} {2013})}\BibitemShut {NoStop}%
\bibitem [{\citenamefont {Chesi}\ \emph {et~al.}(2014)\citenamefont {Chesi}, \citenamefont {Wang},\ and\ \citenamefont {Coish}}]{Chesi2014}%
  \BibitemOpen
  \bibfield  {author} {\bibinfo {author} {\bibfnamefont {S.}~\bibnamefont {Chesi}}, \bibinfo {author} {\bibfnamefont {X.~J.}\ \bibnamefont {Wang}},\ and\ \bibinfo {author} {\bibfnamefont {W.~A.}\ \bibnamefont {Coish}},\ }\bibfield  {title} {\bibinfo {title} {Controlling hole spins in quantum dots and wells},\ }\href {https://doi.org/10.1140/epjp/i2014-14086-2} {\bibfield  {journal} {\bibinfo  {journal} {The European Physical Journal Plus}\ }\textbf {\bibinfo {volume} {129}},\ \bibinfo {pages} {86} (\bibinfo {year} {2014})}\BibitemShut {NoStop}%
\bibitem [{\citenamefont {Dobbie}\ \emph {et~al.}(2012)\citenamefont {Dobbie}, \citenamefont {Myronov}, \citenamefont {Morris}, \citenamefont {Hassan}, \citenamefont {Prest}, \citenamefont {Shah}, \citenamefont {Parker}, \citenamefont {Whall},\ and\ \citenamefont {Leadley}}]{Dobbie2012}%
  \BibitemOpen
  \bibfield  {author} {\bibinfo {author} {\bibfnamefont {A.}~\bibnamefont {Dobbie}}, \bibinfo {author} {\bibfnamefont {M.}~\bibnamefont {Myronov}}, \bibinfo {author} {\bibfnamefont {R.~J.~H.}\ \bibnamefont {Morris}}, \bibinfo {author} {\bibfnamefont {A.~H.~A.}\ \bibnamefont {Hassan}}, \bibinfo {author} {\bibfnamefont {M.~J.}\ \bibnamefont {Prest}}, \bibinfo {author} {\bibfnamefont {V.~A.}\ \bibnamefont {Shah}}, \bibinfo {author} {\bibfnamefont {E.~H.~C.}\ \bibnamefont {Parker}}, \bibinfo {author} {\bibfnamefont {T.~E.}\ \bibnamefont {Whall}},\ and\ \bibinfo {author} {\bibfnamefont {D.~R.}\ \bibnamefont {Leadley}},\ }\bibfield  {title} {\bibinfo {title} {{Ultra-high hole mobility exceeding one million in a strained germanium quantum well}},\ }\href {https://doi.org/10.1063/1.4763476} {\bibfield  {journal} {\bibinfo  {journal} {Applied Physics Letters}\ }\textbf {\bibinfo {volume} {101}},\ \bibinfo {pages} {172108} (\bibinfo {year} {2012})}\BibitemShut {NoStop}%
\bibitem [{\citenamefont {Sammak}\ \emph {et~al.}(2019)\citenamefont {Sammak}, \citenamefont {Sabbagh}, \citenamefont {Hendrickx}, \citenamefont {Lodari}, \citenamefont {Paquelet~Wuetz}, \citenamefont {Tosato}, \citenamefont {Yeoh}, \citenamefont {Bollani}, \citenamefont {Virgilio}, \citenamefont {Schubert}, \citenamefont {Zaumseil}, \citenamefont {Capellini}, \citenamefont {Veldhorst},\ and\ \citenamefont {Scappucci}}]{Sammak2019}%
  \BibitemOpen
  \bibfield  {author} {\bibinfo {author} {\bibfnamefont {A.}~\bibnamefont {Sammak}}, \bibinfo {author} {\bibfnamefont {D.}~\bibnamefont {Sabbagh}}, \bibinfo {author} {\bibfnamefont {N.~W.}\ \bibnamefont {Hendrickx}}, \bibinfo {author} {\bibfnamefont {M.}~\bibnamefont {Lodari}}, \bibinfo {author} {\bibfnamefont {B.}~\bibnamefont {Paquelet~Wuetz}}, \bibinfo {author} {\bibfnamefont {A.}~\bibnamefont {Tosato}}, \bibinfo {author} {\bibfnamefont {L.}~\bibnamefont {Yeoh}}, \bibinfo {author} {\bibfnamefont {M.}~\bibnamefont {Bollani}}, \bibinfo {author} {\bibfnamefont {M.}~\bibnamefont {Virgilio}}, \bibinfo {author} {\bibfnamefont {M.~A.}\ \bibnamefont {Schubert}}, \bibinfo {author} {\bibfnamefont {P.}~\bibnamefont {Zaumseil}}, \bibinfo {author} {\bibfnamefont {G.}~\bibnamefont {Capellini}}, \bibinfo {author} {\bibfnamefont {M.}~\bibnamefont {Veldhorst}},\ and\ \bibinfo {author} {\bibfnamefont {G.}~\bibnamefont {Scappucci}},\ }\bibfield  {title} {\bibinfo {title} {Shallow and undoped germanium quantum wells: A
  playground for spin and hybrid quantum technology},\ }\href {https://doi.org/https://doi.org/10.1002/adfm.201807613} {\bibfield  {journal} {\bibinfo  {journal} {Advanced Functional Materials}\ }\textbf {\bibinfo {volume} {29}},\ \bibinfo {pages} {1807613} (\bibinfo {year} {2019})}\BibitemShut {NoStop}%
\bibitem [{\citenamefont {Lodari}\ \emph {et~al.}(2019)\citenamefont {Lodari}, \citenamefont {Tosato}, \citenamefont {Sabbagh}, \citenamefont {Schubert}, \citenamefont {Capellini}, \citenamefont {Sammak}, \citenamefont {Veldhorst},\ and\ \citenamefont {Scappucci}}]{Lodari2019}%
  \BibitemOpen
  \bibfield  {author} {\bibinfo {author} {\bibfnamefont {M.}~\bibnamefont {Lodari}}, \bibinfo {author} {\bibfnamefont {A.}~\bibnamefont {Tosato}}, \bibinfo {author} {\bibfnamefont {D.}~\bibnamefont {Sabbagh}}, \bibinfo {author} {\bibfnamefont {M.~A.}\ \bibnamefont {Schubert}}, \bibinfo {author} {\bibfnamefont {G.}~\bibnamefont {Capellini}}, \bibinfo {author} {\bibfnamefont {A.}~\bibnamefont {Sammak}}, \bibinfo {author} {\bibfnamefont {M.}~\bibnamefont {Veldhorst}},\ and\ \bibinfo {author} {\bibfnamefont {G.}~\bibnamefont {Scappucci}},\ }\bibfield  {title} {\bibinfo {title} {Light effective hole mass in undoped ge/sige quantum wells},\ }\href {https://doi.org/10.1103/PhysRevB.100.041304} {\bibfield  {journal} {\bibinfo  {journal} {Phys. Rev. B}\ }\textbf {\bibinfo {volume} {100}},\ \bibinfo {pages} {041304} (\bibinfo {year} {2019})}\BibitemShut {NoStop}%
\bibitem [{\citenamefont {Lodari}\ \emph {et~al.}(2021)\citenamefont {Lodari}, \citenamefont {Hendrickx}, \citenamefont {Lawrie}, \citenamefont {Hsiao}, \citenamefont {Vandersypen}, \citenamefont {Sammak}, \citenamefont {Veldhorst},\ and\ \citenamefont {Scappucci}}]{Lodari2021}%
  \BibitemOpen
  \bibfield  {author} {\bibinfo {author} {\bibfnamefont {M.}~\bibnamefont {Lodari}}, \bibinfo {author} {\bibfnamefont {N.~W.}\ \bibnamefont {Hendrickx}}, \bibinfo {author} {\bibfnamefont {W.~I.~L.}\ \bibnamefont {Lawrie}}, \bibinfo {author} {\bibfnamefont {T.-K.}\ \bibnamefont {Hsiao}}, \bibinfo {author} {\bibfnamefont {L.~M.~K.}\ \bibnamefont {Vandersypen}}, \bibinfo {author} {\bibfnamefont {A.}~\bibnamefont {Sammak}}, \bibinfo {author} {\bibfnamefont {M.}~\bibnamefont {Veldhorst}},\ and\ \bibinfo {author} {\bibfnamefont {G.}~\bibnamefont {Scappucci}},\ }\bibfield  {title} {\bibinfo {title} {Low percolation density and charge noise with holes in germanium},\ }\href {https://doi.org/10.1088/2633-4356/abcd82} {\bibfield  {journal} {\bibinfo  {journal} {Materials for Quantum Technology}\ }\textbf {\bibinfo {volume} {1}},\ \bibinfo {pages} {011002} (\bibinfo {year} {2021})}\BibitemShut {NoStop}%
\bibitem [{\citenamefont {Hendrickx}\ \emph {et~al.}(2018)\citenamefont {Hendrickx}, \citenamefont {Franke}, \citenamefont {Sammak}, \citenamefont {Kouwenhoven}, \citenamefont {Sabbagh}, \citenamefont {Yeoh}, \citenamefont {Li}, \citenamefont {Tagliaferri}, \citenamefont {Virgilio}, \citenamefont {Capellini}, \citenamefont {Scappucci},\ and\ \citenamefont {Veldhorst}}]{Hendrickx2018}%
  \BibitemOpen
  \bibfield  {author} {\bibinfo {author} {\bibfnamefont {N.~W.}\ \bibnamefont {Hendrickx}}, \bibinfo {author} {\bibfnamefont {D.~P.}\ \bibnamefont {Franke}}, \bibinfo {author} {\bibfnamefont {A.}~\bibnamefont {Sammak}}, \bibinfo {author} {\bibfnamefont {M.}~\bibnamefont {Kouwenhoven}}, \bibinfo {author} {\bibfnamefont {D.}~\bibnamefont {Sabbagh}}, \bibinfo {author} {\bibfnamefont {L.}~\bibnamefont {Yeoh}}, \bibinfo {author} {\bibfnamefont {R.}~\bibnamefont {Li}}, \bibinfo {author} {\bibfnamefont {M.~L.~V.}\ \bibnamefont {Tagliaferri}}, \bibinfo {author} {\bibfnamefont {M.}~\bibnamefont {Virgilio}}, \bibinfo {author} {\bibfnamefont {G.}~\bibnamefont {Capellini}}, \bibinfo {author} {\bibfnamefont {G.}~\bibnamefont {Scappucci}},\ and\ \bibinfo {author} {\bibfnamefont {M.}~\bibnamefont {Veldhorst}},\ }\bibfield  {title} {\bibinfo {title} {Gate-controlled quantum dots and superconductivity in planar germanium},\ }\href {https://doi.org/10.1038/s41467-018-05299-x} {\bibfield  {journal} {\bibinfo  {journal} {Nature
  Communications}\ }\textbf {\bibinfo {volume} {9}},\ \bibinfo {pages} {2835} (\bibinfo {year} {2018})}\BibitemShut {NoStop}%
\bibitem [{\citenamefont {Terrazos}\ \emph {et~al.}(2021)\citenamefont {Terrazos}, \citenamefont {Marcellina}, \citenamefont {Wang}, \citenamefont {Coppersmith}, \citenamefont {Friesen}, \citenamefont {Hamilton}, \citenamefont {Hu}, \citenamefont {Koiller}, \citenamefont {Saraiva}, \citenamefont {Culcer},\ and\ \citenamefont {Capaz}}]{Terrazos2021}%
  \BibitemOpen
  \bibfield  {author} {\bibinfo {author} {\bibfnamefont {L.~A.}\ \bibnamefont {Terrazos}}, \bibinfo {author} {\bibfnamefont {E.}~\bibnamefont {Marcellina}}, \bibinfo {author} {\bibfnamefont {Z.}~\bibnamefont {Wang}}, \bibinfo {author} {\bibfnamefont {S.~N.}\ \bibnamefont {Coppersmith}}, \bibinfo {author} {\bibfnamefont {M.}~\bibnamefont {Friesen}}, \bibinfo {author} {\bibfnamefont {A.~R.}\ \bibnamefont {Hamilton}}, \bibinfo {author} {\bibfnamefont {X.}~\bibnamefont {Hu}}, \bibinfo {author} {\bibfnamefont {B.}~\bibnamefont {Koiller}}, \bibinfo {author} {\bibfnamefont {A.~L.}\ \bibnamefont {Saraiva}}, \bibinfo {author} {\bibfnamefont {D.}~\bibnamefont {Culcer}},\ and\ \bibinfo {author} {\bibfnamefont {R.~B.}\ \bibnamefont {Capaz}},\ }\bibfield  {title} {\bibinfo {title} {Theory of hole-spin qubits in strained germanium quantum dots},\ }\href {https://doi.org/10.1103/PhysRevB.103.125201} {\bibfield  {journal} {\bibinfo  {journal} {Phys. Rev. B}\ }\textbf {\bibinfo {volume} {103}},\ \bibinfo {pages} {125201}
  (\bibinfo {year} {2021})}\BibitemShut {NoStop}%
\bibitem [{\citenamefont {Gao}\ \emph {et~al.}(2020)\citenamefont {Gao}, \citenamefont {Wang}, \citenamefont {Watzinger}, \citenamefont {Hu}, \citenamefont {Rančić}, \citenamefont {Zhang}, \citenamefont {Wang}, \citenamefont {Yao}, \citenamefont {Wang}, \citenamefont {Kukučka}, \citenamefont {Vukušić}, \citenamefont {Kloeffel}, \citenamefont {Loss}, \citenamefont {Liu}, \citenamefont {Katsaros},\ and\ \citenamefont {Zhang}}]{Gao2020}%
  \BibitemOpen
  \bibfield  {author} {\bibinfo {author} {\bibfnamefont {F.}~\bibnamefont {Gao}}, \bibinfo {author} {\bibfnamefont {J.-H.}\ \bibnamefont {Wang}}, \bibinfo {author} {\bibfnamefont {H.}~\bibnamefont {Watzinger}}, \bibinfo {author} {\bibfnamefont {H.}~\bibnamefont {Hu}}, \bibinfo {author} {\bibfnamefont {M.~J.}\ \bibnamefont {Rančić}}, \bibinfo {author} {\bibfnamefont {J.-Y.}\ \bibnamefont {Zhang}}, \bibinfo {author} {\bibfnamefont {T.}~\bibnamefont {Wang}}, \bibinfo {author} {\bibfnamefont {Y.}~\bibnamefont {Yao}}, \bibinfo {author} {\bibfnamefont {G.-L.}\ \bibnamefont {Wang}}, \bibinfo {author} {\bibfnamefont {J.}~\bibnamefont {Kukučka}}, \bibinfo {author} {\bibfnamefont {L.}~\bibnamefont {Vukušić}}, \bibinfo {author} {\bibfnamefont {C.}~\bibnamefont {Kloeffel}}, \bibinfo {author} {\bibfnamefont {D.}~\bibnamefont {Loss}}, \bibinfo {author} {\bibfnamefont {F.}~\bibnamefont {Liu}}, \bibinfo {author} {\bibfnamefont {G.}~\bibnamefont {Katsaros}},\ and\ \bibinfo {author} {\bibfnamefont {J.-J.}\ \bibnamefont
  {Zhang}},\ }\bibfield  {title} {\bibinfo {title} {Site-controlled uniform ge/si hut wires with electrically tunable spin–orbit coupling},\ }\href {https://doi.org/https://doi.org/10.1002/adma.201906523} {\bibfield  {journal} {\bibinfo  {journal} {Advanced Materials}\ }\textbf {\bibinfo {volume} {32}},\ \bibinfo {pages} {1906523} (\bibinfo {year} {2020})}\BibitemShut {NoStop}%
\bibitem [{\citenamefont {Liu}\ \emph {et~al.}(2022)\citenamefont {Liu}, \citenamefont {Zhang}, \citenamefont {Wang}, \citenamefont {Gao}, \citenamefont {Xu}, \citenamefont {Zhang}, \citenamefont {Li}, \citenamefont {Cao}, \citenamefont {Wang}, \citenamefont {Zhang}, \citenamefont {Hu}, \citenamefont {Li},\ and\ \citenamefont {Guo}}]{Liu2022}%
  \BibitemOpen
  \bibfield  {author} {\bibinfo {author} {\bibfnamefont {H.}~\bibnamefont {Liu}}, \bibinfo {author} {\bibfnamefont {T.}~\bibnamefont {Zhang}}, \bibinfo {author} {\bibfnamefont {K.}~\bibnamefont {Wang}}, \bibinfo {author} {\bibfnamefont {F.}~\bibnamefont {Gao}}, \bibinfo {author} {\bibfnamefont {G.}~\bibnamefont {Xu}}, \bibinfo {author} {\bibfnamefont {X.}~\bibnamefont {Zhang}}, \bibinfo {author} {\bibfnamefont {S.-X.}\ \bibnamefont {Li}}, \bibinfo {author} {\bibfnamefont {G.}~\bibnamefont {Cao}}, \bibinfo {author} {\bibfnamefont {T.}~\bibnamefont {Wang}}, \bibinfo {author} {\bibfnamefont {J.}~\bibnamefont {Zhang}}, \bibinfo {author} {\bibfnamefont {X.}~\bibnamefont {Hu}}, \bibinfo {author} {\bibfnamefont {H.-O.}\ \bibnamefont {Li}},\ and\ \bibinfo {author} {\bibfnamefont {G.-P.}\ \bibnamefont {Guo}},\ }\bibfield  {title} {\bibinfo {title} {Gate-tunable spin-orbit coupling in a germanium hole double quantum dot},\ }\href {https://doi.org/10.1103/PhysRevApplied.17.044052} {\bibfield  {journal} {\bibinfo
  {journal} {Phys. Rev. Appl.}\ }\textbf {\bibinfo {volume} {17}},\ \bibinfo {pages} {044052} (\bibinfo {year} {2022})}\BibitemShut {NoStop}%
\bibitem [{\citenamefont {Liu}\ \emph {et~al.}(2023)\citenamefont {Liu}, \citenamefont {Wang}, \citenamefont {Gao}, \citenamefont {Leng}, \citenamefont {Liu}, \citenamefont {Zhou}, \citenamefont {Cao}, \citenamefont {Wang}, \citenamefont {Zhang}, \citenamefont {Huang}, \citenamefont {Li},\ and\ \citenamefont {Guo}}]{Liu2023}%
  \BibitemOpen
  \bibfield  {author} {\bibinfo {author} {\bibfnamefont {H.}~\bibnamefont {Liu}}, \bibinfo {author} {\bibfnamefont {K.}~\bibnamefont {Wang}}, \bibinfo {author} {\bibfnamefont {F.}~\bibnamefont {Gao}}, \bibinfo {author} {\bibfnamefont {J.}~\bibnamefont {Leng}}, \bibinfo {author} {\bibfnamefont {Y.}~\bibnamefont {Liu}}, \bibinfo {author} {\bibfnamefont {Y.-C.}\ \bibnamefont {Zhou}}, \bibinfo {author} {\bibfnamefont {G.}~\bibnamefont {Cao}}, \bibinfo {author} {\bibfnamefont {T.}~\bibnamefont {Wang}}, \bibinfo {author} {\bibfnamefont {J.}~\bibnamefont {Zhang}}, \bibinfo {author} {\bibfnamefont {P.}~\bibnamefont {Huang}}, \bibinfo {author} {\bibfnamefont {H.-O.}\ \bibnamefont {Li}},\ and\ \bibinfo {author} {\bibfnamefont {G.-P.}\ \bibnamefont {Guo}},\ }\bibfield  {title} {\bibinfo {title} {Ultrafast and electrically tunable rabi frequency in a germanium hut wire hole spin qubit},\ }\href {https://doi.org/10.1021/acs.nanolett.3c00213} {\bibfield  {journal} {\bibinfo  {journal} {Nano Letters}\ }\textbf {\bibinfo
  {volume} {23}},\ \bibinfo {pages} {3810} (\bibinfo {year} {2023})}\BibitemShut {NoStop}%
\bibitem [{\citenamefont {Rodr\'{\i}guez-Mena}\ \emph {et~al.}(2023)\citenamefont {Rodr\'{\i}guez-Mena}, \citenamefont {Abadillo-Uriel}, \citenamefont {Veste}, \citenamefont {Martinez}, \citenamefont {Li}, \citenamefont {Skl\'enard},\ and\ \citenamefont {Niquet}}]{Mena2023}%
  \BibitemOpen
  \bibfield  {author} {\bibinfo {author} {\bibfnamefont {E.~A.}\ \bibnamefont {Rodr\'{\i}guez-Mena}}, \bibinfo {author} {\bibfnamefont {J.~C.}\ \bibnamefont {Abadillo-Uriel}}, \bibinfo {author} {\bibfnamefont {G.}~\bibnamefont {Veste}}, \bibinfo {author} {\bibfnamefont {B.}~\bibnamefont {Martinez}}, \bibinfo {author} {\bibfnamefont {J.}~\bibnamefont {Li}}, \bibinfo {author} {\bibfnamefont {B.}~\bibnamefont {Skl\'enard}},\ and\ \bibinfo {author} {\bibfnamefont {Y.-M.}\ \bibnamefont {Niquet}},\ }\bibfield  {title} {\bibinfo {title} {Linear-in-momentum spin orbit interactions in planar ge/gesi heterostructures and spin qubits},\ }\href {https://doi.org/10.1103/PhysRevB.108.205416} {\bibfield  {journal} {\bibinfo  {journal} {Phys. Rev. B}\ }\textbf {\bibinfo {volume} {108}},\ \bibinfo {pages} {205416} (\bibinfo {year} {2023})}\BibitemShut {NoStop}%
\bibitem [{\citenamefont {Froning}\ \emph {et~al.}(2021{\natexlab{b}})\citenamefont {Froning}, \citenamefont {Ran\ifmmode \check{c}\else \v{c}\fi{}i\ifmmode~\acute{c}\else \'{c}\fi{}}, \citenamefont {Het\'enyi}, \citenamefont {Bosco}, \citenamefont {Rehmann}, \citenamefont {Li}, \citenamefont {Bakkers}, \citenamefont {Zwanenburg}, \citenamefont {Loss}, \citenamefont {Zumb\"uhl},\ and\ \citenamefont {Braakman}}]{Froning2021PRR}%
  \BibitemOpen
  \bibfield  {author} {\bibinfo {author} {\bibfnamefont {F.~N.~M.}\ \bibnamefont {Froning}}, \bibinfo {author} {\bibfnamefont {M.~J.}\ \bibnamefont {Ran\ifmmode \check{c}\else \v{c}\fi{}i\ifmmode~\acute{c}\else \'{c}\fi{}}}, \bibinfo {author} {\bibfnamefont {B.}~\bibnamefont {Het\'enyi}}, \bibinfo {author} {\bibfnamefont {S.}~\bibnamefont {Bosco}}, \bibinfo {author} {\bibfnamefont {M.~K.}\ \bibnamefont {Rehmann}}, \bibinfo {author} {\bibfnamefont {A.}~\bibnamefont {Li}}, \bibinfo {author} {\bibfnamefont {E.~P. A.~M.}\ \bibnamefont {Bakkers}}, \bibinfo {author} {\bibfnamefont {F.~A.}\ \bibnamefont {Zwanenburg}}, \bibinfo {author} {\bibfnamefont {D.}~\bibnamefont {Loss}}, \bibinfo {author} {\bibfnamefont {D.~M.}\ \bibnamefont {Zumb\"uhl}},\ and\ \bibinfo {author} {\bibfnamefont {F.~R.}\ \bibnamefont {Braakman}},\ }\bibfield  {title} {\bibinfo {title} {Strong spin-orbit interaction and $g$-factor renormalization of hole spins in ge/si nanowire quantum dots},\ }\href
  {https://doi.org/10.1103/PhysRevResearch.3.013081} {\bibfield  {journal} {\bibinfo  {journal} {Phys. Rev. Res.}\ }\textbf {\bibinfo {volume} {3}},\ \bibinfo {pages} {013081} (\bibinfo {year} {2021}{\natexlab{b}})}\BibitemShut {NoStop}%
\bibitem [{\citenamefont {Lawrie}\ \emph {et~al.}(2023)\citenamefont {Lawrie}, \citenamefont {Rimbach-Russ}, \citenamefont {Riggelen}, \citenamefont {Hendrickx}, \citenamefont {Snoo}, \citenamefont {Sammak}, \citenamefont {Scappucci}, \citenamefont {Helsen},\ and\ \citenamefont {Veldhorst}}]{Lawrie2023}%
  \BibitemOpen
  \bibfield  {author} {\bibinfo {author} {\bibfnamefont {W.~I.~L.}\ \bibnamefont {Lawrie}}, \bibinfo {author} {\bibfnamefont {M.}~\bibnamefont {Rimbach-Russ}}, \bibinfo {author} {\bibfnamefont {F.~v.}\ \bibnamefont {Riggelen}}, \bibinfo {author} {\bibfnamefont {N.~W.}\ \bibnamefont {Hendrickx}}, \bibinfo {author} {\bibfnamefont {S.~L.~d.}\ \bibnamefont {Snoo}}, \bibinfo {author} {\bibfnamefont {A.}~\bibnamefont {Sammak}}, \bibinfo {author} {\bibfnamefont {G.}~\bibnamefont {Scappucci}}, \bibinfo {author} {\bibfnamefont {J.}~\bibnamefont {Helsen}},\ and\ \bibinfo {author} {\bibfnamefont {M.}~\bibnamefont {Veldhorst}},\ }\bibfield  {title} {\bibinfo {title} {Simultaneous single-qubit driving of semiconductor spin qubits at the fault-tolerant threshold},\ }\href {https://doi.org/10.1038/s41467-023-39334-3} {\bibfield  {journal} {\bibinfo  {journal} {Nature Communications}\ }\textbf {\bibinfo {volume} {14}},\ \bibinfo {pages} {3617} (\bibinfo {year} {2023})}\BibitemShut {NoStop}%
\bibitem [{\citenamefont {van Riggelen-Doelman}\ \emph {et~al.}(2024)\citenamefont {van Riggelen-Doelman}, \citenamefont {Wang}, \citenamefont {de~Snoo}, \citenamefont {Lawrie}, \citenamefont {Hendrickx}, \citenamefont {Rimbach-Russ}, \citenamefont {Sammak}, \citenamefont {Scappucci}, \citenamefont {D{\'e}prez},\ and\ \citenamefont {Veldhorst}}]{Riggelen2024}%
  \BibitemOpen
  \bibfield  {author} {\bibinfo {author} {\bibfnamefont {F.}~\bibnamefont {van Riggelen-Doelman}}, \bibinfo {author} {\bibfnamefont {C.-A.}\ \bibnamefont {Wang}}, \bibinfo {author} {\bibfnamefont {S.~L.}\ \bibnamefont {de~Snoo}}, \bibinfo {author} {\bibfnamefont {W.~I.~L.}\ \bibnamefont {Lawrie}}, \bibinfo {author} {\bibfnamefont {N.~W.}\ \bibnamefont {Hendrickx}}, \bibinfo {author} {\bibfnamefont {M.}~\bibnamefont {Rimbach-Russ}}, \bibinfo {author} {\bibfnamefont {A.}~\bibnamefont {Sammak}}, \bibinfo {author} {\bibfnamefont {G.}~\bibnamefont {Scappucci}}, \bibinfo {author} {\bibfnamefont {C.}~\bibnamefont {D{\'e}prez}},\ and\ \bibinfo {author} {\bibfnamefont {M.}~\bibnamefont {Veldhorst}},\ }\bibfield  {title} {\bibinfo {title} {Coherent spin qubit shuttling through germanium quantum dots},\ }\href {https://doi.org/10.1038/s41467-024-49358-y} {\bibfield  {journal} {\bibinfo  {journal} {Nature Communications}\ }\textbf {\bibinfo {volume} {15}},\ \bibinfo {pages} {5716} (\bibinfo {year} {2024})}\BibitemShut
  {NoStop}%
\bibitem [{\citenamefont {Bosco}\ \emph {et~al.}(2024)\citenamefont {Bosco}, \citenamefont {Zou},\ and\ \citenamefont {Loss}}]{Bosco2024}%
  \BibitemOpen
  \bibfield  {author} {\bibinfo {author} {\bibfnamefont {S.}~\bibnamefont {Bosco}}, \bibinfo {author} {\bibfnamefont {J.}~\bibnamefont {Zou}},\ and\ \bibinfo {author} {\bibfnamefont {D.}~\bibnamefont {Loss}},\ }\bibfield  {title} {\bibinfo {title} {High-fidelity spin qubit shuttling via large spin-orbit interactions},\ }\href {https://doi.org/10.1103/PRXQuantum.5.020353} {\bibfield  {journal} {\bibinfo  {journal} {PRX Quantum}\ }\textbf {\bibinfo {volume} {5}},\ \bibinfo {pages} {020353} (\bibinfo {year} {2024})}\BibitemShut {NoStop}%
\bibitem [{\citenamefont {Lawrie}\ \emph {et~al.}(2020)\citenamefont {Lawrie}, \citenamefont {Hendrickx}, \citenamefont {van Riggelen}, \citenamefont {Russ}, \citenamefont {Petit}, \citenamefont {Sammak}, \citenamefont {Scappucci},\ and\ \citenamefont {Veldhorst}}]{Lawrie2020}%
  \BibitemOpen
  \bibfield  {author} {\bibinfo {author} {\bibfnamefont {W.~I.~L.}\ \bibnamefont {Lawrie}}, \bibinfo {author} {\bibfnamefont {N.~W.}\ \bibnamefont {Hendrickx}}, \bibinfo {author} {\bibfnamefont {F.}~\bibnamefont {van Riggelen}}, \bibinfo {author} {\bibfnamefont {M.}~\bibnamefont {Russ}}, \bibinfo {author} {\bibfnamefont {L.}~\bibnamefont {Petit}}, \bibinfo {author} {\bibfnamefont {A.}~\bibnamefont {Sammak}}, \bibinfo {author} {\bibfnamefont {G.}~\bibnamefont {Scappucci}},\ and\ \bibinfo {author} {\bibfnamefont {M.}~\bibnamefont {Veldhorst}},\ }\bibfield  {title} {\bibinfo {title} {Spin relaxation benchmarks and individual qubit addressability for holes in quantum dots},\ }\href {https://doi.org/10.1021/acs.nanolett.0c02589} {\bibfield  {journal} {\bibinfo  {journal} {Nano Letters}\ }\textbf {\bibinfo {volume} {20}},\ \bibinfo {pages} {7237} (\bibinfo {year} {2020})}\BibitemShut {NoStop}%
\bibitem [{\citenamefont {Hendrickx}\ \emph {et~al.}(2024)\citenamefont {Hendrickx}, \citenamefont {Massai}, \citenamefont {Mergenthaler}, \citenamefont {Schupp}, \citenamefont {Paredes}, \citenamefont {Bedell}, \citenamefont {Salis},\ and\ \citenamefont {Fuhrer}}]{Hendrickx2024}%
  \BibitemOpen
  \bibfield  {author} {\bibinfo {author} {\bibfnamefont {N.~W.}\ \bibnamefont {Hendrickx}}, \bibinfo {author} {\bibfnamefont {L.}~\bibnamefont {Massai}}, \bibinfo {author} {\bibfnamefont {M.}~\bibnamefont {Mergenthaler}}, \bibinfo {author} {\bibfnamefont {F.~J.}\ \bibnamefont {Schupp}}, \bibinfo {author} {\bibfnamefont {S.}~\bibnamefont {Paredes}}, \bibinfo {author} {\bibfnamefont {S.~W.}\ \bibnamefont {Bedell}}, \bibinfo {author} {\bibfnamefont {G.}~\bibnamefont {Salis}},\ and\ \bibinfo {author} {\bibfnamefont {A.}~\bibnamefont {Fuhrer}},\ }\bibfield  {title} {\bibinfo {title} {Sweet-spot operation of a germanium hole spin qubit with highly anisotropic noise sensitivity},\ }\href {https://doi.org/10.1038/s41563-024-01857-5} {\bibfield  {journal} {\bibinfo  {journal} {Nature Materials}\ }\textbf {\bibinfo {volume} {23}},\ \bibinfo {pages} {920} (\bibinfo {year} {2024})}\BibitemShut {NoStop}%
\bibitem [{\citenamefont {Piot}\ \emph {et~al.}(2022)\citenamefont {Piot}, \citenamefont {Brun}, \citenamefont {Schmitt}, \citenamefont {Zihlmann}, \citenamefont {Michal}, \citenamefont {Apra}, \citenamefont {Abadillo-Uriel}, \citenamefont {Jehl}, \citenamefont {Bertrand}, \citenamefont {Niebojewski}, \citenamefont {Hutin}, \citenamefont {Vinet}, \citenamefont {Urdampilleta}, \citenamefont {Meunier}, \citenamefont {Niquet}, \citenamefont {Maurand},\ and\ \citenamefont {Franceschi}}]{Piot2022}%
  \BibitemOpen
  \bibfield  {author} {\bibinfo {author} {\bibfnamefont {N.}~\bibnamefont {Piot}}, \bibinfo {author} {\bibfnamefont {B.}~\bibnamefont {Brun}}, \bibinfo {author} {\bibfnamefont {V.}~\bibnamefont {Schmitt}}, \bibinfo {author} {\bibfnamefont {S.}~\bibnamefont {Zihlmann}}, \bibinfo {author} {\bibfnamefont {V.~P.}\ \bibnamefont {Michal}}, \bibinfo {author} {\bibfnamefont {A.}~\bibnamefont {Apra}}, \bibinfo {author} {\bibfnamefont {J.~C.}\ \bibnamefont {Abadillo-Uriel}}, \bibinfo {author} {\bibfnamefont {X.}~\bibnamefont {Jehl}}, \bibinfo {author} {\bibfnamefont {B.}~\bibnamefont {Bertrand}}, \bibinfo {author} {\bibfnamefont {H.}~\bibnamefont {Niebojewski}}, \bibinfo {author} {\bibfnamefont {L.}~\bibnamefont {Hutin}}, \bibinfo {author} {\bibfnamefont {M.}~\bibnamefont {Vinet}}, \bibinfo {author} {\bibfnamefont {M.}~\bibnamefont {Urdampilleta}}, \bibinfo {author} {\bibfnamefont {T.}~\bibnamefont {Meunier}}, \bibinfo {author} {\bibfnamefont {Y.~M.}\ \bibnamefont {Niquet}}, \bibinfo {author} {\bibfnamefont
  {R.}~\bibnamefont {Maurand}},\ and\ \bibinfo {author} {\bibfnamefont {S.~D.}\ \bibnamefont {Franceschi}},\ }\bibfield  {title} {\bibinfo {title} {A single hole spin with enhanced coherence in natural silicon},\ }\href {https://doi.org/10.1038/s41565-022-01196-z} {\bibfield  {journal} {\bibinfo  {journal} {Nature Nanotechnology}\ }\textbf {\bibinfo {volume} {17}},\ \bibinfo {pages} {1072} (\bibinfo {year} {2022})}\BibitemShut {NoStop}%
\bibitem [{\citenamefont {Shimatani}\ \emph {et~al.}(2020)\citenamefont {Shimatani}, \citenamefont {Yamaoka}, \citenamefont {Ishihara}, \citenamefont {Andreev}, \citenamefont {Williams}, \citenamefont {Oda},\ and\ \citenamefont {Kodera}}]{Shimatani2020}%
  \BibitemOpen
  \bibfield  {author} {\bibinfo {author} {\bibfnamefont {N.}~\bibnamefont {Shimatani}}, \bibinfo {author} {\bibfnamefont {Y.}~\bibnamefont {Yamaoka}}, \bibinfo {author} {\bibfnamefont {R.}~\bibnamefont {Ishihara}}, \bibinfo {author} {\bibfnamefont {A.}~\bibnamefont {Andreev}}, \bibinfo {author} {\bibfnamefont {D.~A.}\ \bibnamefont {Williams}}, \bibinfo {author} {\bibfnamefont {S.}~\bibnamefont {Oda}},\ and\ \bibinfo {author} {\bibfnamefont {T.}~\bibnamefont {Kodera}},\ }\bibfield  {title} {\bibinfo {title} {{Temperature dependence of hole transport properties through physically defined silicon quantum dots}},\ }\href {https://doi.org/10.1063/5.0010981} {\bibfield  {journal} {\bibinfo  {journal} {Applied Physics Letters}\ }\textbf {\bibinfo {volume} {117}},\ \bibinfo {pages} {094001} (\bibinfo {year} {2020})}\BibitemShut {NoStop}%
\bibitem [{\citenamefont {Camenzind}\ \emph {et~al.}(2021)\citenamefont {Camenzind}, \citenamefont {Svab}, \citenamefont {Stano}, \citenamefont {Yu}, \citenamefont {Zimmerman}, \citenamefont {Gossard}, \citenamefont {Loss},\ and\ \citenamefont {Zumb\"uhl}}]{Camenzind2021}%
  \BibitemOpen
  \bibfield  {author} {\bibinfo {author} {\bibfnamefont {L.~C.}\ \bibnamefont {Camenzind}}, \bibinfo {author} {\bibfnamefont {S.}~\bibnamefont {Svab}}, \bibinfo {author} {\bibfnamefont {P.}~\bibnamefont {Stano}}, \bibinfo {author} {\bibfnamefont {L.}~\bibnamefont {Yu}}, \bibinfo {author} {\bibfnamefont {J.~D.}\ \bibnamefont {Zimmerman}}, \bibinfo {author} {\bibfnamefont {A.~C.}\ \bibnamefont {Gossard}}, \bibinfo {author} {\bibfnamefont {D.}~\bibnamefont {Loss}},\ and\ \bibinfo {author} {\bibfnamefont {D.~M.}\ \bibnamefont {Zumb\"uhl}},\ }\bibfield  {title} {\bibinfo {title} {Isotropic and anisotropic $g$-factor corrections in gaas quantum dots},\ }\href {https://doi.org/10.1103/PhysRevLett.127.057701} {\bibfield  {journal} {\bibinfo  {journal} {Phys. Rev. Lett.}\ }\textbf {\bibinfo {volume} {127}},\ \bibinfo {pages} {057701} (\bibinfo {year} {2021})}\BibitemShut {NoStop}%
\bibitem [{\citenamefont {Camenzind}\ \emph {et~al.}(2022)\citenamefont {Camenzind}, \citenamefont {Geyer}, \citenamefont {Fuhrer}, \citenamefont {Warburton}, \citenamefont {Zumb{\"u}hl},\ and\ \citenamefont {Kuhlmann}}]{Camenzind2022}%
  \BibitemOpen
  \bibfield  {author} {\bibinfo {author} {\bibfnamefont {L.~C.}\ \bibnamefont {Camenzind}}, \bibinfo {author} {\bibfnamefont {S.}~\bibnamefont {Geyer}}, \bibinfo {author} {\bibfnamefont {A.}~\bibnamefont {Fuhrer}}, \bibinfo {author} {\bibfnamefont {R.~J.}\ \bibnamefont {Warburton}}, \bibinfo {author} {\bibfnamefont {D.~M.}\ \bibnamefont {Zumb{\"u}hl}},\ and\ \bibinfo {author} {\bibfnamefont {A.~V.}\ \bibnamefont {Kuhlmann}},\ }\bibfield  {title} {\bibinfo {title} {A hole spin qubit in a fin field-effect transistor above 4 kelvin},\ }\href {https://doi.org/10.1038/s41928-022-00722-0} {\bibfield  {journal} {\bibinfo  {journal} {Nature Electronics}\ }\textbf {\bibinfo {volume} {5}},\ \bibinfo {pages} {178} (\bibinfo {year} {2022})}\BibitemShut {NoStop}%
\bibitem [{\citenamefont {van Riggelen}\ \emph {et~al.}(2022)\citenamefont {van Riggelen}, \citenamefont {Lawrie}, \citenamefont {Russ}, \citenamefont {Hendrickx}, \citenamefont {Sammak}, \citenamefont {Rispler}, \citenamefont {Terhal}, \citenamefont {Scappucci},\ and\ \citenamefont {Veldhorst}}]{Riggelen2022}%
  \BibitemOpen
  \bibfield  {author} {\bibinfo {author} {\bibfnamefont {F.}~\bibnamefont {van Riggelen}}, \bibinfo {author} {\bibfnamefont {W.~I.~L.}\ \bibnamefont {Lawrie}}, \bibinfo {author} {\bibfnamefont {M.}~\bibnamefont {Russ}}, \bibinfo {author} {\bibfnamefont {N.~W.}\ \bibnamefont {Hendrickx}}, \bibinfo {author} {\bibfnamefont {A.}~\bibnamefont {Sammak}}, \bibinfo {author} {\bibfnamefont {M.}~\bibnamefont {Rispler}}, \bibinfo {author} {\bibfnamefont {B.~M.}\ \bibnamefont {Terhal}}, \bibinfo {author} {\bibfnamefont {G.}~\bibnamefont {Scappucci}},\ and\ \bibinfo {author} {\bibfnamefont {M.}~\bibnamefont {Veldhorst}},\ }\bibfield  {title} {\bibinfo {title} {Phase flip code with semiconductor spin qubits},\ }\href {https://doi.org/10.1038/s41534-022-00639-8} {\bibfield  {journal} {\bibinfo  {journal} {npj Quantum Information}\ }\textbf {\bibinfo {volume} {8}},\ \bibinfo {pages} {124} (\bibinfo {year} {2022})}\BibitemShut {NoStop}%
\bibitem [{\citenamefont {Wang}\ \emph {et~al.}(2023)\citenamefont {Wang}, \citenamefont {D{\'e}prez}, \citenamefont {Tidjani}, \citenamefont {Lawrie}, \citenamefont {Hendrickx}, \citenamefont {Sammak}, \citenamefont {Scappucci},\ and\ \citenamefont {Veldhorst}}]{Chien2023}%
  \BibitemOpen
  \bibfield  {author} {\bibinfo {author} {\bibfnamefont {C.-A.}\ \bibnamefont {Wang}}, \bibinfo {author} {\bibfnamefont {C.}~\bibnamefont {D{\'e}prez}}, \bibinfo {author} {\bibfnamefont {H.}~\bibnamefont {Tidjani}}, \bibinfo {author} {\bibfnamefont {W.~I.~L.}\ \bibnamefont {Lawrie}}, \bibinfo {author} {\bibfnamefont {N.~W.}\ \bibnamefont {Hendrickx}}, \bibinfo {author} {\bibfnamefont {A.}~\bibnamefont {Sammak}}, \bibinfo {author} {\bibfnamefont {G.}~\bibnamefont {Scappucci}},\ and\ \bibinfo {author} {\bibfnamefont {M.}~\bibnamefont {Veldhorst}},\ }\bibfield  {title} {\bibinfo {title} {Probing resonating valence bonds on a programmable germanium quantum simulator},\ }\href {https://doi.org/10.1038/s41534-023-00727-3} {\bibfield  {journal} {\bibinfo  {journal} {npj Quantum Information}\ }\textbf {\bibinfo {volume} {9}},\ \bibinfo {pages} {58} (\bibinfo {year} {2023})}\BibitemShut {NoStop}%
\bibitem [{\citenamefont {Borsoi}\ \emph {et~al.}(2024)\citenamefont {Borsoi}, \citenamefont {Hendrickx}, \citenamefont {John}, \citenamefont {Meyer}, \citenamefont {Motz}, \citenamefont {van Riggelen}, \citenamefont {Sammak}, \citenamefont {de~Snoo}, \citenamefont {Scappucci},\ and\ \citenamefont {Veldhorst}}]{Borsoi2024}%
  \BibitemOpen
  \bibfield  {author} {\bibinfo {author} {\bibfnamefont {F.}~\bibnamefont {Borsoi}}, \bibinfo {author} {\bibfnamefont {N.~W.}\ \bibnamefont {Hendrickx}}, \bibinfo {author} {\bibfnamefont {V.}~\bibnamefont {John}}, \bibinfo {author} {\bibfnamefont {M.}~\bibnamefont {Meyer}}, \bibinfo {author} {\bibfnamefont {S.}~\bibnamefont {Motz}}, \bibinfo {author} {\bibfnamefont {F.}~\bibnamefont {van Riggelen}}, \bibinfo {author} {\bibfnamefont {A.}~\bibnamefont {Sammak}}, \bibinfo {author} {\bibfnamefont {S.~L.}\ \bibnamefont {de~Snoo}}, \bibinfo {author} {\bibfnamefont {G.}~\bibnamefont {Scappucci}},\ and\ \bibinfo {author} {\bibfnamefont {M.}~\bibnamefont {Veldhorst}},\ }\bibfield  {title} {\bibinfo {title} {Shared control of a 16 semiconductor quantum dot crossbar array},\ }\href {https://doi.org/10.1038/s41565-023-01491-3} {\bibfield  {journal} {\bibinfo  {journal} {Nature Nanotechnology}\ }\textbf {\bibinfo {volume} {19}},\ \bibinfo {pages} {21} (\bibinfo {year} {2024})}\BibitemShut {NoStop}%
\bibitem [{\citenamefont {Ungerer}\ \emph {et~al.}(2023)\citenamefont {Ungerer}, \citenamefont {Kwon}, \citenamefont {Patlatiuk}, \citenamefont {Ridderbos}, \citenamefont {Kononov}, \citenamefont {Sarmah}, \citenamefont {Bakkers}, \citenamefont {Zumbühl},\ and\ \citenamefont {Schönenberger}}]{Ungerer2023}%
  \BibitemOpen
  \bibfield  {author} {\bibinfo {author} {\bibfnamefont {J.~H.}\ \bibnamefont {Ungerer}}, \bibinfo {author} {\bibfnamefont {P.~C.}\ \bibnamefont {Kwon}}, \bibinfo {author} {\bibfnamefont {T.}~\bibnamefont {Patlatiuk}}, \bibinfo {author} {\bibfnamefont {J.}~\bibnamefont {Ridderbos}}, \bibinfo {author} {\bibfnamefont {A.}~\bibnamefont {Kononov}}, \bibinfo {author} {\bibfnamefont {D.}~\bibnamefont {Sarmah}}, \bibinfo {author} {\bibfnamefont {E.~P. A.~M.}\ \bibnamefont {Bakkers}}, \bibinfo {author} {\bibfnamefont {D.}~\bibnamefont {Zumbühl}},\ and\ \bibinfo {author} {\bibfnamefont {C.}~\bibnamefont {Schönenberger}},\ }\bibfield  {title} {\bibinfo {title} {Charge-sensing of a ge/si core/shell nanowire double quantum dot using a high-impedance superconducting resonator},\ }\href {https://doi.org/10.1088/2633-4356/ace2a6} {\bibfield  {journal} {\bibinfo  {journal} {Materials for Quantum Technology}\ }\textbf {\bibinfo {volume} {3}},\ \bibinfo {pages} {031001} (\bibinfo {year} {2023})}\BibitemShut {NoStop}%
\bibitem [{\citenamefont {Rooney}\ \emph {et~al.}(2023)\citenamefont {Rooney}, \citenamefont {Luo}, \citenamefont {Stehouwer}, \citenamefont {Scappucci}, \citenamefont {Veldhorst},\ and\ \citenamefont {Jiang}}]{Rooney2023}%
  \BibitemOpen
  \bibfield  {author} {\bibinfo {author} {\bibfnamefont {J.}~\bibnamefont {Rooney}}, \bibinfo {author} {\bibfnamefont {Z.}~\bibnamefont {Luo}}, \bibinfo {author} {\bibfnamefont {L.~E.~A.}\ \bibnamefont {Stehouwer}}, \bibinfo {author} {\bibfnamefont {G.}~\bibnamefont {Scappucci}}, \bibinfo {author} {\bibfnamefont {M.}~\bibnamefont {Veldhorst}},\ and\ \bibinfo {author} {\bibfnamefont {H.-W.}\ \bibnamefont {Jiang}},\ }\href {https://arxiv.org/abs/2311.10188} {\bibinfo {title} {Gate modulation of the hole singlet-triplet qubit frequency in germanium}} (\bibinfo {year} {2023}),\ \Eprint {https://arxiv.org/abs/2311.10188} {arXiv:2311.10188 [cond-mat.mes-hall]} \BibitemShut {NoStop}%
\bibitem [{\citenamefont {Sarkar}\ \emph {et~al.}(2023)\citenamefont {Sarkar}, \citenamefont {Wang}, \citenamefont {Rendell}, \citenamefont {Hendrickx}, \citenamefont {Veldhorst}, \citenamefont {Scappucci}, \citenamefont {Khalifa}, \citenamefont {Salfi}, \citenamefont {Saraiva}, \citenamefont {Dzurak}, \citenamefont {Hamilton},\ and\ \citenamefont {Culcer}}]{Sarkar2023}%
  \BibitemOpen
  \bibfield  {author} {\bibinfo {author} {\bibfnamefont {A.}~\bibnamefont {Sarkar}}, \bibinfo {author} {\bibfnamefont {Z.}~\bibnamefont {Wang}}, \bibinfo {author} {\bibfnamefont {M.}~\bibnamefont {Rendell}}, \bibinfo {author} {\bibfnamefont {N.~W.}\ \bibnamefont {Hendrickx}}, \bibinfo {author} {\bibfnamefont {M.}~\bibnamefont {Veldhorst}}, \bibinfo {author} {\bibfnamefont {G.}~\bibnamefont {Scappucci}}, \bibinfo {author} {\bibfnamefont {M.}~\bibnamefont {Khalifa}}, \bibinfo {author} {\bibfnamefont {J.}~\bibnamefont {Salfi}}, \bibinfo {author} {\bibfnamefont {A.}~\bibnamefont {Saraiva}}, \bibinfo {author} {\bibfnamefont {A.~S.}\ \bibnamefont {Dzurak}}, \bibinfo {author} {\bibfnamefont {A.~R.}\ \bibnamefont {Hamilton}},\ and\ \bibinfo {author} {\bibfnamefont {D.}~\bibnamefont {Culcer}},\ }\bibfield  {title} {\bibinfo {title} {Electrical operation of planar ge hole spin qubits in an in-plane magnetic field},\ }\href {https://doi.org/10.1103/PhysRevB.108.245301} {\bibfield  {journal} {\bibinfo  {journal} {Phys.
  Rev. B}\ }\textbf {\bibinfo {volume} {108}},\ \bibinfo {pages} {245301} (\bibinfo {year} {2023})}\BibitemShut {NoStop}%
\bibitem [{\citenamefont {Zhang}\ \emph {et~al.}(2024)\citenamefont {Zhang}, \citenamefont {Morozova}, \citenamefont {Rimbach-Russ}, \citenamefont {Jirovec}, \citenamefont {Hsiao}, \citenamefont {Fari{\~n}a}, \citenamefont {Wang}, \citenamefont {Oosterhout}, \citenamefont {Sammak}, \citenamefont {Scappucci}, \citenamefont {Veldhorst},\ and\ \citenamefont {Vandersypen}}]{Xin2024}%
  \BibitemOpen
  \bibfield  {author} {\bibinfo {author} {\bibfnamefont {X.}~\bibnamefont {Zhang}}, \bibinfo {author} {\bibfnamefont {E.}~\bibnamefont {Morozova}}, \bibinfo {author} {\bibfnamefont {M.}~\bibnamefont {Rimbach-Russ}}, \bibinfo {author} {\bibfnamefont {D.}~\bibnamefont {Jirovec}}, \bibinfo {author} {\bibfnamefont {T.-K.}\ \bibnamefont {Hsiao}}, \bibinfo {author} {\bibfnamefont {P.~C.}\ \bibnamefont {Fari{\~n}a}}, \bibinfo {author} {\bibfnamefont {C.-A.}\ \bibnamefont {Wang}}, \bibinfo {author} {\bibfnamefont {S.~D.}\ \bibnamefont {Oosterhout}}, \bibinfo {author} {\bibfnamefont {A.}~\bibnamefont {Sammak}}, \bibinfo {author} {\bibfnamefont {G.}~\bibnamefont {Scappucci}}, \bibinfo {author} {\bibfnamefont {M.}~\bibnamefont {Veldhorst}},\ and\ \bibinfo {author} {\bibfnamefont {L.~M.~K.}\ \bibnamefont {Vandersypen}},\ }\bibfield  {title} {\bibinfo {title} {Universal control of four singlet--triplet qubits},\ }\bibfield  {journal} {\bibinfo  {journal} {Nature Nanotechnology}\ }\href
  {https://doi.org/10.1038/s41565-024-01817-9} {10.1038/s41565-024-01817-9} (\bibinfo {year} {2024})\BibitemShut {NoStop}%
\bibitem [{\citenamefont {Wang}\ \emph {et~al.}(2024)\citenamefont {Wang}, \citenamefont {Sarkar}, \citenamefont {Liles}, \citenamefont {Saraiva}, \citenamefont {Dzurak}, \citenamefont {Hamilton},\ and\ \citenamefont {Culcer}}]{Wang2024}%
  \BibitemOpen
  \bibfield  {author} {\bibinfo {author} {\bibfnamefont {Z.}~\bibnamefont {Wang}}, \bibinfo {author} {\bibfnamefont {A.}~\bibnamefont {Sarkar}}, \bibinfo {author} {\bibfnamefont {S.~D.}\ \bibnamefont {Liles}}, \bibinfo {author} {\bibfnamefont {A.}~\bibnamefont {Saraiva}}, \bibinfo {author} {\bibfnamefont {A.~S.}\ \bibnamefont {Dzurak}}, \bibinfo {author} {\bibfnamefont {A.~R.}\ \bibnamefont {Hamilton}},\ and\ \bibinfo {author} {\bibfnamefont {D.}~\bibnamefont {Culcer}},\ }\bibfield  {title} {\bibinfo {title} {Electrical operation of hole spin qubits in planar mos silicon quantum dots},\ }\href {https://doi.org/10.1103/PhysRevB.109.075427} {\bibfield  {journal} {\bibinfo  {journal} {Phys. Rev. B}\ }\textbf {\bibinfo {volume} {109}},\ \bibinfo {pages} {075427} (\bibinfo {year} {2024})}\BibitemShut {NoStop}%
\bibitem [{\citenamefont {John}\ \emph {et~al.}(2024)\citenamefont {John}, \citenamefont {Borsoi}, \citenamefont {Gy\"orgy}, \citenamefont {Wang}, \citenamefont {Sz\'echenyi}, \citenamefont {van Riggelen-Doelman}, \citenamefont {Lawrie}, \citenamefont {Hendrickx}, \citenamefont {Sammak}, \citenamefont {Scappucci}, \citenamefont {P\'alyi},\ and\ \citenamefont {Veldhorst}}]{John2024}%
  \BibitemOpen
  \bibfield  {author} {\bibinfo {author} {\bibfnamefont {V.}~\bibnamefont {John}}, \bibinfo {author} {\bibfnamefont {F.}~\bibnamefont {Borsoi}}, \bibinfo {author} {\bibfnamefont {Z.}~\bibnamefont {Gy\"orgy}}, \bibinfo {author} {\bibfnamefont {C.-A.}\ \bibnamefont {Wang}}, \bibinfo {author} {\bibfnamefont {G.}~\bibnamefont {Sz\'echenyi}}, \bibinfo {author} {\bibfnamefont {F.}~\bibnamefont {van Riggelen-Doelman}}, \bibinfo {author} {\bibfnamefont {W.~I.~L.}\ \bibnamefont {Lawrie}}, \bibinfo {author} {\bibfnamefont {N.~W.}\ \bibnamefont {Hendrickx}}, \bibinfo {author} {\bibfnamefont {A.}~\bibnamefont {Sammak}}, \bibinfo {author} {\bibfnamefont {G.}~\bibnamefont {Scappucci}}, \bibinfo {author} {\bibfnamefont {A.}~\bibnamefont {P\'alyi}},\ and\ \bibinfo {author} {\bibfnamefont {M.}~\bibnamefont {Veldhorst}},\ }\bibfield  {title} {\bibinfo {title} {Bichromatic rabi control of semiconductor qubits},\ }\href {https://doi.org/10.1103/PhysRevLett.132.067001} {\bibfield  {journal} {\bibinfo  {journal} {Phys. Rev.
  Lett.}\ }\textbf {\bibinfo {volume} {132}},\ \bibinfo {pages} {067001} (\bibinfo {year} {2024})}\BibitemShut {NoStop}%
\bibitem [{\citenamefont {Li}\ \emph {et~al.}(2018)\citenamefont {Li}, \citenamefont {Li}, \citenamefont {Gao}, \citenamefont {Li}, \citenamefont {Xu}, \citenamefont {Wang}, \citenamefont {Liu}, \citenamefont {Cao}, \citenamefont {Xiao}, \citenamefont {Wang}, \citenamefont {Zhang}, \citenamefont {Guo},\ and\ \citenamefont {Guo}}]{Li2018}%
  \BibitemOpen
  \bibfield  {author} {\bibinfo {author} {\bibfnamefont {Y.}~\bibnamefont {Li}}, \bibinfo {author} {\bibfnamefont {S.-X.}\ \bibnamefont {Li}}, \bibinfo {author} {\bibfnamefont {F.}~\bibnamefont {Gao}}, \bibinfo {author} {\bibfnamefont {H.-O.}\ \bibnamefont {Li}}, \bibinfo {author} {\bibfnamefont {G.}~\bibnamefont {Xu}}, \bibinfo {author} {\bibfnamefont {K.}~\bibnamefont {Wang}}, \bibinfo {author} {\bibfnamefont {D.}~\bibnamefont {Liu}}, \bibinfo {author} {\bibfnamefont {G.}~\bibnamefont {Cao}}, \bibinfo {author} {\bibfnamefont {M.}~\bibnamefont {Xiao}}, \bibinfo {author} {\bibfnamefont {T.}~\bibnamefont {Wang}}, \bibinfo {author} {\bibfnamefont {J.-J.}\ \bibnamefont {Zhang}}, \bibinfo {author} {\bibfnamefont {G.-C.}\ \bibnamefont {Guo}},\ and\ \bibinfo {author} {\bibfnamefont {G.-P.}\ \bibnamefont {Guo}},\ }\bibfield  {title} {\bibinfo {title} {Coupling a germanium hut wire hole quantum dot to a superconducting microwave resonator},\ }\href {https://doi.org/10.1021/acs.nanolett.8b00272} {\bibfield  {journal}
  {\bibinfo  {journal} {Nano Letters}\ }\textbf {\bibinfo {volume} {18}},\ \bibinfo {pages} {2091} (\bibinfo {year} {2018})}\BibitemShut {NoStop}%
\bibitem [{\citenamefont {Xu}\ \emph {et~al.}(2020)\citenamefont {Xu}, \citenamefont {Li}, \citenamefont {Gao}, \citenamefont {Li}, \citenamefont {Liu}, \citenamefont {Wang}, \citenamefont {Cao}, \citenamefont {Wang}, \citenamefont {Zhang}, \citenamefont {Guo},\ and\ \citenamefont {Guo}}]{Gang2020}%
  \BibitemOpen
  \bibfield  {author} {\bibinfo {author} {\bibfnamefont {G.}~\bibnamefont {Xu}}, \bibinfo {author} {\bibfnamefont {Y.}~\bibnamefont {Li}}, \bibinfo {author} {\bibfnamefont {F.}~\bibnamefont {Gao}}, \bibinfo {author} {\bibfnamefont {H.-O.}\ \bibnamefont {Li}}, \bibinfo {author} {\bibfnamefont {H.}~\bibnamefont {Liu}}, \bibinfo {author} {\bibfnamefont {K.}~\bibnamefont {Wang}}, \bibinfo {author} {\bibfnamefont {G.}~\bibnamefont {Cao}}, \bibinfo {author} {\bibfnamefont {T.}~\bibnamefont {Wang}}, \bibinfo {author} {\bibfnamefont {J.-J.}\ \bibnamefont {Zhang}}, \bibinfo {author} {\bibfnamefont {G.-C.}\ \bibnamefont {Guo}},\ and\ \bibinfo {author} {\bibfnamefont {G.-P.}\ \bibnamefont {Guo}},\ }\bibfield  {title} {\bibinfo {title} {Dipole coupling of a hole double quantum dot in germanium hut wire to a microwave resonator},\ }\href {https://doi.org/10.1088/1367-2630/aba85a} {\bibfield  {journal} {\bibinfo  {journal} {New Journal of Physics}\ }\textbf {\bibinfo {volume} {22}},\ \bibinfo {pages} {083068} (\bibinfo
  {year} {2020})}\BibitemShut {NoStop}%
\bibitem [{\citenamefont {Spethmann}\ \emph {et~al.}(2024)\citenamefont {Spethmann}, \citenamefont {Bosco}, \citenamefont {Hofmann}, \citenamefont {Klinovaja},\ and\ \citenamefont {Loss}}]{Spethmann2024}%
  \BibitemOpen
  \bibfield  {author} {\bibinfo {author} {\bibfnamefont {M.}~\bibnamefont {Spethmann}}, \bibinfo {author} {\bibfnamefont {S.}~\bibnamefont {Bosco}}, \bibinfo {author} {\bibfnamefont {A.}~\bibnamefont {Hofmann}}, \bibinfo {author} {\bibfnamefont {J.}~\bibnamefont {Klinovaja}},\ and\ \bibinfo {author} {\bibfnamefont {D.}~\bibnamefont {Loss}},\ }\bibfield  {title} {\bibinfo {title} {High-fidelity two-qubit gates of hybrid superconducting-semiconducting singlet-triplet qubits},\ }\href {https://doi.org/10.1103/PhysRevB.109.085303} {\bibfield  {journal} {\bibinfo  {journal} {Phys. Rev. B}\ }\textbf {\bibinfo {volume} {109}},\ \bibinfo {pages} {085303} (\bibinfo {year} {2024})}\BibitemShut {NoStop}%
\bibitem [{\citenamefont {Hu}\ and\ \citenamefont {Das~Sarma}(2006)}]{Xuedong2006}%
  \BibitemOpen
  \bibfield  {author} {\bibinfo {author} {\bibfnamefont {X.}~\bibnamefont {Hu}}\ and\ \bibinfo {author} {\bibfnamefont {S.}~\bibnamefont {Das~Sarma}},\ }\bibfield  {title} {\bibinfo {title} {Charge-fluctuation-induced dephasing of exchange-coupled spin qubits},\ }\href {https://doi.org/10.1103/PhysRevLett.96.100501} {\bibfield  {journal} {\bibinfo  {journal} {Phys. Rev. Lett.}\ }\textbf {\bibinfo {volume} {96}},\ \bibinfo {pages} {100501} (\bibinfo {year} {2006})}\BibitemShut {NoStop}%
\bibitem [{\citenamefont {Prechtel}\ \emph {et~al.}(2016)\citenamefont {Prechtel}, \citenamefont {Kuhlmann}, \citenamefont {Houel}, \citenamefont {Ludwig}, \citenamefont {Valentin}, \citenamefont {Wieck},\ and\ \citenamefont {Warburton}}]{Prechtel2016}%
  \BibitemOpen
  \bibfield  {author} {\bibinfo {author} {\bibfnamefont {J.~H.}\ \bibnamefont {Prechtel}}, \bibinfo {author} {\bibfnamefont {A.~V.}\ \bibnamefont {Kuhlmann}}, \bibinfo {author} {\bibfnamefont {J.}~\bibnamefont {Houel}}, \bibinfo {author} {\bibfnamefont {A.}~\bibnamefont {Ludwig}}, \bibinfo {author} {\bibfnamefont {S.~R.}\ \bibnamefont {Valentin}}, \bibinfo {author} {\bibfnamefont {A.~D.}\ \bibnamefont {Wieck}},\ and\ \bibinfo {author} {\bibfnamefont {R.~J.}\ \bibnamefont {Warburton}},\ }\bibfield  {title} {\bibinfo {title} {Decoupling a hole spin qubit from the nuclear spins},\ }\href {https://doi.org/10.1038/nmat4704} {\bibfield  {journal} {\bibinfo  {journal} {Nature Materials}\ }\textbf {\bibinfo {volume} {15}},\ \bibinfo {pages} {981} (\bibinfo {year} {2016})}\BibitemShut {NoStop}%
\bibitem [{\citenamefont {Li}\ \emph {et~al.}(2020)\citenamefont {Li}, \citenamefont {Venitucci},\ and\ \citenamefont {Niquet}}]{Li2020}%
  \BibitemOpen
  \bibfield  {author} {\bibinfo {author} {\bibfnamefont {J.}~\bibnamefont {Li}}, \bibinfo {author} {\bibfnamefont {B.}~\bibnamefont {Venitucci}},\ and\ \bibinfo {author} {\bibfnamefont {Y.-M.}\ \bibnamefont {Niquet}},\ }\bibfield  {title} {\bibinfo {title} {Hole-phonon interactions in quantum dots: Effects of phonon confinement and encapsulation materials on spin-orbit qubits},\ }\href {https://doi.org/10.1103/PhysRevB.102.075415} {\bibfield  {journal} {\bibinfo  {journal} {Phys. Rev. B}\ }\textbf {\bibinfo {volume} {102}},\ \bibinfo {pages} {075415} (\bibinfo {year} {2020})}\BibitemShut {NoStop}%
\bibitem [{\citenamefont {Bellentani}\ \emph {et~al.}(2021)\citenamefont {Bellentani}, \citenamefont {Bina}, \citenamefont {Bonen}, \citenamefont {Secchi}, \citenamefont {Bertoni}, \citenamefont {Voinigescu}, \citenamefont {Padovani}, \citenamefont {Larcher},\ and\ \citenamefont {Troiani}}]{Bellentani2021}%
  \BibitemOpen
  \bibfield  {author} {\bibinfo {author} {\bibfnamefont {L.}~\bibnamefont {Bellentani}}, \bibinfo {author} {\bibfnamefont {M.}~\bibnamefont {Bina}}, \bibinfo {author} {\bibfnamefont {S.}~\bibnamefont {Bonen}}, \bibinfo {author} {\bibfnamefont {A.}~\bibnamefont {Secchi}}, \bibinfo {author} {\bibfnamefont {A.}~\bibnamefont {Bertoni}}, \bibinfo {author} {\bibfnamefont {S.~P.}\ \bibnamefont {Voinigescu}}, \bibinfo {author} {\bibfnamefont {A.}~\bibnamefont {Padovani}}, \bibinfo {author} {\bibfnamefont {L.}~\bibnamefont {Larcher}},\ and\ \bibinfo {author} {\bibfnamefont {F.}~\bibnamefont {Troiani}},\ }\bibfield  {title} {\bibinfo {title} {Toward hole-spin qubits in $\mathrm{Si}$ $p$-mosfets within a planar cmos foundry technology},\ }\href {https://doi.org/10.1103/PhysRevApplied.16.054034} {\bibfield  {journal} {\bibinfo  {journal} {Phys. Rev. Appl.}\ }\textbf {\bibinfo {volume} {16}},\ \bibinfo {pages} {054034} (\bibinfo {year} {2021})}\BibitemShut {NoStop}%
\bibitem [{\citenamefont {Froning}\ \emph {et~al.}(2021{\natexlab{c}})\citenamefont {Froning}, \citenamefont {Camenzind}, \citenamefont {van~der Molen}, \citenamefont {Li}, \citenamefont {Bakkers}, \citenamefont {Zumb{\"u}hl},\ and\ \citenamefont {Braakman}}]{Froning2021}%
  \BibitemOpen
  \bibfield  {author} {\bibinfo {author} {\bibfnamefont {F.~N.~M.}\ \bibnamefont {Froning}}, \bibinfo {author} {\bibfnamefont {L.~C.}\ \bibnamefont {Camenzind}}, \bibinfo {author} {\bibfnamefont {O.~A.~H.}\ \bibnamefont {van~der Molen}}, \bibinfo {author} {\bibfnamefont {A.}~\bibnamefont {Li}}, \bibinfo {author} {\bibfnamefont {E.~P. A.~M.}\ \bibnamefont {Bakkers}}, \bibinfo {author} {\bibfnamefont {D.~M.}\ \bibnamefont {Zumb{\"u}hl}},\ and\ \bibinfo {author} {\bibfnamefont {F.~R.}\ \bibnamefont {Braakman}},\ }\bibfield  {title} {\bibinfo {title} {Ultrafast hole spin qubit with gate-tunable spin--orbit switch functionality},\ }\href {https://doi.org/10.1038/s41565-020-00828-6} {\bibfield  {journal} {\bibinfo  {journal} {Nature Nanotechnology}\ }\textbf {\bibinfo {volume} {16}},\ \bibinfo {pages} {308} (\bibinfo {year} {2021}{\natexlab{c}})}\BibitemShut {NoStop}%
\bibitem [{\citenamefont {Shalak}\ \emph {et~al.}(2023)\citenamefont {Shalak}, \citenamefont {Delerue},\ and\ \citenamefont {Niquet}}]{Shalak2023}%
  \BibitemOpen
  \bibfield  {author} {\bibinfo {author} {\bibfnamefont {B.}~\bibnamefont {Shalak}}, \bibinfo {author} {\bibfnamefont {C.}~\bibnamefont {Delerue}},\ and\ \bibinfo {author} {\bibfnamefont {Y.-M.}\ \bibnamefont {Niquet}},\ }\bibfield  {title} {\bibinfo {title} {Modeling of spin decoherence in a si hole qubit perturbed by a single charge fluctuator},\ }\href {https://doi.org/10.1103/PhysRevB.107.125415} {\bibfield  {journal} {\bibinfo  {journal} {Phys. Rev. B}\ }\textbf {\bibinfo {volume} {107}},\ \bibinfo {pages} {125415} (\bibinfo {year} {2023})}\BibitemShut {NoStop}%
\bibitem [{\citenamefont {Tahan}\ \emph {et~al.}(2002)\citenamefont {Tahan}, \citenamefont {Friesen},\ and\ \citenamefont {Joynt}}]{Tahan2002}%
  \BibitemOpen
  \bibfield  {author} {\bibinfo {author} {\bibfnamefont {C.}~\bibnamefont {Tahan}}, \bibinfo {author} {\bibfnamefont {M.}~\bibnamefont {Friesen}},\ and\ \bibinfo {author} {\bibfnamefont {R.}~\bibnamefont {Joynt}},\ }\bibfield  {title} {\bibinfo {title} {Decoherence of electron spin qubits in si-based quantum computers},\ }\href {https://doi.org/10.1103/PhysRevB.66.035314} {\bibfield  {journal} {\bibinfo  {journal} {Phys. Rev. B}\ }\textbf {\bibinfo {volume} {66}},\ \bibinfo {pages} {035314} (\bibinfo {year} {2002})}\BibitemShut {NoStop}%
\bibitem [{\citenamefont {Itakura}\ and\ \citenamefont {Tokura}(2003)}]{Itakura2003}%
  \BibitemOpen
  \bibfield  {author} {\bibinfo {author} {\bibfnamefont {T.}~\bibnamefont {Itakura}}\ and\ \bibinfo {author} {\bibfnamefont {Y.}~\bibnamefont {Tokura}},\ }\bibfield  {title} {\bibinfo {title} {Dephasing due to background charge fluctuations},\ }\href {https://doi.org/10.1103/PhysRevB.67.195320} {\bibfield  {journal} {\bibinfo  {journal} {Phys. Rev. B}\ }\textbf {\bibinfo {volume} {67}},\ \bibinfo {pages} {195320} (\bibinfo {year} {2003})}\BibitemShut {NoStop}%
\bibitem [{\citenamefont {Tahan}\ and\ \citenamefont {Joynt}(2005)}]{Tahan2005}%
  \BibitemOpen
  \bibfield  {author} {\bibinfo {author} {\bibfnamefont {C.}~\bibnamefont {Tahan}}\ and\ \bibinfo {author} {\bibfnamefont {R.}~\bibnamefont {Joynt}},\ }\bibfield  {title} {\bibinfo {title} {Rashba spin-orbit coupling and spin relaxation in silicon quantum wells},\ }\href {https://doi.org/10.1103/PhysRevB.71.075315} {\bibfield  {journal} {\bibinfo  {journal} {Phys. Rev. B}\ }\textbf {\bibinfo {volume} {71}},\ \bibinfo {pages} {075315} (\bibinfo {year} {2005})}\BibitemShut {NoStop}%
\bibitem [{\citenamefont {Tahan}\ and\ \citenamefont {Joynt}(2014)}]{Tahan2014}%
  \BibitemOpen
  \bibfield  {author} {\bibinfo {author} {\bibfnamefont {C.}~\bibnamefont {Tahan}}\ and\ \bibinfo {author} {\bibfnamefont {R.}~\bibnamefont {Joynt}},\ }\bibfield  {title} {\bibinfo {title} {Relaxation of excited spin, orbital, and valley qubit states in ideal silicon quantum dots},\ }\href {https://doi.org/10.1103/PhysRevB.89.075302} {\bibfield  {journal} {\bibinfo  {journal} {Phys. Rev. B}\ }\textbf {\bibinfo {volume} {89}},\ \bibinfo {pages} {075302} (\bibinfo {year} {2014})}\BibitemShut {NoStop}%
\bibitem [{\citenamefont {Cywi\ifmmode~\acute{n}\else \'{n}\fi{}ski}\ \emph {et~al.}(2008)\citenamefont {Cywi\ifmmode~\acute{n}\else \'{n}\fi{}ski}, \citenamefont {Lutchyn}, \citenamefont {Nave},\ and\ \citenamefont {Das~Sarma}}]{Cywinski2008}%
  \BibitemOpen
  \bibfield  {author} {\bibinfo {author} {\bibfnamefont {L.}~\bibnamefont {Cywi\ifmmode~\acute{n}\else \'{n}\fi{}ski}}, \bibinfo {author} {\bibfnamefont {R.~M.}\ \bibnamefont {Lutchyn}}, \bibinfo {author} {\bibfnamefont {C.~P.}\ \bibnamefont {Nave}},\ and\ \bibinfo {author} {\bibfnamefont {S.}~\bibnamefont {Das~Sarma}},\ }\bibfield  {title} {\bibinfo {title} {How to enhance dephasing time in superconducting qubits},\ }\href {https://doi.org/10.1103/PhysRevB.77.174509} {\bibfield  {journal} {\bibinfo  {journal} {Phys. Rev. B}\ }\textbf {\bibinfo {volume} {77}},\ \bibinfo {pages} {174509} (\bibinfo {year} {2008})}\BibitemShut {NoStop}%
\bibitem [{\citenamefont {Lutchyn}\ \emph {et~al.}(2008)\citenamefont {Lutchyn}, \citenamefont {Cywi\ifmmode~\acute{n}\else \'{n}\fi{}ski}, \citenamefont {Nave},\ and\ \citenamefont {Das~Sarma}}]{Lutchyn2008}%
  \BibitemOpen
  \bibfield  {author} {\bibinfo {author} {\bibfnamefont {R.~M.}\ \bibnamefont {Lutchyn}}, \bibinfo {author} {\bibfnamefont {L.}~\bibnamefont {Cywi\ifmmode~\acute{n}\else \'{n}\fi{}ski}}, \bibinfo {author} {\bibfnamefont {C.~P.}\ \bibnamefont {Nave}},\ and\ \bibinfo {author} {\bibfnamefont {S.}~\bibnamefont {Das~Sarma}},\ }\bibfield  {title} {\bibinfo {title} {Quantum decoherence of a charge qubit in a spin-fermion model},\ }\href {https://doi.org/10.1103/PhysRevB.78.024508} {\bibfield  {journal} {\bibinfo  {journal} {Phys. Rev. B}\ }\textbf {\bibinfo {volume} {78}},\ \bibinfo {pages} {024508} (\bibinfo {year} {2008})}\BibitemShut {NoStop}%
\bibitem [{\citenamefont {Bergli}\ \emph {et~al.}(2009)\citenamefont {Bergli}, \citenamefont {Galperin},\ and\ \citenamefont {Altshuler}}]{Bergli2009}%
  \BibitemOpen
  \bibfield  {author} {\bibinfo {author} {\bibfnamefont {J.}~\bibnamefont {Bergli}}, \bibinfo {author} {\bibfnamefont {Y.~M.}\ \bibnamefont {Galperin}},\ and\ \bibinfo {author} {\bibfnamefont {B.~L.}\ \bibnamefont {Altshuler}},\ }\bibfield  {title} {\bibinfo {title} {Decoherence in qubits due to low-frequency noise},\ }\href {https://doi.org/10.1088/1367-2630/11/2/025002} {\bibfield  {journal} {\bibinfo  {journal} {New Journal of Physics}\ }\textbf {\bibinfo {volume} {11}},\ \bibinfo {pages} {025002} (\bibinfo {year} {2009})}\BibitemShut {NoStop}%
\bibitem [{\citenamefont {Paladino}\ \emph {et~al.}(2014)\citenamefont {Paladino}, \citenamefont {Galperin}, \citenamefont {Falci},\ and\ \citenamefont {Altshuler}}]{Paladino2014}%
  \BibitemOpen
  \bibfield  {author} {\bibinfo {author} {\bibfnamefont {E.}~\bibnamefont {Paladino}}, \bibinfo {author} {\bibfnamefont {Y.~M.}\ \bibnamefont {Galperin}}, \bibinfo {author} {\bibfnamefont {G.}~\bibnamefont {Falci}},\ and\ \bibinfo {author} {\bibfnamefont {B.~L.}\ \bibnamefont {Altshuler}},\ }\bibfield  {title} {\bibinfo {title} {$1/f$ noise: Implications for solid-state quantum information},\ }\href {https://doi.org/10.1103/RevModPhys.86.361} {\bibfield  {journal} {\bibinfo  {journal} {Rev. Mod. Phys.}\ }\textbf {\bibinfo {volume} {86}},\ \bibinfo {pages} {361} (\bibinfo {year} {2014})}\BibitemShut {NoStop}%
\bibitem [{\citenamefont {Koch}\ \emph {et~al.}(2007)\citenamefont {Koch}, \citenamefont {Yu}, \citenamefont {Gambetta}, \citenamefont {Houck}, \citenamefont {Schuster}, \citenamefont {Majer}, \citenamefont {Blais}, \citenamefont {Devoret}, \citenamefont {Girvin},\ and\ \citenamefont {Schoelkopf}}]{Koch2007}%
  \BibitemOpen
  \bibfield  {author} {\bibinfo {author} {\bibfnamefont {J.}~\bibnamefont {Koch}}, \bibinfo {author} {\bibfnamefont {T.~M.}\ \bibnamefont {Yu}}, \bibinfo {author} {\bibfnamefont {J.}~\bibnamefont {Gambetta}}, \bibinfo {author} {\bibfnamefont {A.~A.}\ \bibnamefont {Houck}}, \bibinfo {author} {\bibfnamefont {D.~I.}\ \bibnamefont {Schuster}}, \bibinfo {author} {\bibfnamefont {J.}~\bibnamefont {Majer}}, \bibinfo {author} {\bibfnamefont {A.}~\bibnamefont {Blais}}, \bibinfo {author} {\bibfnamefont {M.~H.}\ \bibnamefont {Devoret}}, \bibinfo {author} {\bibfnamefont {S.~M.}\ \bibnamefont {Girvin}},\ and\ \bibinfo {author} {\bibfnamefont {R.~J.}\ \bibnamefont {Schoelkopf}},\ }\bibfield  {title} {\bibinfo {title} {Charge-insensitive qubit design derived from the cooper pair box},\ }\href {https://doi.org/10.1103/PhysRevA.76.042319} {\bibfield  {journal} {\bibinfo  {journal} {Phys. Rev. A}\ }\textbf {\bibinfo {volume} {76}},\ \bibinfo {pages} {042319} (\bibinfo {year} {2007})}\BibitemShut {NoStop}%
\bibitem [{\citenamefont {You}\ \emph {et~al.}(2007)\citenamefont {You}, \citenamefont {Hu}, \citenamefont {Ashhab},\ and\ \citenamefont {Nori}}]{You2007}%
  \BibitemOpen
  \bibfield  {author} {\bibinfo {author} {\bibfnamefont {J.~Q.}\ \bibnamefont {You}}, \bibinfo {author} {\bibfnamefont {X.}~\bibnamefont {Hu}}, \bibinfo {author} {\bibfnamefont {S.}~\bibnamefont {Ashhab}},\ and\ \bibinfo {author} {\bibfnamefont {F.}~\bibnamefont {Nori}},\ }\bibfield  {title} {\bibinfo {title} {Low-decoherence flux qubit},\ }\href {https://doi.org/10.1103/PhysRevB.75.140515} {\bibfield  {journal} {\bibinfo  {journal} {Phys. Rev. B}\ }\textbf {\bibinfo {volume} {75}},\ \bibinfo {pages} {140515} (\bibinfo {year} {2007})}\BibitemShut {NoStop}%
\bibitem [{\citenamefont {Bergli}\ \emph {et~al.}(2006)\citenamefont {Bergli}, \citenamefont {Galperin},\ and\ \citenamefont {Altshuler}}]{Bergli2006}%
  \BibitemOpen
  \bibfield  {author} {\bibinfo {author} {\bibfnamefont {J.}~\bibnamefont {Bergli}}, \bibinfo {author} {\bibfnamefont {Y.~M.}\ \bibnamefont {Galperin}},\ and\ \bibinfo {author} {\bibfnamefont {B.~L.}\ \bibnamefont {Altshuler}},\ }\bibfield  {title} {\bibinfo {title} {Decoherence of a qubit by non-gaussian noise at an arbitrary working point},\ }\href {https://doi.org/10.1103/PhysRevB.74.024509} {\bibfield  {journal} {\bibinfo  {journal} {Phys. Rev. B}\ }\textbf {\bibinfo {volume} {74}},\ \bibinfo {pages} {024509} (\bibinfo {year} {2006})}\BibitemShut {NoStop}%
\bibitem [{\citenamefont {Dutta}\ and\ \citenamefont {Horn}(1981)}]{Dutta1981}%
  \BibitemOpen
  \bibfield  {author} {\bibinfo {author} {\bibfnamefont {P.}~\bibnamefont {Dutta}}\ and\ \bibinfo {author} {\bibfnamefont {P.~M.}\ \bibnamefont {Horn}},\ }\bibfield  {title} {\bibinfo {title} {Low-frequency fluctuations in solids: $\frac{1}{f}$ noise},\ }\href {https://doi.org/10.1103/RevModPhys.53.497} {\bibfield  {journal} {\bibinfo  {journal} {Rev. Mod. Phys.}\ }\textbf {\bibinfo {volume} {53}},\ \bibinfo {pages} {497} (\bibinfo {year} {1981})}\BibitemShut {NoStop}%
\bibitem [{\citenamefont {Shnirman}\ \emph {et~al.}(2005)\citenamefont {Shnirman}, \citenamefont {Sch\"on}, \citenamefont {Martin},\ and\ \citenamefont {Makhlin}}]{Shnirman2005}%
  \BibitemOpen
  \bibfield  {author} {\bibinfo {author} {\bibfnamefont {A.}~\bibnamefont {Shnirman}}, \bibinfo {author} {\bibfnamefont {G.}~\bibnamefont {Sch\"on}}, \bibinfo {author} {\bibfnamefont {I.}~\bibnamefont {Martin}},\ and\ \bibinfo {author} {\bibfnamefont {Y.}~\bibnamefont {Makhlin}},\ }\bibfield  {title} {\bibinfo {title} {Low- and high-frequency noise from coherent two-level systems},\ }\href {https://doi.org/10.1103/PhysRevLett.94.127002} {\bibfield  {journal} {\bibinfo  {journal} {Phys. Rev. Lett.}\ }\textbf {\bibinfo {volume} {94}},\ \bibinfo {pages} {127002} (\bibinfo {year} {2005})}\BibitemShut {NoStop}%
\bibitem [{\citenamefont {K{\c{e}}pa}\ \emph {et~al.}(2023)\citenamefont {K{\c{e}}pa}, \citenamefont {Focke}, \citenamefont {Cywi{\'n}ski},\ and\ \citenamefont {Krzywda}}]{Focke2023}%
  \BibitemOpen
  \bibfield  {author} {\bibinfo {author} {\bibfnamefont {M.}~\bibnamefont {K{\c{e}}pa}}, \bibinfo {author} {\bibfnamefont {N.}~\bibnamefont {Focke}}, \bibinfo {author} {\bibfnamefont {{\L}.}~\bibnamefont {Cywi{\'n}ski}},\ and\ \bibinfo {author} {\bibfnamefont {J.~A.}\ \bibnamefont {Krzywda}},\ }\bibfield  {title} {\bibinfo {title} {{Simulation of 1 / f charge noise affecting a quantum dot in a Si/SiGe structure}},\ }\href {https://doi.org/10.1063/5.0151029} {\bibfield  {journal} {\bibinfo  {journal} {Applied Physics Letters}\ }\textbf {\bibinfo {volume} {123}},\ \bibinfo {pages} {034005} (\bibinfo {year} {2023})}\BibitemShut {NoStop}%
\bibitem [{\citenamefont {Shehata}\ \emph {et~al.}(2023)\citenamefont {Shehata}, \citenamefont {Simion}, \citenamefont {Li}, \citenamefont {Mohiyaddin}, \citenamefont {Wan}, \citenamefont {Mongillo}, \citenamefont {Govoreanu}, \citenamefont {Radu}, \citenamefont {De~Greve},\ and\ \citenamefont {Van~Dorpe}}]{Shehata2023}%
  \BibitemOpen
  \bibfield  {author} {\bibinfo {author} {\bibfnamefont {M.~M. E.~K.}\ \bibnamefont {Shehata}}, \bibinfo {author} {\bibfnamefont {G.}~\bibnamefont {Simion}}, \bibinfo {author} {\bibfnamefont {R.}~\bibnamefont {Li}}, \bibinfo {author} {\bibfnamefont {F.~A.}\ \bibnamefont {Mohiyaddin}}, \bibinfo {author} {\bibfnamefont {D.}~\bibnamefont {Wan}}, \bibinfo {author} {\bibfnamefont {M.}~\bibnamefont {Mongillo}}, \bibinfo {author} {\bibfnamefont {B.}~\bibnamefont {Govoreanu}}, \bibinfo {author} {\bibfnamefont {I.}~\bibnamefont {Radu}}, \bibinfo {author} {\bibfnamefont {K.}~\bibnamefont {De~Greve}},\ and\ \bibinfo {author} {\bibfnamefont {P.}~\bibnamefont {Van~Dorpe}},\ }\bibfield  {title} {\bibinfo {title} {Modeling semiconductor spin qubits and their charge noise environment for quantum gate fidelity estimation},\ }\href {https://doi.org/10.1103/PhysRevB.108.045305} {\bibfield  {journal} {\bibinfo  {journal} {Phys. Rev. B}\ }\textbf {\bibinfo {volume} {108}},\ \bibinfo {pages} {045305} (\bibinfo {year}
  {2023})}\BibitemShut {NoStop}%
\bibitem [{\citenamefont {Martinez}\ \emph {et~al.}(2024)\citenamefont {Martinez}, \citenamefont {de~Franceschi},\ and\ \citenamefont {Niquet}}]{Martinez2024}%
  \BibitemOpen
  \bibfield  {author} {\bibinfo {author} {\bibfnamefont {B.}~\bibnamefont {Martinez}}, \bibinfo {author} {\bibfnamefont {S.}~\bibnamefont {de~Franceschi}},\ and\ \bibinfo {author} {\bibfnamefont {Y.-M.}\ \bibnamefont {Niquet}},\ }\bibfield  {title} {\bibinfo {title} {Mitigating variability in epitaxial-heterostructure-based spin-qubit devices by optimizing gate layout},\ }\href {https://doi.org/10.1103/PhysRevApplied.22.024030} {\bibfield  {journal} {\bibinfo  {journal} {Phys. Rev. Appl.}\ }\textbf {\bibinfo {volume} {22}},\ \bibinfo {pages} {024030} (\bibinfo {year} {2024})}\BibitemShut {NoStop}%
\bibitem [{\citenamefont {Martinez}\ and\ \citenamefont {Niquet}(2022)}]{Martinez2022PRA}%
  \BibitemOpen
  \bibfield  {author} {\bibinfo {author} {\bibfnamefont {B.}~\bibnamefont {Martinez}}\ and\ \bibinfo {author} {\bibfnamefont {Y.-M.}\ \bibnamefont {Niquet}},\ }\bibfield  {title} {\bibinfo {title} {Variability of electron and hole spin qubits due to interface roughness and charge traps},\ }\href {https://doi.org/10.1103/PhysRevApplied.17.024022} {\bibfield  {journal} {\bibinfo  {journal} {Phys. Rev. Appl.}\ }\textbf {\bibinfo {volume} {17}},\ \bibinfo {pages} {024022} (\bibinfo {year} {2022})}\BibitemShut {NoStop}%
\bibitem [{\citenamefont {Luo}\ \emph {et~al.}(2015)\citenamefont {Luo}, \citenamefont {Bester},\ and\ \citenamefont {Zunger}}]{JunWei2015}%
  \BibitemOpen
  \bibfield  {author} {\bibinfo {author} {\bibfnamefont {J.-W.}\ \bibnamefont {Luo}}, \bibinfo {author} {\bibfnamefont {G.}~\bibnamefont {Bester}},\ and\ \bibinfo {author} {\bibfnamefont {A.}~\bibnamefont {Zunger}},\ }\bibfield  {title} {\bibinfo {title} {Supercoupling between heavy-hole and light-hole states in nanostructures},\ }\href {https://doi.org/10.1103/PhysRevB.92.165301} {\bibfield  {journal} {\bibinfo  {journal} {Phys. Rev. B}\ }\textbf {\bibinfo {volume} {92}},\ \bibinfo {pages} {165301} (\bibinfo {year} {2015})}\BibitemShut {NoStop}%
\bibitem [{\citenamefont {Martinez}\ \emph {et~al.}(2022)\citenamefont {Martinez}, \citenamefont {Abadillo-Uriel}, \citenamefont {Rodr\'{\i}guez-Mena},\ and\ \citenamefont {Niquet}}]{Martinez2022}%
  \BibitemOpen
  \bibfield  {author} {\bibinfo {author} {\bibfnamefont {B.}~\bibnamefont {Martinez}}, \bibinfo {author} {\bibfnamefont {J.~C.}\ \bibnamefont {Abadillo-Uriel}}, \bibinfo {author} {\bibfnamefont {E.~A.}\ \bibnamefont {Rodr\'{\i}guez-Mena}},\ and\ \bibinfo {author} {\bibfnamefont {Y.-M.}\ \bibnamefont {Niquet}},\ }\bibfield  {title} {\bibinfo {title} {Hole spin manipulation in inhomogeneous and nonseparable electric fields},\ }\href {https://doi.org/10.1103/PhysRevB.106.235426} {\bibfield  {journal} {\bibinfo  {journal} {Phys. Rev. B}\ }\textbf {\bibinfo {volume} {106}},\ \bibinfo {pages} {235426} (\bibinfo {year} {2022})}\BibitemShut {NoStop}%
\bibitem [{\citenamefont {Mauro}\ \emph {et~al.}(2024{\natexlab{a}})\citenamefont {Mauro}, \citenamefont {Rodr\'{\i}guez-Mena}, \citenamefont {Bassi}, \citenamefont {Schmitt},\ and\ \citenamefont {Niquet}}]{Mauro2024}%
  \BibitemOpen
  \bibfield  {author} {\bibinfo {author} {\bibfnamefont {L.}~\bibnamefont {Mauro}}, \bibinfo {author} {\bibfnamefont {E.~A.}\ \bibnamefont {Rodr\'{\i}guez-Mena}}, \bibinfo {author} {\bibfnamefont {M.}~\bibnamefont {Bassi}}, \bibinfo {author} {\bibfnamefont {V.}~\bibnamefont {Schmitt}},\ and\ \bibinfo {author} {\bibfnamefont {Y.-M.}\ \bibnamefont {Niquet}},\ }\bibfield  {title} {\bibinfo {title} {Geometry of the dephasing sweet spots of spin-orbit qubits},\ }\href {https://doi.org/10.1103/PhysRevB.109.155406} {\bibfield  {journal} {\bibinfo  {journal} {Phys. Rev. B}\ }\textbf {\bibinfo {volume} {109}},\ \bibinfo {pages} {155406} (\bibinfo {year} {2024}{\natexlab{a}})}\BibitemShut {NoStop}%
\bibitem [{\citenamefont {Bermeister}\ \emph {et~al.}(2014)\citenamefont {Bermeister}, \citenamefont {Keith},\ and\ \citenamefont {Culcer}}]{Bermeister2014}%
  \BibitemOpen
  \bibfield  {author} {\bibinfo {author} {\bibfnamefont {A.}~\bibnamefont {Bermeister}}, \bibinfo {author} {\bibfnamefont {D.}~\bibnamefont {Keith}},\ and\ \bibinfo {author} {\bibfnamefont {D.}~\bibnamefont {Culcer}},\ }\bibfield  {title} {\bibinfo {title} {{Charge noise, spin-orbit coupling, and dephasing of single-spin qubits}},\ }\href {https://doi.org/10.1063/1.4901162} {\bibfield  {journal} {\bibinfo  {journal} {Applied Physics Letters}\ }\textbf {\bibinfo {volume} {105}},\ \bibinfo {pages} {192102} (\bibinfo {year} {2014})}\BibitemShut {NoStop}%
\bibitem [{\citenamefont {Culcer}\ \emph {et~al.}(2006)\citenamefont {Culcer}, \citenamefont {Lechner},\ and\ \citenamefont {Winkler}}]{Culcer2006}%
  \BibitemOpen
  \bibfield  {author} {\bibinfo {author} {\bibfnamefont {D.}~\bibnamefont {Culcer}}, \bibinfo {author} {\bibfnamefont {C.}~\bibnamefont {Lechner}},\ and\ \bibinfo {author} {\bibfnamefont {R.}~\bibnamefont {Winkler}},\ }\bibfield  {title} {\bibinfo {title} {Spin precession and alternating spin polarization in spin-$3/2$ hole systems},\ }\href {https://doi.org/10.1103/PhysRevLett.97.106601} {\bibfield  {journal} {\bibinfo  {journal} {Phys. Rev. Lett.}\ }\textbf {\bibinfo {volume} {97}},\ \bibinfo {pages} {106601} (\bibinfo {year} {2006})}\BibitemShut {NoStop}%
\bibitem [{\citenamefont {Liu}\ \emph {et~al.}(2018)\citenamefont {Liu}, \citenamefont {Marcellina}, \citenamefont {Hamilton},\ and\ \citenamefont {Culcer}}]{Liu2018}%
  \BibitemOpen
  \bibfield  {author} {\bibinfo {author} {\bibfnamefont {H.}~\bibnamefont {Liu}}, \bibinfo {author} {\bibfnamefont {E.}~\bibnamefont {Marcellina}}, \bibinfo {author} {\bibfnamefont {A.~R.}\ \bibnamefont {Hamilton}},\ and\ \bibinfo {author} {\bibfnamefont {D.}~\bibnamefont {Culcer}},\ }\bibfield  {title} {\bibinfo {title} {Strong spin-orbit contribution to the hall coefficient of two-dimensional hole systems},\ }\href {https://doi.org/10.1103/PhysRevLett.121.087701} {\bibfield  {journal} {\bibinfo  {journal} {Phys. Rev. Lett.}\ }\textbf {\bibinfo {volume} {121}},\ \bibinfo {pages} {087701} (\bibinfo {year} {2018})}\BibitemShut {NoStop}%
\bibitem [{\citenamefont {Abadillo-Uriel}\ \emph {et~al.}(2018)\citenamefont {Abadillo-Uriel}, \citenamefont {Salfi}, \citenamefont {Hu}, \citenamefont {Rogge}, \citenamefont {Calderón},\ and\ \citenamefont {Culcer}}]{Abadillo2018}%
  \BibitemOpen
  \bibfield  {author} {\bibinfo {author} {\bibfnamefont {J.~C.}\ \bibnamefont {Abadillo-Uriel}}, \bibinfo {author} {\bibfnamefont {J.}~\bibnamefont {Salfi}}, \bibinfo {author} {\bibfnamefont {X.}~\bibnamefont {Hu}}, \bibinfo {author} {\bibfnamefont {S.}~\bibnamefont {Rogge}}, \bibinfo {author} {\bibfnamefont {M.~J.}\ \bibnamefont {Calderón}},\ and\ \bibinfo {author} {\bibfnamefont {D.}~\bibnamefont {Culcer}},\ }\bibfield  {title} {\bibinfo {title} {{Entanglement control and magic angles for acceptor qubits in Si}},\ }\href {https://doi.org/10.1063/1.5036521} {\bibfield  {journal} {\bibinfo  {journal} {Applied Physics Letters}\ }\textbf {\bibinfo {volume} {113}},\ \bibinfo {pages} {012102} (\bibinfo {year} {2018})}\BibitemShut {NoStop}%
\bibitem [{\citenamefont {Cullen}\ \emph {et~al.}(2021)\citenamefont {Cullen}, \citenamefont {Bhalla}, \citenamefont {Marcellina}, \citenamefont {Hamilton},\ and\ \citenamefont {Culcer}}]{Cullen2021}%
  \BibitemOpen
  \bibfield  {author} {\bibinfo {author} {\bibfnamefont {J.~H.}\ \bibnamefont {Cullen}}, \bibinfo {author} {\bibfnamefont {P.}~\bibnamefont {Bhalla}}, \bibinfo {author} {\bibfnamefont {E.}~\bibnamefont {Marcellina}}, \bibinfo {author} {\bibfnamefont {A.~R.}\ \bibnamefont {Hamilton}},\ and\ \bibinfo {author} {\bibfnamefont {D.}~\bibnamefont {Culcer}},\ }\bibfield  {title} {\bibinfo {title} {Generating a topological anomalous hall effect in a nonmagnetic conductor: An in-plane magnetic field as a direct probe of the berry curvature},\ }\href {https://doi.org/10.1103/PhysRevLett.126.256601} {\bibfield  {journal} {\bibinfo  {journal} {Phys. Rev. Lett.}\ }\textbf {\bibinfo {volume} {126}},\ \bibinfo {pages} {256601} (\bibinfo {year} {2021})}\BibitemShut {NoStop}%
\bibitem [{\citenamefont {Salfi}\ \emph {et~al.}(2016{\natexlab{a}})\citenamefont {Salfi}, \citenamefont {Tong}, \citenamefont {Rogge},\ and\ \citenamefont {Culcer}}]{Salfi2016}%
  \BibitemOpen
  \bibfield  {author} {\bibinfo {author} {\bibfnamefont {J.}~\bibnamefont {Salfi}}, \bibinfo {author} {\bibfnamefont {M.}~\bibnamefont {Tong}}, \bibinfo {author} {\bibfnamefont {S.}~\bibnamefont {Rogge}},\ and\ \bibinfo {author} {\bibfnamefont {D.}~\bibnamefont {Culcer}},\ }\bibfield  {title} {\bibinfo {title} {Quantum computing with acceptor spins in silicon},\ }\href {https://doi.org/10.1088/0957-4484/27/24/244001} {\bibfield  {journal} {\bibinfo  {journal} {Nanotechnology}\ }\textbf {\bibinfo {volume} {27}},\ \bibinfo {pages} {244001} (\bibinfo {year} {2016}{\natexlab{a}})}\BibitemShut {NoStop}%
\bibitem [{\citenamefont {Salfi}\ \emph {et~al.}(2016{\natexlab{b}})\citenamefont {Salfi}, \citenamefont {Mol}, \citenamefont {Culcer},\ and\ \citenamefont {Rogge}}]{Salfi2016PRL}%
  \BibitemOpen
  \bibfield  {author} {\bibinfo {author} {\bibfnamefont {J.}~\bibnamefont {Salfi}}, \bibinfo {author} {\bibfnamefont {J.~A.}\ \bibnamefont {Mol}}, \bibinfo {author} {\bibfnamefont {D.}~\bibnamefont {Culcer}},\ and\ \bibinfo {author} {\bibfnamefont {S.}~\bibnamefont {Rogge}},\ }\bibfield  {title} {\bibinfo {title} {Charge-insensitive single-atom spin-orbit qubit in silicon},\ }\href {https://doi.org/10.1103/PhysRevLett.116.246801} {\bibfield  {journal} {\bibinfo  {journal} {Phys. Rev. Lett.}\ }\textbf {\bibinfo {volume} {116}},\ \bibinfo {pages} {246801} (\bibinfo {year} {2016}{\natexlab{b}})}\BibitemShut {NoStop}%
\bibitem [{\citenamefont {Kloeffel}\ \emph {et~al.}(2018)\citenamefont {Kloeffel}, \citenamefont {Ran\ifmmode \check{c}\else \v{c}\fi{}i\ifmmode~\acute{c}\else \'{c}\fi{}},\ and\ \citenamefont {Loss}}]{Kloeffel2018}%
  \BibitemOpen
  \bibfield  {author} {\bibinfo {author} {\bibfnamefont {C.}~\bibnamefont {Kloeffel}}, \bibinfo {author} {\bibfnamefont {M.~J.}\ \bibnamefont {Ran\ifmmode \check{c}\else \v{c}\fi{}i\ifmmode~\acute{c}\else \'{c}\fi{}}},\ and\ \bibinfo {author} {\bibfnamefont {D.}~\bibnamefont {Loss}},\ }\bibfield  {title} {\bibinfo {title} {Direct rashba spin-orbit interaction in si and ge nanowires with different growth directions},\ }\href {https://doi.org/10.1103/PhysRevB.97.235422} {\bibfield  {journal} {\bibinfo  {journal} {Phys. Rev. B}\ }\textbf {\bibinfo {volume} {97}},\ \bibinfo {pages} {235422} (\bibinfo {year} {2018})}\BibitemShut {NoStop}%
\bibitem [{\citenamefont {Wang}\ \emph {et~al.}(2021)\citenamefont {Wang}, \citenamefont {Marcellina}, \citenamefont {Hamilton}, \citenamefont {Cullen}, \citenamefont {Rogge}, \citenamefont {Salfi},\ and\ \citenamefont {Culcer}}]{Wang2021}%
  \BibitemOpen
  \bibfield  {author} {\bibinfo {author} {\bibfnamefont {Z.}~\bibnamefont {Wang}}, \bibinfo {author} {\bibfnamefont {E.}~\bibnamefont {Marcellina}}, \bibinfo {author} {\bibfnamefont {A.~R.}\ \bibnamefont {Hamilton}}, \bibinfo {author} {\bibfnamefont {J.~H.}\ \bibnamefont {Cullen}}, \bibinfo {author} {\bibfnamefont {S.}~\bibnamefont {Rogge}}, \bibinfo {author} {\bibfnamefont {J.}~\bibnamefont {Salfi}},\ and\ \bibinfo {author} {\bibfnamefont {D.}~\bibnamefont {Culcer}},\ }\bibfield  {title} {\bibinfo {title} {Optimal operation points for ultrafast, highly coherent ge hole spin-orbit qubits},\ }\href {https://doi.org/10.1038/s41534-021-00386-2} {\bibfield  {journal} {\bibinfo  {journal} {npj Quantum Information}\ }\textbf {\bibinfo {volume} {7}},\ \bibinfo {pages} {54} (\bibinfo {year} {2021})}\BibitemShut {NoStop}%
\bibitem [{\citenamefont {Bosco}\ \emph {et~al.}(2021{\natexlab{a}})\citenamefont {Bosco}, \citenamefont {Het\'enyi},\ and\ \citenamefont {Loss}}]{Bosco2021PRX}%
  \BibitemOpen
  \bibfield  {author} {\bibinfo {author} {\bibfnamefont {S.}~\bibnamefont {Bosco}}, \bibinfo {author} {\bibfnamefont {B.}~\bibnamefont {Het\'enyi}},\ and\ \bibinfo {author} {\bibfnamefont {D.}~\bibnamefont {Loss}},\ }\bibfield  {title} {\bibinfo {title} {Hole spin qubits in $\mathrm{Si}$ finfets with fully tunable spin-orbit coupling and sweet spots for charge noise},\ }\href {https://doi.org/10.1103/PRXQuantum.2.010348} {\bibfield  {journal} {\bibinfo  {journal} {PRX Quantum}\ }\textbf {\bibinfo {volume} {2}},\ \bibinfo {pages} {010348} (\bibinfo {year} {2021}{\natexlab{a}})}\BibitemShut {NoStop}%
\bibitem [{\citenamefont {Bosco}\ and\ \citenamefont {Loss}(2022)}]{Bosco2022}%
  \BibitemOpen
  \bibfield  {author} {\bibinfo {author} {\bibfnamefont {S.}~\bibnamefont {Bosco}}\ and\ \bibinfo {author} {\bibfnamefont {D.}~\bibnamefont {Loss}},\ }\bibfield  {title} {\bibinfo {title} {Hole spin qubits in thin curved quantum wells},\ }\href {https://doi.org/10.1103/PhysRevApplied.18.044038} {\bibfield  {journal} {\bibinfo  {journal} {Phys. Rev. Appl.}\ }\textbf {\bibinfo {volume} {18}},\ \bibinfo {pages} {044038} (\bibinfo {year} {2022})}\BibitemShut {NoStop}%
\bibitem [{\citenamefont {Malkoc}\ \emph {et~al.}(2022)\citenamefont {Malkoc}, \citenamefont {Stano},\ and\ \citenamefont {Loss}}]{Malkoc2022}%
  \BibitemOpen
  \bibfield  {author} {\bibinfo {author} {\bibfnamefont {O.}~\bibnamefont {Malkoc}}, \bibinfo {author} {\bibfnamefont {P.}~\bibnamefont {Stano}},\ and\ \bibinfo {author} {\bibfnamefont {D.}~\bibnamefont {Loss}},\ }\bibfield  {title} {\bibinfo {title} {Charge-noise-induced dephasing in silicon hole-spin qubits},\ }\href {https://doi.org/10.1103/PhysRevLett.129.247701} {\bibfield  {journal} {\bibinfo  {journal} {Phys. Rev. Lett.}\ }\textbf {\bibinfo {volume} {129}},\ \bibinfo {pages} {247701} (\bibinfo {year} {2022})}\BibitemShut {NoStop}%
\bibitem [{\citenamefont {Shnirman}\ \emph {et~al.}(2002)\citenamefont {Shnirman}, \citenamefont {Makhlin},\ and\ \citenamefont {Schön}}]{Shnirman2002}%
  \BibitemOpen
  \bibfield  {author} {\bibinfo {author} {\bibfnamefont {A.}~\bibnamefont {Shnirman}}, \bibinfo {author} {\bibfnamefont {Y.}~\bibnamefont {Makhlin}},\ and\ \bibinfo {author} {\bibfnamefont {G.}~\bibnamefont {Schön}},\ }\bibfield  {title} {\bibinfo {title} {Noise and decoherence in quantum two-level systems},\ }\href {https://doi.org/10.1238/Physica.Topical.102a00147} {\bibfield  {journal} {\bibinfo  {journal} {Physica Scripta}\ }\textbf {\bibinfo {volume} {2002}},\ \bibinfo {pages} {147} (\bibinfo {year} {2002})}\BibitemShut {NoStop}%
\bibitem [{\citenamefont {Makhlin}\ \emph {et~al.}(2004)\citenamefont {Makhlin}, \citenamefont {Schön},\ and\ \citenamefont {Shnirman}}]{Makhlin2004}%
  \BibitemOpen
  \bibfield  {author} {\bibinfo {author} {\bibfnamefont {Y.}~\bibnamefont {Makhlin}}, \bibinfo {author} {\bibfnamefont {G.}~\bibnamefont {Schön}},\ and\ \bibinfo {author} {\bibfnamefont {A.}~\bibnamefont {Shnirman}},\ }\bibfield  {title} {\bibinfo {title} {Dissipative effects in josephson qubits},\ }\href {https://doi.org/https://doi.org/10.1016/j.chemphys.2003.09.025} {\bibfield  {journal} {\bibinfo  {journal} {Chemical Physics}\ }\textbf {\bibinfo {volume} {296}},\ \bibinfo {pages} {315} (\bibinfo {year} {2004})},\ \bibinfo {note} {the Spin-Boson Problem: From Electron Transfer to Quantum Computing ... to the 60th Birthday of Professor Ulrich Weiss}\BibitemShut {NoStop}%
\bibitem [{\citenamefont {Mauro}\ \emph {et~al.}(2024{\natexlab{b}})\citenamefont {Mauro}, \citenamefont {Rodríguez-Mena}, \citenamefont {Martinez},\ and\ \citenamefont {Niquet}}]{Mauro2024arxiv}%
  \BibitemOpen
  \bibfield  {author} {\bibinfo {author} {\bibfnamefont {L.}~\bibnamefont {Mauro}}, \bibinfo {author} {\bibfnamefont {E.~A.}\ \bibnamefont {Rodríguez-Mena}}, \bibinfo {author} {\bibfnamefont {B.}~\bibnamefont {Martinez}},\ and\ \bibinfo {author} {\bibfnamefont {Y.-M.}\ \bibnamefont {Niquet}},\ }\href {https://arxiv.org/abs/2407.19854} {\bibinfo {title} {Strain engineering in ge/gesi spin qubits heterostructures}} (\bibinfo {year} {2024}{\natexlab{b}}),\ \Eprint {https://arxiv.org/abs/2407.19854} {arXiv:2407.19854 [cond-mat.mes-hall]} \BibitemShut {NoStop}%
\bibitem [{\citenamefont {Afonin}\ \emph {et~al.}(2002)\citenamefont {Afonin}, \citenamefont {Bergli}, \citenamefont {Galperin}, \citenamefont {Gurevich},\ and\ \citenamefont {Kozub}}]{Afonin2002}%
  \BibitemOpen
  \bibfield  {author} {\bibinfo {author} {\bibfnamefont {V.~V.}\ \bibnamefont {Afonin}}, \bibinfo {author} {\bibfnamefont {J.}~\bibnamefont {Bergli}}, \bibinfo {author} {\bibfnamefont {Y.~M.}\ \bibnamefont {Galperin}}, \bibinfo {author} {\bibfnamefont {V.~L.}\ \bibnamefont {Gurevich}},\ and\ \bibinfo {author} {\bibfnamefont {V.~I.}\ \bibnamefont {Kozub}},\ }\bibfield  {title} {\bibinfo {title} {Possible weak temperature dependence of electron dephasing},\ }\href {https://doi.org/10.1103/PhysRevB.66.165326} {\bibfield  {journal} {\bibinfo  {journal} {Phys. Rev. B}\ }\textbf {\bibinfo {volume} {66}},\ \bibinfo {pages} {165326} (\bibinfo {year} {2002})}\BibitemShut {NoStop}%
\bibitem [{\citenamefont {Pekola}\ \emph {et~al.}(2015)\citenamefont {Pekola}, \citenamefont {Masuyama}, \citenamefont {Nakamura}, \citenamefont {Bergli},\ and\ \citenamefont {Galperin}}]{Pekola2015}%
  \BibitemOpen
  \bibfield  {author} {\bibinfo {author} {\bibfnamefont {J.~P.}\ \bibnamefont {Pekola}}, \bibinfo {author} {\bibfnamefont {Y.}~\bibnamefont {Masuyama}}, \bibinfo {author} {\bibfnamefont {Y.}~\bibnamefont {Nakamura}}, \bibinfo {author} {\bibfnamefont {J.}~\bibnamefont {Bergli}},\ and\ \bibinfo {author} {\bibfnamefont {Y.~M.}\ \bibnamefont {Galperin}},\ }\bibfield  {title} {\bibinfo {title} {Dephasing and dissipation in qubit thermodynamics},\ }\href {https://doi.org/10.1103/PhysRevE.91.062109} {\bibfield  {journal} {\bibinfo  {journal} {Phys. Rev. E}\ }\textbf {\bibinfo {volume} {91}},\ \bibinfo {pages} {062109} (\bibinfo {year} {2015})}\BibitemShut {NoStop}%
\bibitem [{\citenamefont {Ziel}(1950)}]{VanDerZiel1950}%
  \BibitemOpen
  \bibfield  {author} {\bibinfo {author} {\bibfnamefont {A.~V.~D.}\ \bibnamefont {Ziel}},\ }\bibfield  {title} {\bibinfo {title} {On the noise spectra of semi-conductor noise and of flicker effect},\ }\href {https://doi.org/https://doi.org/10.1016/0031-8914(50)90078-4} {\bibfield  {journal} {\bibinfo  {journal} {Physica}\ }\textbf {\bibinfo {volume} {16}},\ \bibinfo {pages} {359} (\bibinfo {year} {1950})}\BibitemShut {NoStop}%
\bibitem [{\citenamefont {Landauer}\ and\ \citenamefont {Martin}(1994)}]{Landauer1994}%
  \BibitemOpen
  \bibfield  {author} {\bibinfo {author} {\bibfnamefont {R.}~\bibnamefont {Landauer}}\ and\ \bibinfo {author} {\bibfnamefont {T.}~\bibnamefont {Martin}},\ }\bibfield  {title} {\bibinfo {title} {Barrier interaction time in tunneling},\ }\href {https://doi.org/10.1103/RevModPhys.66.217} {\bibfield  {journal} {\bibinfo  {journal} {Rev. Mod. Phys.}\ }\textbf {\bibinfo {volume} {66}},\ \bibinfo {pages} {217} (\bibinfo {year} {1994})}\BibitemShut {NoStop}%
\bibitem [{\citenamefont {F{\'e}vrier}\ and\ \citenamefont {Gabelli}(2018)}]{Pierre2018}%
  \BibitemOpen
  \bibfield  {author} {\bibinfo {author} {\bibfnamefont {P.}~\bibnamefont {F{\'e}vrier}}\ and\ \bibinfo {author} {\bibfnamefont {J.}~\bibnamefont {Gabelli}},\ }\bibfield  {title} {\bibinfo {title} {Tunneling time probed by quantum shot noise},\ }\href {https://doi.org/10.1038/s41467-018-07369-6} {\bibfield  {journal} {\bibinfo  {journal} {Nature Communications}\ }\textbf {\bibinfo {volume} {9}},\ \bibinfo {pages} {4940} (\bibinfo {year} {2018})}\BibitemShut {NoStop}%
\bibitem [{\citenamefont {Machlup}(1954)}]{Machlup1954}%
  \BibitemOpen
  \bibfield  {author} {\bibinfo {author} {\bibfnamefont {S.}~\bibnamefont {Machlup}},\ }\bibfield  {title} {\bibinfo {title} {Noise in semiconductors: Spectrum of a two‐parameter random signal},\ }\href {https://doi.org/10.1063/1.1721637} {\bibfield  {journal} {\bibinfo  {journal} {Journal of Applied Physics}\ }\textbf {\bibinfo {volume} {25}},\ \bibinfo {pages} {341} (\bibinfo {year} {1954})},\ \Eprint {https://arxiv.org/abs/https://pubs.aip.org/aip/jap/article-pdf/25/3/341/18313317/341\_1\_online.pdf} {https://pubs.aip.org/aip/jap/article-pdf/25/3/341/18313317/341\_1\_online.pdf} \BibitemShut {NoStop}%
\bibitem [{\citenamefont {Petit}\ \emph {et~al.}(2018)\citenamefont {Petit}, \citenamefont {Boter}, \citenamefont {Eenink}, \citenamefont {Droulers}, \citenamefont {Tagliaferri}, \citenamefont {Li}, \citenamefont {Franke}, \citenamefont {Singh}, \citenamefont {Clarke}, \citenamefont {Schouten}, \citenamefont {Dobrovitski}, \citenamefont {Vandersypen},\ and\ \citenamefont {Veldhorst}}]{Petit2018}%
  \BibitemOpen
  \bibfield  {author} {\bibinfo {author} {\bibfnamefont {L.}~\bibnamefont {Petit}}, \bibinfo {author} {\bibfnamefont {J.~M.}\ \bibnamefont {Boter}}, \bibinfo {author} {\bibfnamefont {H.~G.~J.}\ \bibnamefont {Eenink}}, \bibinfo {author} {\bibfnamefont {G.}~\bibnamefont {Droulers}}, \bibinfo {author} {\bibfnamefont {M.~L.~V.}\ \bibnamefont {Tagliaferri}}, \bibinfo {author} {\bibfnamefont {R.}~\bibnamefont {Li}}, \bibinfo {author} {\bibfnamefont {D.~P.}\ \bibnamefont {Franke}}, \bibinfo {author} {\bibfnamefont {K.~J.}\ \bibnamefont {Singh}}, \bibinfo {author} {\bibfnamefont {J.~S.}\ \bibnamefont {Clarke}}, \bibinfo {author} {\bibfnamefont {R.~N.}\ \bibnamefont {Schouten}}, \bibinfo {author} {\bibfnamefont {V.~V.}\ \bibnamefont {Dobrovitski}}, \bibinfo {author} {\bibfnamefont {L.~M.~K.}\ \bibnamefont {Vandersypen}},\ and\ \bibinfo {author} {\bibfnamefont {M.}~\bibnamefont {Veldhorst}},\ }\bibfield  {title} {\bibinfo {title} {Spin lifetime and charge noise in hot silicon quantum dot qubits},\ }\href
  {https://doi.org/10.1103/PhysRevLett.121.076801} {\bibfield  {journal} {\bibinfo  {journal} {Phys. Rev. Lett.}\ }\textbf {\bibinfo {volume} {121}},\ \bibinfo {pages} {076801} (\bibinfo {year} {2018})}\BibitemShut {NoStop}%
\bibitem [{\citenamefont {Connors}\ \emph {et~al.}(2019)\citenamefont {Connors}, \citenamefont {Nelson}, \citenamefont {Qiao}, \citenamefont {Edge},\ and\ \citenamefont {Nichol}}]{Connors2019}%
  \BibitemOpen
  \bibfield  {author} {\bibinfo {author} {\bibfnamefont {E.~J.}\ \bibnamefont {Connors}}, \bibinfo {author} {\bibfnamefont {J.}~\bibnamefont {Nelson}}, \bibinfo {author} {\bibfnamefont {H.}~\bibnamefont {Qiao}}, \bibinfo {author} {\bibfnamefont {L.~F.}\ \bibnamefont {Edge}},\ and\ \bibinfo {author} {\bibfnamefont {J.~M.}\ \bibnamefont {Nichol}},\ }\bibfield  {title} {\bibinfo {title} {Low-frequency charge noise in si/sige quantum dots},\ }\href {https://doi.org/10.1103/PhysRevB.100.165305} {\bibfield  {journal} {\bibinfo  {journal} {Phys. Rev. B}\ }\textbf {\bibinfo {volume} {100}},\ \bibinfo {pages} {165305} (\bibinfo {year} {2019})}\BibitemShut {NoStop}%
\bibitem [{\citenamefont {Rojas-Arias}\ \emph {et~al.}(2023)\citenamefont {Rojas-Arias}, \citenamefont {Noiri}, \citenamefont {Stano}, \citenamefont {Nakajima}, \citenamefont {Yoneda}, \citenamefont {Takeda}, \citenamefont {Kobayashi}, \citenamefont {Sammak}, \citenamefont {Scappucci}, \citenamefont {Loss},\ and\ \citenamefont {Tarucha}}]{Rojas-Arias2023}%
  \BibitemOpen
  \bibfield  {author} {\bibinfo {author} {\bibfnamefont {J.}~\bibnamefont {Rojas-Arias}}, \bibinfo {author} {\bibfnamefont {A.}~\bibnamefont {Noiri}}, \bibinfo {author} {\bibfnamefont {P.}~\bibnamefont {Stano}}, \bibinfo {author} {\bibfnamefont {T.}~\bibnamefont {Nakajima}}, \bibinfo {author} {\bibfnamefont {J.}~\bibnamefont {Yoneda}}, \bibinfo {author} {\bibfnamefont {K.}~\bibnamefont {Takeda}}, \bibinfo {author} {\bibfnamefont {T.}~\bibnamefont {Kobayashi}}, \bibinfo {author} {\bibfnamefont {A.}~\bibnamefont {Sammak}}, \bibinfo {author} {\bibfnamefont {G.}~\bibnamefont {Scappucci}}, \bibinfo {author} {\bibfnamefont {D.}~\bibnamefont {Loss}},\ and\ \bibinfo {author} {\bibfnamefont {S.}~\bibnamefont {Tarucha}},\ }\bibfield  {title} {\bibinfo {title} {Spatial noise correlations beyond nearest neighbors in ${}^{28}\mathrm{Si}/$si-ge spin qubits},\ }\href {https://doi.org/10.1103/PhysRevApplied.20.054024} {\bibfield  {journal} {\bibinfo  {journal} {Phys. Rev. Appl.}\ }\textbf {\bibinfo {volume} {20}},\ \bibinfo
  {pages} {054024} (\bibinfo {year} {2023})}\BibitemShut {NoStop}%
\bibitem [{\citenamefont {de~Sousa}(2009)}]{deSousa2009}%
  \BibitemOpen
  \bibfield  {author} {\bibinfo {author} {\bibfnamefont {R.}~\bibnamefont {de~Sousa}},\ }\bibinfo {title} {Electron spin as a spectrometer of nuclear-spin noise and other fluctuations},\ in\ \href {https://doi.org/10.1007/978-3-540-79365-6_10} {\emph {\bibinfo {booktitle} {Electron Spin Resonance and Related Phenomena in Low-Dimensional Structures}}},\ \bibinfo {editor} {edited by\ \bibinfo {editor} {\bibfnamefont {M.}~\bibnamefont {Fanciulli}}}\ (\bibinfo  {publisher} {Springer Berlin Heidelberg},\ \bibinfo {address} {Berlin, Heidelberg},\ \bibinfo {year} {2009})\ pp.\ \bibinfo {pages} {183--220}\BibitemShut {NoStop}%
\bibitem [{\citenamefont {de~Sousa}\ and\ \citenamefont {Das~Sarma}(2003)}]{deSousa2003}%
  \BibitemOpen
  \bibfield  {author} {\bibinfo {author} {\bibfnamefont {R.}~\bibnamefont {de~Sousa}}\ and\ \bibinfo {author} {\bibfnamefont {S.}~\bibnamefont {Das~Sarma}},\ }\bibfield  {title} {\bibinfo {title} {Theory of nuclear-induced spectral diffusion: Spin decoherence of phosphorus donors in si and gaas quantum dots},\ }\href {https://doi.org/10.1103/PhysRevB.68.115322} {\bibfield  {journal} {\bibinfo  {journal} {Phys. Rev. B}\ }\textbf {\bibinfo {volume} {68}},\ \bibinfo {pages} {115322} (\bibinfo {year} {2003})}\BibitemShut {NoStop}%
\bibitem [{\citenamefont {Culcer}\ \emph {et~al.}(2009)\citenamefont {Culcer}, \citenamefont {Hu},\ and\ \citenamefont {Das~Sarma}}]{Culcer2009}%
  \BibitemOpen
  \bibfield  {author} {\bibinfo {author} {\bibfnamefont {D.}~\bibnamefont {Culcer}}, \bibinfo {author} {\bibfnamefont {X.}~\bibnamefont {Hu}},\ and\ \bibinfo {author} {\bibfnamefont {S.}~\bibnamefont {Das~Sarma}},\ }\bibfield  {title} {\bibinfo {title} {{Dephasing of Si spin qubits due to charge noise}},\ }\href {https://doi.org/10.1063/1.3194778} {\bibfield  {journal} {\bibinfo  {journal} {Applied Physics Letters}\ }\textbf {\bibinfo {volume} {95}},\ \bibinfo {pages} {073102} (\bibinfo {year} {2009})}\BibitemShut {NoStop}%
\bibitem [{\citenamefont {Corley-Wiciak}\ \emph {et~al.}(2023)\citenamefont {Corley-Wiciak}, \citenamefont {Richter}, \citenamefont {Zoellner}, \citenamefont {Zaitsev}, \citenamefont {Manganelli}, \citenamefont {Zatterin}, \citenamefont {Sch{\"u}lli}, \citenamefont {Corley-Wiciak}, \citenamefont {Katzer}, \citenamefont {Reichmann}, \citenamefont {Klesse}, \citenamefont {Hendrickx}, \citenamefont {Sammak}, \citenamefont {Veldhorst}, \citenamefont {Scappucci}, \citenamefont {Virgilio},\ and\ \citenamefont {Capellini}}]{Corley2023}%
  \BibitemOpen
  \bibfield  {author} {\bibinfo {author} {\bibfnamefont {C.}~\bibnamefont {Corley-Wiciak}}, \bibinfo {author} {\bibfnamefont {C.}~\bibnamefont {Richter}}, \bibinfo {author} {\bibfnamefont {M.~H.}\ \bibnamefont {Zoellner}}, \bibinfo {author} {\bibfnamefont {I.}~\bibnamefont {Zaitsev}}, \bibinfo {author} {\bibfnamefont {C.~L.}\ \bibnamefont {Manganelli}}, \bibinfo {author} {\bibfnamefont {E.}~\bibnamefont {Zatterin}}, \bibinfo {author} {\bibfnamefont {T.~U.}\ \bibnamefont {Sch{\"u}lli}}, \bibinfo {author} {\bibfnamefont {A.~A.}\ \bibnamefont {Corley-Wiciak}}, \bibinfo {author} {\bibfnamefont {J.}~\bibnamefont {Katzer}}, \bibinfo {author} {\bibfnamefont {F.}~\bibnamefont {Reichmann}}, \bibinfo {author} {\bibfnamefont {W.~M.}\ \bibnamefont {Klesse}}, \bibinfo {author} {\bibfnamefont {N.~W.}\ \bibnamefont {Hendrickx}}, \bibinfo {author} {\bibfnamefont {A.}~\bibnamefont {Sammak}}, \bibinfo {author} {\bibfnamefont {M.}~\bibnamefont {Veldhorst}}, \bibinfo {author} {\bibfnamefont {G.}~\bibnamefont {Scappucci}},
  \bibinfo {author} {\bibfnamefont {M.}~\bibnamefont {Virgilio}},\ and\ \bibinfo {author} {\bibfnamefont {G.}~\bibnamefont {Capellini}},\ }\bibfield  {title} {\bibinfo {title} {Nanoscale mapping of the 3d strain tensor in a germanium quantum well hosting a functional spin qubit device},\ }\href {https://doi.org/10.1021/acsami.2c17395} {\bibfield  {journal} {\bibinfo  {journal} {ACS Applied Materials \& Interfaces}\ }\textbf {\bibinfo {volume} {15}},\ \bibinfo {pages} {3119} (\bibinfo {year} {2023})}\BibitemShut {NoStop}%
\bibitem [{\citenamefont {Bosco}\ \emph {et~al.}(2021{\natexlab{b}})\citenamefont {Bosco}, \citenamefont {Benito}, \citenamefont {Adelsberger},\ and\ \citenamefont {Loss}}]{Bosco2021PRB}%
  \BibitemOpen
  \bibfield  {author} {\bibinfo {author} {\bibfnamefont {S.}~\bibnamefont {Bosco}}, \bibinfo {author} {\bibfnamefont {M.}~\bibnamefont {Benito}}, \bibinfo {author} {\bibfnamefont {C.}~\bibnamefont {Adelsberger}},\ and\ \bibinfo {author} {\bibfnamefont {D.}~\bibnamefont {Loss}},\ }\bibfield  {title} {\bibinfo {title} {Squeezed hole spin qubits in ge quantum dots with ultrafast gates at low power},\ }\href {https://doi.org/10.1103/PhysRevB.104.115425} {\bibfield  {journal} {\bibinfo  {journal} {Phys. Rev. B}\ }\textbf {\bibinfo {volume} {104}},\ \bibinfo {pages} {115425} (\bibinfo {year} {2021}{\natexlab{b}})}\BibitemShut {NoStop}%
\bibitem [{\citenamefont {Geyer}\ \emph {et~al.}(2024)\citenamefont {Geyer}, \citenamefont {Het{\'e}nyi}, \citenamefont {Bosco}, \citenamefont {Camenzind}, \citenamefont {Eggli}, \citenamefont {Fuhrer}, \citenamefont {Loss}, \citenamefont {Warburton}, \citenamefont {Zumb{\"u}hl},\ and\ \citenamefont {Kuhlmann}}]{Geyer2024}%
  \BibitemOpen
  \bibfield  {author} {\bibinfo {author} {\bibfnamefont {S.}~\bibnamefont {Geyer}}, \bibinfo {author} {\bibfnamefont {B.}~\bibnamefont {Het{\'e}nyi}}, \bibinfo {author} {\bibfnamefont {S.}~\bibnamefont {Bosco}}, \bibinfo {author} {\bibfnamefont {L.~C.}\ \bibnamefont {Camenzind}}, \bibinfo {author} {\bibfnamefont {R.~S.}\ \bibnamefont {Eggli}}, \bibinfo {author} {\bibfnamefont {A.}~\bibnamefont {Fuhrer}}, \bibinfo {author} {\bibfnamefont {D.}~\bibnamefont {Loss}}, \bibinfo {author} {\bibfnamefont {R.~J.}\ \bibnamefont {Warburton}}, \bibinfo {author} {\bibfnamefont {D.~M.}\ \bibnamefont {Zumb{\"u}hl}},\ and\ \bibinfo {author} {\bibfnamefont {A.~V.}\ \bibnamefont {Kuhlmann}},\ }\bibfield  {title} {\bibinfo {title} {Anisotropic exchange interaction of two hole-spin qubits},\ }\href {https://doi.org/10.1038/s41567-024-02481-5} {\bibfield  {journal} {\bibinfo  {journal} {Nature Physics}\ }\textbf {\bibinfo {volume} {20}},\ \bibinfo {pages} {1152} (\bibinfo {year} {2024})}\BibitemShut {NoStop}%
\bibitem [{\citenamefont {Mickelsen}\ \emph {et~al.}(2023)\citenamefont {Mickelsen}, \citenamefont {Carruzzo},\ and\ \citenamefont {Yu}}]{Mickelsen2023}%
  \BibitemOpen
  \bibfield  {author} {\bibinfo {author} {\bibfnamefont {D.~L.}\ \bibnamefont {Mickelsen}}, \bibinfo {author} {\bibfnamefont {H.~M.}\ \bibnamefont {Carruzzo}},\ and\ \bibinfo {author} {\bibfnamefont {C.~C.}\ \bibnamefont {Yu}},\ }\bibfield  {title} {\bibinfo {title} {Interacting two-level systems as a source of $1/f$ charge noise in quantum dot qubits},\ }\href {https://doi.org/10.1103/PhysRevB.108.195307} {\bibfield  {journal} {\bibinfo  {journal} {Phys. Rev. B}\ }\textbf {\bibinfo {volume} {108}},\ \bibinfo {pages} {195307} (\bibinfo {year} {2023})}\BibitemShut {NoStop}%
\bibitem [{\citenamefont {Mutter}\ and\ \citenamefont {Burkard}(2021)}]{Mutter2021}%
  \BibitemOpen
  \bibfield  {author} {\bibinfo {author} {\bibfnamefont {P.~M.}\ \bibnamefont {Mutter}}\ and\ \bibinfo {author} {\bibfnamefont {G.}~\bibnamefont {Burkard}},\ }\bibfield  {title} {\bibinfo {title} {Natural heavy-hole flopping mode qubit in germanium},\ }\href {https://doi.org/10.1103/PhysRevResearch.3.013194} {\bibfield  {journal} {\bibinfo  {journal} {Phys. Rev. Res.}\ }\textbf {\bibinfo {volume} {3}},\ \bibinfo {pages} {013194} (\bibinfo {year} {2021})}\BibitemShut {NoStop}%
\bibitem [{\citenamefont {Zwanenburg}\ \emph {et~al.}(2013)\citenamefont {Zwanenburg}, \citenamefont {Dzurak}, \citenamefont {Morello}, \citenamefont {Simmons}, \citenamefont {Hollenberg}, \citenamefont {Klimeck}, \citenamefont {Rogge}, \citenamefont {Coppersmith},\ and\ \citenamefont {Eriksson}}]{Zwanenburg2013}%
  \BibitemOpen
  \bibfield  {author} {\bibinfo {author} {\bibfnamefont {F.~A.}\ \bibnamefont {Zwanenburg}}, \bibinfo {author} {\bibfnamefont {A.~S.}\ \bibnamefont {Dzurak}}, \bibinfo {author} {\bibfnamefont {A.}~\bibnamefont {Morello}}, \bibinfo {author} {\bibfnamefont {M.~Y.}\ \bibnamefont {Simmons}}, \bibinfo {author} {\bibfnamefont {L.~C.~L.}\ \bibnamefont {Hollenberg}}, \bibinfo {author} {\bibfnamefont {G.}~\bibnamefont {Klimeck}}, \bibinfo {author} {\bibfnamefont {S.}~\bibnamefont {Rogge}}, \bibinfo {author} {\bibfnamefont {S.~N.}\ \bibnamefont {Coppersmith}},\ and\ \bibinfo {author} {\bibfnamefont {M.~A.}\ \bibnamefont {Eriksson}},\ }\bibfield  {title} {\bibinfo {title} {Silicon quantum electronics},\ }\href {https://doi.org/10.1103/RevModPhys.85.961} {\bibfield  {journal} {\bibinfo  {journal} {Rev. Mod. Phys.}\ }\textbf {\bibinfo {volume} {85}},\ \bibinfo {pages} {961} (\bibinfo {year} {2013})}\BibitemShut {NoStop}%
\end{thebibliography}

\end{document}